\documentclass[12pt]{article}%
\usepackage[nosort]{cite}
\usepackage[usenames, dvipsnames]{xcolor}
\usepackage{graphicx}
\usepackage{multicol}
\usepackage{amsfonts}
\usepackage{amssymb}
\usepackage{amsmath}
\usepackage{heck}
\usepackage{afterpage}
\usepackage{setspace}
\usepackage{verbatim}
\usepackage{color}
\usepackage{longtable}
\usepackage{float}
\usepackage{subcaption}
\usepackage{epsfig}
\usepackage{enumerate}
\usepackage{epstopdf}
\usepackage[enableskew, vcentermath]{youngtab}
\usepackage{adjustbox}
\usepackage{multirow}
\usepackage{tikz}
\usepackage[margin=1in]{geometry}
\usepackage{titletoc}
\usepackage[percent]{overpic}
\usepackage{gensymb}
\usepackage{mathtools}
\usepackage{tikz-cd}%
\providecommand{\U}[1]{\protect\rule{.1in}{.1in}}
\pdfoutput=1
\newsavebox{\mysavebox}

\usetikzlibrary{decorations.markings}

\numberwithin{equation}{section}

\usetikzlibrary{chains}
\allowdisplaybreaks
\tikzset{node distance=2em, ch/.style={circle,draw,on chain,inner sep=2pt},chj/.style={ch,join},every path/.style={shorten >=4pt,shorten <=4pt},line width=1pt,baseline=-1ex}

\newcommand{\ba}{\begin{eqnarray}}
\newcommand{\ea}{\end{eqnarray}}

\newcommand{\be}{\begin{equation}}
\newcommand{\ee}{\end{equation}}

\newcommand{\al}[1]{\begin{align}#1\end{align}}
\tikzstyle{startstop} = [rectangle, rounded corners, minimum width=3cm, minimum height=1cm,text centered, draw=black, fill=blue!10]
\tikzstyle{startstop} = [rectangle, rounded corners, minimum width=3cm, minimum height=1cm,text centered, draw=black, fill=blue!10]
\tikzstyle{io} = [trapezium, trapezium left angle=70, trapezium right angle=110, minimum width=3cm, minimum height=1cm, text centered, draw=black, fill=blue!30]
\tikzstyle{process} = [rectangle, minimum width=3cm, minimum height=1cm, text centered, draw=black, fill=orange!30]
\tikzstyle{decision} = [diamond, minimum width=3cm, minimum height=1cm, text centered, draw=black, fill=green!30]
\tikzstyle{arrow} = [thick,->,>=stealth]
\tikzset{->-/.style={decoration={
  markings,
  mark=at position #1 with {\arrow[scale=2.4]{>}}},postaction={decorate}}}
\makeatletter \@addtoreset{equation}{section} \makeatother

\newcommand{\p}{\partial}

\renewcommand{\[}{\left[}

\colorlet{darkblue}{blue!70!black}
\colorlet{darkgreen}{green!70!black}

\usepackage[colorlinks=true,urlcolor=darkblue,linktocpage=true,linkcolor=darkblue,citecolor=darkblue]{hyperref}
\hypersetup{
colorlinks=true,
linkcolor=blue,
citecolor=magenta,
}
\begin{document}


\date{July 2021}

\title{Reflections on the Matter of 3d $\mathcal{N} = 1$ Vacua\\[4mm] and Local $Spin(7)$ Compactifications}

\institution{PENN}{\centerline{${}^{1}$Department of Physics and Astronomy, University of Pennsylvania, Philadelphia, PA 19104, USA}}

\institution{PENNMATH}{\centerline{${}^{2}$Department of Mathematics, University of Pennsylvania, Philadelphia, PA 19104, USA}}

\institution{MARIBOR}{\centerline{${}^{3}$Center for Applied Mathematics and Theoretical Physics, University of Maribor, Maribor, Slovenia}}

\authors{
Mirjam Cveti\v{c}\worksat{\PENN, \PENNMATH, \MARIBOR}\footnote{e-mail: {\tt cvetic@physics.upenn.edu}},
Jonathan J. Heckman\worksat{\PENN, \PENNMATH}\footnote{e-mail: {\tt jheckman@sas.upenn.edu}},\\[4mm]
Ethan Torres\worksat{\PENN}\footnote{e-mail: {\tt emtorres@sas.upenn.edu}},
and Gianluca Zoccarato\worksat{\PENN}\footnote{e-mail: {\tt gzoc@sas.upenn.edu}}}

\abstract{We use Higgs bundles to study the 3d $\mathcal{N} = 1$ vacua obtained from
M-theory compactified on a local $Spin(7)$ space given as a four-manifold $M_4$ of ADE singularities
with further generic enhancements in the singularity type along one-dimensional subspaces.
There can be strong quantum corrections to the massless degrees of freedom in the low energy effective field theory,
but topologically robust quantities such as ``parity'' anomalies are still calculable.
We show how geometric reflections of the compactification space descend to 3d reflections
and discrete symmetries.  The ``parity'' anomalies of
the effective field theory descend from topological data of the compactification.
The geometric perspective also allows us to track various perturbative and
non-perturbative corrections to the 3d effective field theory.
We also provide some explicit constructions of well-known 3d theories,
including those which arise as edge modes of 4d topological insulators,
and 3d $\mathcal{N} = 1$ analogs of grand unified theories.
An additional result of our analysis is that we are able to track
the spectrum of extended objects and their transformations
under higher-form symmetries.}

{\small \texttt{\hfill UPR-1312-T}}

\maketitle

\setcounter{tocdepth}{2}

\tableofcontents


\newpage

\section{Introduction \label{sec:INTRO}}

Geometric engineering provides a promising way to recast
difficult questions in quantum field theory (QFT) in terms
of the geometry of extra dimensions in string theory.
One of the underlying themes in much of this progress has been the use of
holomorphic structures, and its close connection with supersymmetric QFT
in flat space.

But there are also many QFTs where we cannot rely on holomorphy considerations.
In general, this makes it difficult to study strong coupling
dynamics in such systems. Anomalies of symmetries provide a potentially promising
route for constraining strong coupling dynamics because they are among
the most robust ``topological'' aspects of a QFT.

From the perspective of string theory, it is natural to ask whether there is a geometric lift of these
``bottom up'' considerations which can be used to constrain the dynamics of string compactification geometries
with little or no supersymmetry. In the putative effective field theory, strong coupling effects
lead to modifications of the classical internal geometry. Conversely, by studying non-perturbative contributions
in a string compactification, one can hope to pinpoint the onset of strong coupling effects in a given QFT.

In this paper we study 3d $\mathcal{N} = 1$ vacua as obtained from
M-theory on a local $Spin(7)$ space. Such compactifications are closely
related to 4d $\mathcal{N} = 1$ vacua as obtained from M-theory on local $G_2$ spaces,
and F-theory on local elliptically fibered Calabi-Yau fourfolds,
where holomorphic structures play a more prominent role
\cite{Braun:2018joh, Cvetic:2020piw}. These systems are also potentially
of interest in the study of ``4d $\mathcal{N} = 1/2$'' F-theory based models of dark energy
(see e.g. \cite{Witten:1994cga, Vafa:1996xn, Heckman:2018mxl, Heckman:2019dsj}). For earlier
work on M-theory compactified on a $Spin(7)$ space, see e.g. \cite{Becker:2000jc, Cvetic:2001pga,
Cvetic:2001zx, Cvetic:2001ye, Cvetic:2001sr, Gukov:2001hf, Gukov:2002es, Gukov:2002zg}.

In particular, we focus on the local $Spin(7)$ space
defined by a four-manifold $M_4$ of ADE singularities, with further enhancement in the singularity type
along subspaces. As advocated in \cite{Heckman:2018mxl}, a fruitful way to analyze such geometries is via the
associated 7d gauge theory of a spacetime filling six-brane wrapping $M_4$. Local intersections
with other six-branes can be modelled in terms of background internal profiles for fields of the 7d gauge theory.
These background fields satisfy the Vafa-Witten equations on a non-K\"ahler four-manifold \cite{Vafa:1994tf},
and specify a stable Higgs bundle.

In these models, the field content descends from modes spread over the entirety of $M_4$, namely ``bulk modes,'' and
those which are trapped along a subspace, i.e. ``local matter''. Depending on the choice of $M_4$
and localization profile, one can envision engineering a rich class of possible 3d $\mathcal{N} = 1$
gauge theories coupled to matter. A general feature of these QFTs is that the localized matter actually
realizes a 3d $\mathcal{N} = 2$ sector,
and the bulk modes of the system can be packaged in terms of 3d $\mathcal{N} =1$ multiplets.
We can also use this geometric starting point to analyze the spectrum of extended objects, as obtained
from M2-branes and M5-branes wrapped on non-compact cycles of the geometry, as well as the resulting
higher-form symmetries acting on these objects.

Performing a general analysis of localized zero modes,
we show that they generically reside on codimension three subspaces inside of $M_4$. This is very much in line with what we would have gotten
from the circle reduction of M-theory on a local $G_2$ space, as obtained from a three-manifold $M_3$ of ADE singularities, where chiral matter is localized at points of the three-manifold (see e.g. \cite{Acharya:2001gy, Pantev:2009de, Braun:2018vhk, Barbosa:2019bgh}). It is, however, a bit surprising from the perspective of F-theory compactified on a local Calabi-Yau fourfold given by a K\"ahler surface of ADE singularities, where chiral matter is obtained from 6D hypermultiplets in the presence of a background gauge field
flux \cite{Donagi:2008ca, Beasley:2008dc}.\footnote{In F-theory models, this leads to zero modes which are localized in all four directions of the internal K\"ahler manifold. The reason this generically does not occur in the $Spin(7)$ setting is that there are three rather than two adjoint valued degrees of freedom in the Higgs field.}

We give some general methods of construction centered on building up four-manifolds $M_4$ from connected sums with summands $M^{(i)}_3 \times S^1$, where $M^{(i)}_3$ is a three-manifold. Local $G_2$ systems with matter at points of $M^{(i)}_{3}$ give rise to 4d $\mathcal{N} = 1$ matter, and this general gluing construction produces 3d $\mathcal{N} = 2$ matter coupled together by 3d $\mathcal{N} = 1$ ``bulk modes'' spread over all of $M_4$. A related construction involving gluing non-K\"ahler four-manifolds to K\"ahler manifolds with matter on holomorphic curves leads to 3d $\mathcal{N} = 4$ matter coupled via 3d $\mathcal{N} = 1$ matter. We also present a complementary quotient construction based on elliptically fibered Calabi-Yau fourfolds in an Appendix.

Provided one is interested in generating \textit{classical} 3d $\mathcal{N} = 1$ field theories, the geometry of
singular $Spin(7)$ spaces provides a promising way to engineer examples. In the flow to the deep infrared, however,
there can potentially be strong corrections to these classical results due to the absence of any constraint from holomorphy.
Indeed, this is just the string compactification analog of the standard difficulties faced in analyzing
3d $\mathcal{N} = 1$ QFTs!

Faced with these difficulties, we can ask whether robust quantities such as anomalies can constrain
possible quantum corrections. For 3d systems, this has proven to be a powerful way to study the phase structure
in the infrared, even in the absence of supersymmetry (see e.g. \cite{Aharony:2015mjs, Cordova:2017kue}).
From this perspective, it is tempting to just take the classical 3d system engineered from the large volume $Spin(7)$ geometry, and use this as a starting point for a purely 3d analysis. The difficulty here is that in addition to all of the classical zero modes, there is the entire spectrum of Kaluza-Klein states which need to be integrated out. To perform a proper analysis of such systems, we therefore need to track the higher-dimensional origin (when available) of these 3d symmetries.

Our aim here will be to focus on symmetries which can be analyzed
for any candidate 3d theory, namely the action of 3d spatial
reflections on physical fields, and its interplay with other
symmetry transformations coming from gauge symmetries or global symmetries.
Anomalies of these symmetries, including mixed gravitational-parity and gauge-parity anomalies
then provide non-perturbative control over some aspects of these systems.\footnote{These
are sometimes referred to as ``parity anomalies'' but as explained in \cite{Witten:2016cio},
it is more appropriate to view them as associated with various reflections.
In a reflection symmetric background, the latter must vanish but the former (when gravity is decoupled) can be non-zero.}
The corresponding anomalies are often calculable, and provide us with a
sharp tool in constraining the resulting 3d $\mathcal{N} = 1$ vacua obtained from local $Spin(7)$ spaces.
More precisely, we shall be interested in the action of reflections for the 3d theory in a
Euclidean spacetime:
\begin{equation}
\mathsf{R}^{3d}_i : x^i \rightarrow -x^i, \,\,\,\text{and} \,\,\, \mathsf{R}^{3d}_i : x^{j} \rightarrow x^{j} \,\,\,\text{for} \, j \neq i.
\end{equation}
and the corresponding transformations on our physical fields. This turns out to be a bit subtle, because
the physical content of our 3d system descends from a higher-dimensional starting point,
so reflections in 3d may end up being composed with other internal reflection symmetries.

With this in mind, we first track how reflection assignments of 7d SYM theory descend from
M-theory on an ADE singularity and 10d SYM theory on a $T^3$. Doing so, we show that the
7d reflection action on the physical fields is determined by a composition of spacetime and internal
reflections of the higher-dimensional theory. Similarly, in the 3d effective field theory,
the reflection assignments come about as compositions of reflections:
\begin{equation}
\mathsf{R}_{i}^{3d} = \mathsf{R}_{i}^{D} \underset{int}{\prod} \mathsf{R}_{int}^{D},
\end{equation}
for a higher-dimensional theory in $D$ spacetime dimensions, where
$\mathsf{R}_{i}^{D}$ denotes the reflection action on the
$D$-dimensional fields in the 3d spacetime direction, and
$\mathsf{R}_{int}^{D}$ is a reflection in the internal directions.
This leads to additional possible discrete symmetries which can
be arranged by tuning the moduli of the internal geometry.

The classical zero modes of a given local model each make contributions
to anomalies associated with 3d reflection symmetries which we can
explicitly evaluate. Strong coupling effects can potentially gap
the system or at least remove some of the candidate zero modes,
but a remnant of the ``topological order'' associated with these
anomalies will still persist.

We also develop some examples illustrating these general points.
As a preliminary example which we repeatedly return to throughout the paper,
we show that matter localized on a one-dimensional subspace of $M_4$, can, in the limit where $M_4$ is
decompactified, be understood as 4d matter with a position dependent mass, as in the standard
topological insulator construction (see e.g.
\cite{PhysRevLett.95.226801, PhysRevLett.96.106802, Seiberg:2016rsg}). As another general class of examples,
we show how to take a chiral 4d $\mathcal{N} = 1$ system engineered from the Pantev-Wijnholt (PW)
system compactified on a further $S^1$,
and glue it to its reflected image, resulting in a reflection symmetric 3d $\mathcal{N} = 1$ theory.
As an additional set of examples, we consider analogs of grand unified theories (GUTs) in three dimensions
engineered from related gluing constructions.

The rest of this paper is organized as follows. In section \ref{sec:Spin7Review} we discuss aspects of 7d Super Yang-Mills theory coupled to defects, and explain its relation to local $Spin(7)$ spaces. We show in section \ref{sec:NOLOCO} that localized matter fields in local $Spin(7)$ geometries generically lie on one-dimensional subspaces. In section \ref{sec:TOPOINS}
we show that local matter can be interpreted as edge modes of a 4d topological insulator. In section \ref{sec:FIRSTEXAMPLES} we
give some gluing constructions for how to produce 3d $\mathcal{N} = 1$ vacua starting from 3d $\mathcal{N} = 2$ building blocks.
We then turn to quantum effects, beginning in section \ref{sec:CRT} with a study of reflection symmetries in 7d SYM and their higher-dimensional origins. In section \ref{sec:4D3D} we track the resulting reflection assignments for 4d and 3d field theories obtained from
compactification of 7d SYM theory. We then turn to the computation of various parity anomalies
in section \ref{sec:ANOMO}. Section \ref{sec:QUANTUM} analyzes various
quantum corrections to the classical backgrounds. In section \ref{sec:EXAMPLES} we turn to some examples illustrating where we engineer various 3d $\mathcal{N} = 1$ field theories and compute the associated parity anomalies. Section \ref{sec:CONC} contains our conclusions and directions of further investigation.

We defer several additional technical items to the Appendices. In Appendix \ref{app:MOREZERO} we discuss some additional aspects of classical zero mode localization in local $Spin(7)$ systems. In Appendix \ref{app:QUOT} we demonstrate a new construction of $Spin(7)$ manifolds as a quotient of Calabi-Yau four-folds in a local patch that we conjecture extends to compact cases. Appendix \ref{app:SPINOR} reviews our various conventions for 10d and 7d spinors. In Appendix \ref{app:SQM} we provide details on a super quantum mechanics construction of Euclidean M2-branes in our local $Spin(7)$ models which illuminates the appearance of a twisted differential operator in the 3d $\mathcal{N}=1$ superpotential. Appendix \ref{app:CHARGEREF} covers the dimensional reduction of reflection transformations from 7d to 4d and 3d.  Appendix \ref{app:SO10PW} covers a $G_2$-spectral cover construction of 4d $\mathcal{N}=1$ $\mathfrak{so}(10)$ gauge theories, which, after dimensional reduction on a circle, is an example of a building block for constructing 3d $\mathcal{N}=1$ $Spin(7)$ systems in Section $\ref{sec:EXAMPLES}$. Finally, Appendix \ref{app:BORDISM} reviews the various 3d ``parity" anomalies discusses in the bulk of this paper and how they arise as the phase ambiguity of a 3d theory's partition function after placing it on an non-orientable $\mathsf{Pin}^+$ manifold.

\section{Higgs Bundle Approach to Local $Spin(7)$ Geometries \label{sec:Spin7Review}}

As mentioned in the Introduction, our interest in this paper is the study of 3d $\mathcal{N} = 1$ vacua
as engineered by M-theory on manifolds of $Spin(7)$ holonomy. In particular, we shall be interested in
spacetimes given by a (possibly warped) product of 3d Minkowski space with this internal geometry. Early work in this direction
primarily focused on the case of smooth $Spin(7)$ spaces, see e.g. \cite{Becker:2000jc, Gukov:2001hf, Gukov:2002zg}. Our general aim in this paper will be to study the comparatively less explored case of a four-manifold $M_4$ of ADE singularities which give rise to singular non-compact $Spin(7)$ spaces. Much as in \cite{Heckman:2018mxl}, our approach will be to analyze the local M-theory dynamics in terms of the worldvolume theory of 7d super Yang-Mills (SYM) theory filling the 3d spacetime and wrapping the four-manifold $M_4$.
This is just the Vafa-Witten twist of $\mathcal{N} = 4$ Super Yang-Mills theory on a four-manifold \cite{Vafa:1994tf}.

Recall that M-theory on the background $\mathbb{C}^2 / \Gamma$ an ADE singularity gives
rise to 7d Super Yang-Mills (SYM) theory with gauge group of ADE type. The bosonic content of this theory consists of a 7d gauge connection and three scalars $\phi_a$ which transform in the triplet representation of the $SU(2)$ R-symmetry. Some of this structure can be seen from the classical geometry of the ADE singularity. For example upon resolving the ADE singularity, we get a collection of $S^2$'s which intersect according to the corresponding ADE Dynkin diagram. Integrating the M-theory three-form potential over each such $S^2$ gives rise to a $U(1)$ vector potential. The ``off-diagonal'' components of the vector potential for the 7d SYM theory come from M2-branes wrapped on the other simple roots obtained from the homology lattice of the resolved space. Fibering this further over a four-manifold $M_4$ just amounts to considering 7d SYM wrapped on $M_4$. The assumption that this fibration is a local $Spin(7)$ space means we retain 3d $\mathcal{N} = 1$ supersymmetry in the transverse 3d Minkowski directions. In the 7d SYM theory, this is enforced by taking a topological twist of the theory so that we retain at least one real doublet of supercharges in the 3d theory.\footnote{More precisely, we start with the $SU(2)_R$ R-symmetry of the 7d theory in flat space, and consider a homomorphism $\mathfrak{su}(2)_R \rightarrow \mathfrak{so}(2,1) \oplus \mathfrak{so}(4)_{M_4}$ which is trivial on the $\mathfrak{so}(2,1)$ factor.} This turns out to be the same as the Vafa-Witten topological twist \cite{Vafa:1994tf} for 4d $\mathcal{N }=4$ Super Yang-Mills theory on a four-manifold $M_4$.

After the topological twist, the triplet of scalars assemble into adjoint-valued self-dual two-forms which we write as $\Phi_{\mathrm{SD}}$. The field content of the theory also includes the 7d gauge connection $A_{7d}$, as well as their fermionic superpartners. Observe that $\Lambda^{2}_{\mathrm{SD}} \rightarrow M_4$ the bundle of self-dual two-forms over $M_4$ gives rise to a local $G_2$ space (with a possibly incomplete metric). Indeed, we could also consider type IIA string theory on this seven-dimensional background, where we wrap D6-branes on $M_4$. The point is that we expect to arrive at the same 3d theory from either starting point. Because we have a local $G_2$ space, we know it comes equipped with a distinguished associative three-form. This also furnishes us with a ``cross-product'' operation on the self-dual two-forms, where we have:
\begin{equation}\label{CROSS}
(\Phi \times \Phi)_{a} = \varepsilon_{abc} \Phi_{b} \Phi_{c},
\end{equation}
where $\varepsilon_{abc}$ is the canonical three-index ``volume form'' of the fiber in $\Lambda^{2}_{\mathrm{SD}} \rightarrow M_4$. For adjoint-valued forms, the multiplication on the right-hand side is replaced by a commutator. This implies that the product of $\Phi$ with itself may be non-zero.

It is convenient to package the 7d fields in terms of suitable 3d $\mathcal{N} = 1$ supermultiplets, sorted according to their Lorentz quantum numbers in the 3d spacetime. For example, the 7d gauge connection decomposes as $A_{7d} \rightarrow A_{3d} \oplus A_{M_4}$, so from $A_{3d}$ we get a 3d $\mathcal{N} = 1$ vector multiplet, while from $A_{M_4}$ we get additional adjoint-valued scalar multiplets. By similar reasoning, the self-dual two-forms can be viewed as a collection of 3d $\mathcal{N} = 1$ scalar multiplets labelled by points of $M_4$.

The vacua of this compactified system are captured by the critical points of a corresponding 3d $\mathcal{N} = 1$ superpotential
\begin{equation}\label{W}
W_{3d}=\int_{M_4} \textnormal{Tr}\; \Phi_{\mathrm{SD}}\wedge (F + \frac{1}{3}\Phi_{\mathrm{SD}}\times \Phi_{\mathrm{SD}}),
\end{equation}
where $\Phi_{\mathrm{SD}} \times \Phi_{\mathrm{SD}}$ is shorthand for the cross product of equation (\ref{CROSS}),
$F$ denotes the superfield associated with the curvature of the field strength on $M_4$, and by abuse
of notation $\Phi_{\mathrm{SD}}$ also labels a superfield. The BPS equations of motion are:
\begin{equation}
F_{\mathrm{SD}} + \Phi_{\mathrm{SD}} \times \Phi_{\mathrm{SD}} = 0\,, \quad d_{A} \Phi_{\mathrm{SD}} = 0\,,
\end{equation}
and vacua are given by solutions to the BPS equations of motion modulo gauge transformations. In the above, $F_{\mathrm{SD}}$
is the self-dual part of the field strength, i.e. $F_{\mathrm{SD}} = \frac{1}{2} (F + \ast F)$. These equations define a stable
Higgs bundle on $M_4$.

From this starting point, we can also analyze backgrounds where the singularity type enhances along subspaces of $M_4$. In general terms, we can consider some ``parent gauge theory'' with gauge group $\widetilde{G}$, and take
a background solution which leaves some residual gauge symmetry unbroken.
We get massless matter fields in the 3d theory from the first order
fluctuations around this background:
\begin{align}
A & = a + \langle A \rangle \\
\Phi & = \varphi + \langle \Phi \rangle .
\end{align}
The matter fields will transform in representations of the unbroken gauge symmetry $H \subset \widetilde{G}$.

For ease of exposition, here we primarily focus on the case where the gauge connection is switched off, so in other words, we take
$\langle A \rangle = 0$, so that only $\langle \Phi \rangle$ has a non-zero background value. In this case, the BPS equations of motion enforce the condition that $\Phi_{\mathrm{SD}}$ takes values in the Cartan subalgebra of $\widetilde{G}$. We can speak of a collection of independent self-dual two-forms which satisfy the equation of motion $d \Phi_{\mathrm{SD}} = 0$. Now, to get a solution on $M_4$ which is globally valid, we sometimes need to allow for singularities along various subspaces. In physical terms, these are ``sources'' which are needed to satisfy a Gauss' law constraint on an otherwise compact space. For example, if we assume that all two-cycle periods of $\Phi_{\mathrm{SD}}$ are zero, there must be these defect sources (flavor brane intersections) as otherwise $\Phi_{\mathrm{SD}}=0$. In this case, the singularities we consider are of the form:
\begin{equation}
d\Phi_{\mathrm{SD}}=\sum_i v_i \delta_{L_i},
\end{equation}
where $L_{i}$ denotes a one-dimensional subspace of $M_4$, and $\delta_{L_i}$ denotes a delta function three-form
with support on $L_{i}$. Treating $\Phi_{\mathrm{SD}}$ as an abelian flux, one can see that Gauss' law requires
\begin{equation}
\sum_i v_i [L_i]=0\in H^1(M_4,\mathbb{R})
\end{equation}
to be satisfied. Notice that since we are free to change the orientations of the $L_i$'s, this condition implies that $\sum v_i=0 \; \textnormal{mod} \; 2$.

This background initiates a breaking pattern to a subgroup $G \times U(1)^{n} \subset \widetilde{G}$ (for now we shall be a bit cavalier with the global topology of the group). A zero mode in a particular representation of $G \times U(1)^n$ comes from decomposition of the adjoint representation of $\widetilde{G}$ into irreducible representations of $G \times U(1)^n$. Focussing on the simplest non-trivial case where we have a single $U(1)$ factor, and \textit{candidate} localized modes with charge $q$ with respect to this $U(1)$, we get the linearized approximation to the BPS equations of motion:
\begin{equation}\label{zmeqab}
(da)_{\mathrm{SD}} + q\Phi_{\mathrm{SD}}\times \varphi=0\,,
\end{equation}
\begin{equation}
d\varphi-q\Phi_{\mathrm{SD}}\wedge a=0\,.
\end{equation}
We can package this in terms of a twisted differential operator:
\begin{equation}
D_{q\Phi} \equiv \begin{pmatrix}
q\Phi_{\mathrm{SD}}\times & D_{sig.} \\
D_{sig.} & -q\Phi_{\mathrm{SD}}\wedge
\end{pmatrix}
\end{equation}
where $D_{sig.}=d+d^\dagger$ is the signature operator\footnote{Note that in four Euclidean dimensions $d^\dagger=-*d*$.}.
For additional discussion on the operator $D_{q\Phi}$ in the context of the associated supersymmetric quantum
mechanics theory, see Appendix \ref{app:SQM}. In terms of this, we can speak of a zero mode equation for a given representation:
\begin{equation}
D_{q \Phi} \Psi_{q} = 0,
\end{equation}
where $\Psi_q = (\varphi , a - \ast a)$ is a two-component vector with mixed form degree.

At a practical level, we can search for localized solutions by considering the rescaled limit $\Phi_{\mathrm{SD}} \rightarrow t \Phi_{\mathrm{SD}}$, and taking $t$ to be large. In this case, our candidate zero modes amount to those loci where $\Phi_{\mathrm{SD}}$ has a zero. From the point of view of a spectral cover construction, these zeros represent pairwise intersections of sheets. For example, let $\widetilde{G}=SU(N)$ and fix a
section $v\in\Lambda^2_{\mathrm{SD}}$.
The spectral equation in the fundamental representation is then:
\begin{equation}
\textnormal{det}(v\mathsf{1}_{N\times N}-\Phi_{\mathrm{SD}})=0
\end{equation}
which defines a four-manifold inside the bundle $\Lambda^2_{\mathrm{SD}}$ which is a finite cover of $M_4$. This is of course directly related to the IIA dual picture of the same system which is a local $G_2$ model on the total space of $\Lambda^2_{\mathrm{SD}}$ with a collection of intersecting D6-branes supported along the spectral cover.

An important caveat here is that zeros of the Higgs field are really just \textit{candidate} zero modes, and in principle, these solutions can receive small mass terms due to Euclidean M2-branes which stretch between pairs of these loci. This is very analogous to what happens in the Pantev-Wijnholt (PW) system \cite{Pantev:2009de}. In the rescaling limit $\Phi_{\mathrm{SD}} \rightarrow t \Phi_{\mathrm{SD}}$, we can see such corrections by performing an expansion in $1 / t$.

With this caveats in mind, let us now turn to the counting of both bulk zero modes and localized zero modes with respect to the background Higgs field profile. Consider first the zero modes with charge $q = 0 $. In this case, we find $b^2_{\mathrm{SD}} + b^1 + 1$ zero modes for each representation uncharged under $U(1)$ (we assume the vector bundle is switched off), where the contribution from $1$ is just the contribution from the 3d $\mathcal{N} = 1$ vector multiplet. These modes will not be localized by the profile of $\Phi_{\mathrm{SD}}$. Even if these modes are massless at the classical level in the 3d effective field theory the actual zero mode count can potentially be different, due to some of these modes pairing up or additional strong coupling effects. Later on, we will show that there is an anomaly capable of ``detecting'' a contribution from $\vert b^2_{\mathrm{SD}} - b^1 + 1 \vert$ such zero modes.

Next, consider the zero modes with charge $q \neq 0$. One issue we encounter is that as far as we are aware,
the counting of zero modes does not cleanly reduce to a cohomology group calculation. One simpler problem is to look for the zero locus of $\Phi_{\mathrm{SD}}$ given that zero modes localize around this locus in the limit of large vev of $\Phi_{\mathrm{SD}}$. The zero locus of $\Phi_{\mathrm{SD}}$ can be characterized quite nicely as comprising a set of $N_{\Phi}$ circles $\ell_m$ where the integer $N_\Phi$ is congruent to $(b^+_2-b_1+1) \; \textnormal{mod} \; 2$ \cite{perutz2006zero}. As in the case of delocalized zero modes, the mod 2 reduction of $N_\Phi$ will be related to the presence of an anomaly in the 3d effective field theory that will be discussed later in the paper. Another useful property is that the zero locus of $\Phi_{\mathrm{SD}}$ is homologically trivial as it is dual to the Euler class of $\Lambda_{\mathrm{SD}}^2$, which is zero \cite{perutz2006zero}. Around any zero of $\Phi_{\mathrm{SD}}$ we can model its profile locally as
\begin{equation}\label{Spin(7)PW}
\Phi_{\mathrm{SD}}|_{\ell_m}=df_m\wedge dt_m+*_3 df_m\,.
\end{equation}
Here $t_m$ is a local coordinate on the circle $\ell_m$, $f_m$ is a function that does not depend on $t_m$ and the Hodge star operation $*_3$ is taken in the directions normal to $\ell_m$ inside $M_4$. The fact that the profile of $\Phi_{\mathrm{SD}}$ can be written in terms of some function on a three-manifold is reminiscent of the PW system \cite{Pantev:2009de}. The localized mode will actually behave as a 4d mode on $\mathbb R^{1,2} \times \ell_m$ with chirality determined by the sign of the determinant of the Hessian of the function $f_m$, much as in reference \cite{Pantev:2009de}.

The leading order interaction terms for these candidate zero modes are obtained by expanding our 3d superpotential
of equation (\ref{W}) about a fixed background. This leads to the interaction terms for candidate zero modes:
\begin{equation}
W_{3d}= \int_{M_4} \sum_{q_i+q_j+q_k=0}\big(\varphi_{i}\wedge (a_{j}\wedge a_{k})  + \varphi_{i}\wedge (\varphi_{j}\times \varphi_{k})\big) + ...
\end{equation}
where the ``...'' refers to contributions which include both mass terms and additional compactification effects which are suppressed
in the large volume limit.

It is also of interest to consider \textit{massive} modes, namely those for which $D_{q \Phi} \Psi_q = m \Psi_q$, which, before twisting, derives from the usual 4d massive Dirac equation written as (turning off the scalars associated to $\Phi_{\mathrm{SD}}$ for simplicity)
\begin{equation}
\begin{pmatrix}
0 & \slashed{D} \\
\slashed{D} & 0
\end{pmatrix}
\begin{pmatrix}
\psi_R \\
\psi_L
\end{pmatrix}=
m\begin{pmatrix}
\psi_R \\
\psi_L
\end{pmatrix}
\end{equation}
where here $\psi_R$ and $\psi_L$ are right/left-handed Weyl fermions
on $M_4$ which are the fermionic components of \textit{massive} fluctuations.
The important point for us is that this Dirac equation
explicitly references the \textit{sign} of a real mass, $m$.
In the context of reducing to a 3d theory, integrating out positive
mass and negative mass terms can shift the Chern-Simons level(s) of
the 3d gauge theory with a contribution which depends on
the sign of the mass term.\footnote{For a related discussion on
the relation between 4d anomalies and their circle compactification
in the context of M- / F-theory duality, see e.g. reference
\cite{Grimm:2011fx, Cvetic:2012xn, Corvilain:2017luj}.}

\subsection{Defects and Higher-Form Symmetries}

Although our main emphasis will be on various aspects of reflection symmetries,
it is worthwhile to also discuss how our considerations fit with more global structures
of the resulting 3d effective field theory, and in particular various
higher-form symmetries which might be present. See e.g. references \cite{Gaiotto:2014kfa,
DelZotto:2015isa, Eckhard:2019jgg, Morrison:2020ool, Albertini:2020mdx,
Closset:2020scj, DelZotto:2020esg, Bhardwaj:2020phs, Buican:2021xhs} for additional
details on aspects of higher-form symmetries
and their relation to string compactification.

For starters, we can ask about the global nature of the 7d SYM gauge group. Indeed, a priori
we could have a simply connected $G$ or a quotient by some subgroup $C_G \subset Z_G$,
with $Z_G$ the center of $G$. Geometrically, the center $\mathcal{C} = \Lambda^{\ast} / \Lambda$,
where $\Lambda = H_{2}^{cpct}(\widetilde{\mathbb{C}^{2}/\Gamma},\mathbb{Z})$ is just the
homology group for the resolution of the ADE singularity $\mathbb{C}^2 / \Gamma$, and $\Lambda^{\ast}$ is the dual lattice.
Indeed, we can consider M2-branes stretched on the non-compact two-cycles
of $\Lambda^{\ast}$, and these descend to non-dynamical Wilson lines.
This is basically just the computation of a ``defect group'' as in references
\cite{Tachikawa:2013hya, DelZotto:2015isa} (see also \cite{Albertini:2020mdx, Cvetic:2021sxm}).
The specific choice of gauge group is then dictated by boundary conditions for discrete fluxes on a bounding $S^3 / \Gamma$ at infinity in $\mathbb{C}^2 / \Gamma$, and the analysis of section 5 in
reference \cite{DelZotto:2015isa} can be re-purposed to recover the different possible options.

Closely related to this is the spectrum of extended objects generated by M5-branes wrapping these same non-compact two-cycles in the fiber direction. For example, we can consider wrapping an M5-brane over all of $M_4$ and a non-compact two-cycle of $\Lambda^{\ast}$, and
this gives rise to objects which are charged under a magnetic zero-form symmetry in three dimensions.
If we have a two-cycle $\Sigma \in H_{2}(M_4, \mathbb{Z})$ (neglecting torsion),
then we get a corresponding string-like defect i.e. domain wall from also wrapping on a non-compact two-cycle of $\Lambda^{\ast}$.
In 3d field theory terms, this is an object which is charged under a magnetic two-form symmetry. The associated extended
objects then have charges labelled by a maximal isotropic sublattice of $H_{2}(M_4 , \mathcal{C})$ (see \cite{DelZotto:2015isa}).

In actual local models, of course, we typically have a more complicated configuration of intersecting six-branes. For example,
with local matter we often speak of a parent gauge group $\widetilde{G}$ and its unfolding to a descendant $H \subset \widetilde{G}$.
Moreover, this local enhancement can be different at different locations of $M_4$. Based on this, we need to be able to globally fit together all the different possible choices. While in general this is a challenging problem, the key point is that in spectral cover constructions
where we start with a single $\widetilde{G}$, we already know that the appropriate notion of $\Lambda$ and $\Lambda^{\ast}$
is given by $\Lambda = H_{2}^{cpct}(\widetilde{\mathbb{C}^{2}/\Gamma_{\widetilde{G}}},\mathbb{Z})$. For example,
if we consider adjoint breaking of $E_8 \supset E_6 \times SU(3) / \mathbb{Z}_3 \supset E_6 \times S(U(1)^3) / \mathbb{Z}_3 $,
we know that we have matter in the $\mathbf{27}$ of $E_6$, so the gauge group cannot be $E_6 / \mathbb{Z}_3$.
The point is that we also need to account for the contribution from the flavor-branes of the model. Indeed, since the center of $E_8$ is trivial (since its Cartan matrix is unimodular), the resulting defect group(s) associated with $\Lambda^{\ast} / \Lambda$ are all trivial. That being said, we can of course consider other starting points for our $\widetilde{G}$ which have a non-trivial center. In such situations, we can use the geometry to quickly ascertain the candidate higher-form symmetries in the 3d effective field theory. A final comment is that there can potentially be an additional contribution from the non-compact four-cycle Poincar\'{e} dual to $M_4$ itself. It would be interesting to study this universal contribution, but we defer this to future work.

\section{Ultra-Local $Spin(7)$ Matter Fields\label{sec:NOLOCO}}

In the previous section we discussed in general terms the use of Higgs bundles on $M_4$ as a tool in understanding localized matter in $Spin(7)$ compactifications of M-theory. Our aim in this section will be study in more detail the profile of these zero mode solutions. In particular, we would like to understand to what extent these zero modes can actually be localized.

As is common in the string compactification literature, there are distinct notions of localization. First we can work in an ``ultra-local'' patch of the four-manifold diffeomorphic to $\mathbb{R}^4$. We can also work in the context of a local model where $M_4$ is compact (up to deleting various subspaces to satisfy Gauss' law constraints), but where the $Spin(7)$ geometry is non-compact. Finally, we can consider what happens when we take a compact $Spin(7)$ geometry, in which case the 3d effective field theory will be coupled to gravity.

In this section we study the ultra-local properties of localized matter fields in $Spin(7)$ spaces. Our analysis in this section will be purely classical, since we neglect all quantum corrections coming from dynamics of the 3d effective field theory (and its lift to non-perturbative instanton corrections to the classical $Spin(7)$ geometry). Later, we will analyze some features which are robust against such corrections.
Throughout, for ease of exposition we assume that the metric on our patch of $M_4$ is just the standard flat space metric on $\mathbb{R}^4$.

To frame the discussion to follow, recall that the local $Spin(7)$ equations of motion amount to a hybrid of the Pantev-Wijnholt (PW) and Beasley-Heckman-Vafa (BHV) Higgs bundle systems \cite{Cvetic:2020piw}, so we expect that the profile of localized zero modes will share some similarities with these cases. Starting from our local $Spin(7)$ system on $M_3 \times S^1$ for $M_3$ a three-manifold, we reach the PW system by assuming all fields are constant along the $S^1$ direction, and contracting $\Phi_{\mathrm{SD}}$ with the one-form on $S^1$, and also dropping the contribution from the gauge field along the circle. We can reach the BHV system by specializing $M_4$ to be a K\"ahler surface, and in this case, $\Phi_{\mathrm{SD}}$ splits up as an adjoint-valued $(2,0)$-form, as well as a $(1,1)$-form proportional to the K\"ahler form which decouples from the rest of the system. We reference the Higgs fields for these special cases as $\Phi_{\mathrm{PW}}$ and $\Phi_{\mathrm{BHV}}$ in what follows.

Now, in the context of local $G_2$ compactifications, chiral matter is localized along codimension seven subspaces, namely at points \cite{Acharya:2001gy, Acharya:2002vs}. This can also be seen by a direct analysis of the corresponding local Higgs bundle, and the vanishing locus of the PW Higgs field \cite{Pantev:2009de, Braun:2018vhk, Barbosa:2019bgh}. In that case, the Higgs field is an adjoint-valued one-form on a three-manifold, and since we have three distinct scalar degrees of freedom, we expect to generically localize matter along a codimension three subspace, that is, a point. Observe that this is precisely what we also expect in the context of local $Spin(7)$ systems on a four-manifold $M_4$. Indeed, in that case as well, we have three independent components of $\Phi_{\mathrm{SD}}$, and this generates a codimension three subspace inside $M_4$, namely a distinguished set of one-dimensional subspaces.

In the case of compactification on elliptically fibered Calabi-Yau fourfolds, localization of matter actually descends from two separate sources. From geometry, we get localization along complex codimension three subspaces. This is in accord with the fact that in the BHV system on a K\"ahler surface $S$, we have a single $(2,0)$-form, and the zeroes of this cut out complex codimension one subspaces. Further localization is possible once we include the effects of flux \cite{Beasley:2008dc, Beasley:2008kw, Heckman:2008qa, Font:2012wq, Font:2013ida}. Generically, we get an equation of motion on a complex curve of the form $\overline{\partial}_{A} \psi = 0$, and so when the curvature associated with the gauge connection is non-trivial, we see a further localization to a real codimension four subspace in $S$.\footnote{We comment here that although this localization appears to be gauge dependent, this is an artifact of only working in a local patch. On a compact manifold, there are peaks and valleys to the associated zero mode, which are determined by the holonomy of the gauge connection.} In the context of local $Spin(7)$ systems, we are of course free to consider the special case of BHV backgrounds, and so in this sense we can cut out a real codimension two subspace in $M_4$ with the Higgs field, and then introduce a suitable gauge field flux to produce further localization at a point.

Comparing the generic expectations from the PW and BHV systems, we see that the former would appear to predict matter localization along codimension seven subspaces, while the BHV system would appear to predict matter localization along codimension eight subspaces. What then, should we expect in the full fledged local $Spin(7)$ system?

Our general claim is that in this setting, matter is generically localized along codimension seven subspaces \textit{even when gauge field flux is switched on}. This is in accord with what we expect from the PW system, but appears to be in tension with expectations from the BHV system. The main idea is that if we again attempt to analyze matter localization by first considering the zeros of the Higgs field, we get a codimension three subspace inside the four-manifold $M_4$. Now, if we attempt to further localize by switching on a gauge field flux, we face the awkward situation that we are attempting to pullback this flux onto the matter one-cycle. The best that can happen is that we can pullback the gauge field to a holonomy on the matter one-cycle, but this does not produce any further localization in the zero mode, a fact which can be explicitly checked by analyzing the 1d ``Dirac equation'' on this subspace.\footnote{The effect of the gauge field background is to produce a non-trivial holonomy around the circle where the mode is localized, and unless the component of the gauge field along the circle $A\in 2 \pi \mathbb Z$ the mode is lifted with mass of order $\left \lfloor{\frac{1}{2\pi}\iota^*_{\ell_i}A} \right \rfloor$.} The perhaps counterintuitive conclusion is that inside $M_4$, generic backgrounds will lead to localization along one-cycles.

Our plan in the rest of this section will be to explicitly illustrate how matter localization works in this setting. For our purposes, it suffices to work in a local patch of $M_4$ which is diffeomorphic to $\mathbb{R}^4$. We first present the explicit local equations of motion, and then turn to the special cases of localized matter in BHV and PW systems, illustrating that merging these solutions involves a clash in the choice of complex structure, which in turn obstructs further localization in the zero modes. We then turn to more general local $Spin(7)$ backgrounds.
Additional technical details are given in Appendix \ref{app:MOREZERO}.

\subsection{Localization in a Patch}

Before delving into the details of the examples, let us write down the system of zero mode equations with respect to a local coordinate system
$(x^1, x^2, x^3, x^4) = (x,y,u,v)$. In the following the background fields are $A = \{A_1,A_2,A_3,A_4\}$ and $\Phi = \{\Phi_1,\Phi_2,\Phi_3\}$. The fluctuations are $a = \{a_1,a_2,a_3,a_4\}$ and $\varphi = \{\varphi_1,\varphi_2,\varphi_3\}$, where the subscript on $\Phi$ and $\varphi$ refers to the directions in the bundle of self-dual two-forms.\footnote{One convenient basis of self-dual two-forms on $\mathbb{R}^4$ are the 't Hooft matrices $\eta^{i}_{\mu \nu}$.}

The zero mode equations for the Spin(7) system are:
\al{D_1 a_4 - D_4 a_1 +D_2 a_3 - D_3 a_2 -[\Phi_2,\varphi_3]+[\Phi_3,\varphi_2]&=0\,,\\
D_1 a_2 - D_2 a_1 + D_3 a_4 - D_4 a_3 -[\Phi_1,\varphi_2]+[\Phi_2,\varphi_1]&=0\,,\\
D_1 a_3 - D_3 a_1 + D_4 a_2 - D_2 a_4 -[\Phi_1,\varphi_3]+[\Phi_3,\varphi_1]&=0\,,\\
D_1 \varphi_1 - [\Phi_1,a_1] + D_2 \varphi_2 - [\Phi_2,a_2] +D_3 \varphi_3 -[\Phi_3,a_3]&=0\,,\\
D_4 \varphi_3 -[\Phi_3,a_4] + D_1 \varphi_2 - [\Phi_2,a_1]-D_2 \varphi_1 +[\Phi_1,a_2]&=0\,,\\
D_1 \varphi_3 -[\Phi_3,a_1]-D_4 \varphi_2+[\Phi_2,a_4]-D_3 \varphi_1+[\Phi_1,a_3]&=0\,,\\
D_2 \varphi_3-[\Phi_3,a_2]-D_3 \varphi_2 + [\Phi_2,a_3]+D_4 \varphi_1-[\Phi_1,a_4]&=0\,.
}
In the examples we take a gauge theory with Lie algebra $\mathfrak{su}(2)$ and align the background fields in the Cartan $P = i/2 \sigma_3$. The fluctuations will be in the generator $Q = i/2 (\sigma_1 + i \sigma_2)$ of the complexified Lie algebra. In the following we will omit matrices for the sake of simplicity. Our aim will be to study localization properties of zero modes by looking at some examples.

Let us now expand on the general expectation that we do not expect to find full localization in codimension four but will only be able to generically achieve localization in codimension three. To this end, we separate the localization effect provided by the Higgs field and by the gauge flux. In order to do so we can take the limit of large value for $\Phi$ via the rescaling $\Phi\rightarrow t\Phi$. One can solve the equations to zeroth-order in $1/t$ by first supposing that $(\varphi,a)$ is a PW solution localized on the line (actually a circle so we identify $x_4\sim x_4+2\pi R$) $x_1=x_2=x_3=0$ but with possible extra dependence on the $x_4$ direction. In other words our ansatz will be $(\varphi,a)=(f_1(x_4)\varphi_{\mathrm{PW}}(x_i), f_2(x_4)a_{\mathrm{PW}}(x_i))$ for $1\leq i \leq 3$. The equations above then become
\begin{align}\label{scalinglim}
&iA_1 a_4-D_4a_1+iA_2 a_3-iA_3 a_2=0\,,\\
&iA_1a_2-iA_2a_1+iA_3a_4-D_4a_3=0\,,\\
&iA_1a_3-iA_3a_1+D_4a_2-iA_2a_4=0\,,\\
&iA_1\varphi_1+iA_2\varphi_2+iA_3\varphi_3=0\,,\\
&D_4\varphi_3+iA_1\varphi_2-iA_2\varphi_1=0\,,\\
&iA_1\varphi_3-D_4\varphi_2-iA_3\varphi_1=0\,,\\
&iA_2\varphi_3-iA_3\varphi_2+D_4\varphi_1=0\,.
\end{align}
If we now undo the scaling of $\Phi$ by a coordinate rescaling, then since $F$ is the same form degree of $\Phi$ we get instead $F\rightarrow \frac{1}{t}F$. This means that in equation (\ref{scalinglim}), it is possible to gauge away $A_1$, $A_2$, and $A_3$ in the $t\rightarrow \infty$ limit since the connection is becoming flat. Hence, $f_1=f_2=Ce^{-iA_4x_4}$ for some constant (vector) $C$. Effectively this means that the effect of the gauge background is to produce a non-trivial holonomy around the circle where the mode is localized, but this does not produce any further localization.

We will now discuss localization for the BHV and PW systems separately and then try to combine both and see that no fully localized mode exists.

\subsection{BHV versus PW Localization}

We now turn to some examples of matter localization in BHV and PW systems, and in particular, illustrate some of the difficulties in simply combining these solutions to produce further localization in $Spin(7)$ backgrounds.

\subsubsection{BHV Localization}

To discuss the BHV mode, let us consider a solution on $\mathbb{R}^4$ with complex coordinates $x+iy$ and $u+iv$ with a K\"ahler form $-\frac{i}{2}(dx\wedge dy+du\wedge dv$). We can parameterize a general abelian background as
\al{ A_1 = N y\,, \ A_2 = - N x\,, \ A_3 = - N v\,, \ A_4 = N u\,, \ \Phi_1 = -  \mu x\,, \ \Phi_2 = \mu y\,, \ \Phi_3 = 0\,.
}
This background admits a fully localized mode. The solution is
\al{ a_1 &= -\left( N + \sqrt{N^2 + \mu^2} \right)\,e^{-\frac{1}{2} N (u^2 + v^2) -\frac{1}{2} (x^2 + y^2 ) \sqrt{N^2 + \mu^2}}\,,\\
 a_2 &= i \left( N + \sqrt{N^2 + \mu^2} \right)\,e^{-\frac{1}{2} N (u^2 + v^2) -\frac{1}{2} (x^2 + y^2 ) \sqrt{N^2 + \mu^2}}\,,\\
  \varphi_1 &= i \mu \,e^{-\frac{1}{2} N (u^2 + v^2) -\frac{1}{2} (x^2 + y^2 ) \sqrt{N^2 + \mu^2}}\,,\\
    \varphi_2 &=- \mu \,e^{-\frac{1}{2} N (u^2 + v^2) -\frac{1}{2} (x^2 + y^2 ) \sqrt{N^2 + \mu^2}}\,,
}
with all the remaining components set to zero. Notice in particular the component $a_1 + i a_2 $ is localized and so is $\varphi_1 - i \varphi_2$. This is a familiar situation in F-theory models  \cite{Beasley:2008dc, Beasley:2008kw, Heckman:2008qa, Font:2012wq, Font:2013ida} where a 6d (in our case 5d) hypermultiplet can be first localized on a complex matter curve and full localization can be provided by threading an abelian flux through the curve. Here the matter curve is located at $x=y=0$.

\subsubsection{PW Localization}

Let us now turn to localization of PW matter in our local $Spin(7)$ system.
We isolate one of the four directions in our local patch by writing: $\mathbb{R}^4=\mathbb{R}^3\times \mathbb{R}$.
As an example, consider the abelian background:
\al{ A_1 = 0\,, \ A_2 =  0\,, \ A_3 = 0\,, \ A_4 =0\,,  \ \Phi_1 = -\lambda x\,, \ \Phi_2 = -\lambda y\,, \ \Phi_3 = 2 \lambda u\,.
}
This admits a localized mode in codimension three of the form
\al{ a_3 = e^{- \frac{1}{2}\lambda \left(x^2 + y^2 + 2 u^2 \right)}\,, \quad \varphi_3 = -i e^{- \frac{1}{2}\lambda \left(x^2 + y^2 + 2 u^2 \right)}\,,
}
with all the remaining components set to zero. Notice in particular that this background localizes the component $a_3 + i \varphi_3$.
This background realizes a chiral multiplet along the matter line $x=y=u=0$.

\subsubsection{Obstructions to Further Localization}

One issue when comparing the two localized set of modes in general is the following: the complex structures of the BHV and PW localized modes seem to be incompatible. This is a stumbling block to further localization. More specifically, BHV tends to localize modes of the form\footnote{Here BHV is intended to be identified with a solution of the local $Spin(7)$ equations
such that $\Phi_3 = 0$. Other choices are equivalent to this one.} $a_1 \pm i a_2$, $a_3 \pm i a_3 $ and $\varphi_1 \mp i \varphi_2$ (the upper sign refers to a 4d chiral mode and the lower one to a 4d anti-chiral mode). By this we mean that if for example $a_1 + i a_2$ is localized then it must happen that $a_2 = - i a_1$. On the other hand PW tends to localize one component of the fluctuation of the gauge field and its associated fluctuation of the Higgs field.\footnote{Here PW is a solution that is invariant in one direction $t$ where $t$ can be any of the coordinates on the four-manifold. Given a gauge field $A_i$ the associated component of the Higgs field is $\iota_t \iota_ i \Phi_{\text{SD}}$.} This means that the PW Higgs field localizes only one component of the gauge field and one component of the Higgs field, whereas BHV needs either two components of the gauge field or two components of the Higgs field, and na{\"\i}vely it seems that both conditions are incompatible at the same time.

\subsection{Spin(7) Backgrounds}

Having dealt with the special cases of BHV and PW backgrounds embedded in the local $Spin(7)$ system of equations, we now turn to a more general class of examples. As a first example, we consider a background in which the Higgs field and gauge field flux are both switched on. This is essentially a ``hybrid'' of the BHV and PW backgrounds considered previously. As another example, we show that for backgrounds with just the
Higgs field switched on, localization can always be repackaged in terms of a closely related PW system. In both of these examples, zero modes localize on a one-dimensional subspace, i.e. they are codimension three in the four-manifold $M_4$, or equivalently, codimension seven in the local $Spin(7)$ geometry.

\subsubsection{Hyrbid BHV and PW Example}

We now construct an example which is a hybrid of a BHV and PW background:
\al{ A_1 = N y\,, \ A_2 = - N x\,, \ A_3 = - N v\,, \ A_4 = N u\,, \ \Phi_1 = -  \mu x-2\lambda x\,, \ \Phi_2 = \mu y+\lambda y \,, \ \Phi_3 =  \lambda u\,.
}
The solution of the zero mode equations has the form
\al{ \{ a_1,a_2,a_3,a_4,\varphi_1,\varphi_2,\varphi_3\} = \{\alpha_1,\alpha_2,\alpha_3,\alpha_4,\eta_1,\eta_2,\eta_3\} e^{-\frac{1}{2} x.M.x}\,.
}
where we refer to Appendix \ref{app:MOREZERO} for further details of the matrix $M$ in equation (\ref{eq:M}). We have:
\al{&\{\alpha_1,\alpha_2,\alpha_3,\alpha_4,\eta_1,\eta_2,\eta_3\} =\\& \{
\frac{(3 \lambda +2 \mu ) \left(\sqrt{\lambda ^2+4 N ^2} \sqrt{(\lambda +\mu )^2
   \left(\frac{4 N ^2}{(3 \lambda +2 \mu )^2}+1\right)}+\lambda  (\lambda +\mu
   )\right)}{4 N }+N  (\lambda +\mu ) ,\\& -\frac{1}{2} i (\lambda +\mu ) \sqrt{\lambda ^2+4 N ^2}-\frac{1}{2} i (3 \lambda +2 \mu ) \sqrt{(\lambda +\mu )^2 \left(\frac{4 N ^2}{(3
   \lambda +2 \mu )^2}+1\right)} ,0 , 0 ,\\& -\frac{i (3 \lambda +2 \mu ) \left(\mu
   \sqrt{\lambda ^2+4 N ^2}+\lambda  \left(\sqrt{\lambda ^2+4 N ^2}+\sqrt{(\lambda
   +\mu )^2 \left(\frac{4 N ^2}{(3 \lambda +2 \mu )^2}+1\right)}\right)\right)}{4
   N } , (\lambda +\mu )^2 , 0  \}\,.
}
Note that since the real part of $M$ has one zero eigenvalue,
the modes will not be fully localized in codimension four,
confirming our earlier heuristic result.

\subsubsection{Pure Higgs Field Example}

The next background we shall discuss is the generic Spin(7) pure Higgs field background, i.e.
we leave the gauge field flux switched off. Our aim will be to show that this can be recast in terms of a
quite similar analysis in terms of a local PW system. We can write the Higgs field as
\al{ \Phi_i = L_{ij} x_j\,.
}
The background equations become the conditions:
\al{ &L_{11}+L_{22}+L_{33} = 0\,,\\
&L_{12}-L_{21}+L_{34} = 0\,,\\
&L_{13}-L_{31}+L_{24}=0\,,\\
&L_{23}-L_{32}-L_{14} =0\,.
}
To solve for the zero mode equations we make the ansatz
\al{ \{ a_1,a_2,a_3,a_4,\varphi_1,\varphi_2,\varphi_3\} = \{\alpha_1,\alpha_2,\alpha_3,\alpha_4,\eta_1,\eta_2,\eta_3\} e^{-\frac{1}{2} x.M.x}\,.
}
Here $\alpha_i $ and $\eta_i$ are complex numbers. The equations are going to fix these coefficients as well as the matrix $M$. The equations become
\al{ \alpha_4 M_{1j} x_j -\alpha_1 M_{4j} x_j +\alpha_3 M_{2j} x^j- \alpha_4 M_{3j}x_j -i L_{2j}x_j \eta_3 +i L_{3j}x_j \eta_2 &=0\,,\\
\alpha_2 M_{1j} x_j -\alpha_1 M_{2j}x_j +\alpha_4 M_{3j}x_j-\alpha_3 M_{4j}x_j -i L_{1j}x_j \eta_2 +i L_{2j} x_j \eta_1&=0\,,\\
\alpha_3 M_{1j} x_j - \alpha_1 M_{3j} x_j+\alpha_2 M_{4j} x_j -\alpha_4 M_{2j} x_j -i L_{1j} x_j \eta_3 +i L_{3j} x_j\eta_1&=0\,,\\
 M_{1j} x_j \eta_1 -i L_{1j} x_j \alpha_1 + M_{2j} x_j \eta_2 -i L_{2j} x_j \alpha_2 + M_{3j} x_j \eta_3 -i L_{3j}x_j \alpha_3&=0\,,\\
M_{4j} x_j \eta_3 -i L_{3j}x_j \alpha_4 + M_{1j} x_j \eta_2 -i L_{2j}x_j \alpha_1 - M_{2j} x_j \eta_1 +i L_{1j} x_j \alpha_2 &=0\,,\\
M_{1j} x_j \eta_3 -i L_{3j} x_j \alpha_1 - M_{4j}x_j \eta_2 +i L_{2j} x_j \alpha_4 -M_{3j}x_j \eta_1 +i L_{1j} x_j \alpha_3 &=0\,,\\
M_{2j} x_j \eta_3 -i L_{3j} x_j \alpha_2 -M_{3j} x_j \eta_2 +i L_{2j} x_j \alpha_3 + M_{4j} x_j \eta_1 -i L_{1j} x_j \alpha_4 &=0\,.
}
One may might wonder if it is possible to rotate this solution to a PW solution. This requires the $\Phi_{\text{SD}}$ to be independent of one coordinate. In a coordinate independent manner we require thee to exist a vector $X$ such that the Lie derivative $\mathcal L_X \Phi_{\text{SD}} = 0$. We write generically $X = \alpha \p_x + \beta \p_y + \gamma \p_u + \delta \p_v$ and compute the Lie derivative. Recall that when acting on a differential form $\omega$ one has that
\al{ \mathcal L_X \omega = \iota_X d \omega + d \iota_X \omega\,.
}
In our case since $d\Phi_{\text{SD}} = 0$ the first term drops, so it is only necessary to compute the second one. We will spare the details of the computation and simply say that the vanishing of $\mathcal L_X \Phi_{\text{SD}} $ can be written as the linear system
\al{ \left[
\begin{array}{cccc}
 L_{31} & L_{32} & -L_2-L_{11} & L_{12}-L_{21} \\
 -L_{21} & -L_2 & -L_{23} & L_{13}-L_{31} \\
 L_{11} & L_{12} & L_{13} & L_{23}-L_{32} \\
 L_{11} & L_{12} & L_{13} & L_{23}-L_{32} \\
 L_{21} & L_2 & L_{23} & L_{31}-L_{13} \\
 L_{31} & L_{32} & -L_2-L_{11} & L_{12}-L_{21} \\
\end{array}
\right] \left[\begin{array}{c}\alpha \\ \beta \\ \gamma \\ \delta \end{array}\right] = 0\,.
}
Here we put the system on-shell, requiring that the background solves the equations of motion written above. Quite interestingly there is always a solution to this system (the matrix has rank three) so it is always possible to find a vector $X$ with $\mathcal L_X \phi_{\text{SD}} = 0$, implying that for this background we can always make a choice of coordinates that brings us back to a PW system.
This means that a background $\Phi_{\text{SD}}$ that is linear in the coordinates is always equivalent to a PW background.

\section{Local Matter and Topological Insulators} \label{sec:TOPOINS}

Although our eventual aim is the study of 3d matter fields coupled to dynamical gauge fields, in this section we show that
we can already use what we have developed to
engineer a rich class of 3d systems coupled to a 4d topological bulk,
which we can think of as various instances of topological insulators (see e.g. \cite{PhysRevLett.95.226801, PhysRevLett.96.106802, Qi2011}).
An interesting feature of these systems is that many features can be deduced from primarily topological considerations. Since we anticipate strong quantum corrections in local $Spin(7)$ compactifications with 3d $\mathcal{N} = 1$ supersymmetry, our eventual aim in this paper will be to extract robust topological quantities. The case of topological insulators engineered via local $Spin(7)$ compactifications thus serves as a useful ``warmup exercise'' for the full problem.

Recall that a simple instance of this sort of configuration arises from a 4d Dirac fermion on $\mathbb{R}^{2,1} \times \mathbb{R}_{t}$ with a position dependent mass $m(t)$ such that $m >0$ for $t >0$ and $m <0$ for $t< 0$, with $m = 0$ at $t = 0$. In the bulk, we have a $U(1)$ global symmetry, and if we consider the topological term $\theta F \wedge F$, with $F$ the background field strength for this $U(1)$, then we can think of the localized mode as being trapped at the interface between a $\theta = 0$ and $\theta = \pi$ phase.

In this section we illustrate that matter localization in local $Spin(7)$ systems
can be viewed as engineering a class of topological insulators, but where now
the global symmetry in the bulk is more general. One way for us to proceed is to actually start with M-theory on a Calabi-Yau threefold given by a curve $\Sigma$ of ADE singularities. This gives rise to a 5d gauge theory with gauge group $G$ of ADE type. If we allow further enhancements in the singularity type at points of the curve, we get 5d hypermultiplets. To give an explicit example, we engineer a 5d $SU(N)$ gauge theory coupled to a single 5d hypermultiplet in the fundamental representation of $SU(N)$. This is generated by the local Calabi-Yau hypersurface:
\begin{equation}
y^2 = x^2 + z^N (z - u)
\end{equation}
where $z = 0$ denotes the location of the curve $\Sigma$, and $u$ denotes a local coordinate along the curve. At $u = 0$, we get a matter field trapped along the curve.

Returning to the general case, if we further compactify on a circle, we wind up with a 4d $\mathcal{N} = 2$ hypermultiplet in a representation $\mathbf{R}$ of $G$. In the limit where $\Sigma$ is non-compact, we just have a global symmetry $G$ coupled to our 4d field. Observe that the fermionic content of this system consists of a 4d Dirac fermion. Switching on a position dependent mass term of the kind used in the topological insulator just discussed, we get further localization to a 3d Dirac fermion.

This localization can be understood in terms of a PW system on the three-manifold $M_3 = \mathbb{R}_t \times \Sigma$, or equivalently, in terms of a local $Spin(7)$ system on $\mathbb{R}_t \times \Sigma \times S^1$. The Higgs field on the entire $M_4$ is given by:
\begin{equation}
\Phi_{\mathrm{SD}}=\phi_{\mathrm{PW}}\wedge d\theta +*_3 \phi_{\mathrm{PW}}
\end{equation}
where $*_3$ is the Hodge star operator on $M_3$, $\phi_{\mathrm{PW}}$ is a harmonic one-form on $\Sigma\times \mathbb{R}_t$, and
$d \theta$ is the volume form on the $S^1$ factor. We can set
up $\phi_{\mathrm{PW}}$ to have a zero-locus at a point localized at $t=0$ and a point in $\Sigma$ (which in local coordinates we take to be $x_1=x_2=0$), and will locally be of the form
\begin{equation}\label{insul1}
\phi_{\mathrm{PW}}=d[ (1 - \kappa) x_1^2 - (1 + \kappa) x_2^2 + 2 \kappa t^2].
\end{equation}
Observe that when $\kappa = 0$ the matter field is delocalized on $\mathbb{R}_t \times S^1$,
but is still trapped at $x_1 = x_2 = 0$.
The relative minus signs on our configuration are necessary to
ensure that we can actually solve the equations of motion on $\Sigma$ in this limit. When $\kappa \neq 0$,
we can interpret our construction in terms of a 4d domain wall fermion trapped at $t = 0$. The relative strength of localization in the different directions depends on the metric data. For example, if $\Sigma$ is very small, then there is a sense in which the mode delocalizes on $\Sigma$, but remains tightly trapped in the $\mathbb{R}_t$. Explicitly, the zero mode profile for our domain wall fermion is of the form:
\begin{equation}\label{insul2}
\psi \sim \exp\left( -\frac{1}{2} \left( \vert \kappa \vert t^2- \vert 1 - \kappa \vert x_1^2- \vert 1 + \kappa \vert x_2^2 \right) \right),
\end{equation}
where to avoid clutter we have suppressed the explicit form content.\footnote{Recall from section \ref{sec:Spin7Review} that the $Spin(7)$ wavefunctions on $\Sigma\times S^1\times \mathbb{R}_t$ follow from equation (\ref{insul2}) as $a = \psi$, $\varphi=\psi\wedge d\theta +*_3 \psi$.}
We have therefore engineered a 3d $\mathcal{N}=2$ chiral matter multiplet localized along the $S^1$-factor,
or its 4d $\mathcal{N}=1$ analog if we decompactify the $S^1$ factor.
Observe that on $\mathbb{R}_t$, our matter field $\psi$ satisfies the equation of motion:
\begin{equation}\label{eq:4dtdepmass}
(\partial_t+ \vert \kappa \vert t)\psi (t)=0
\end{equation}
which can be seen as a Dirac equation for the localized chiral multiplet with a position-dependent mass, $m(t)= \vert \kappa \vert t$.

A further comment here is that because our internal space is non-compact, we do not have a 3d dynamical gauge field.
The 4d gauge coupling is given by:
\begin{equation}
\frac{1}{g^2_{4d}} = \frac{1}{g^{2}_{7d}} \times \mathrm{Vol}(\Sigma \times S^1),
\end{equation}
namely we think of M-theory on an ADE singularity as engineering 7d Super Yang-Mills theory, and further compactification relates this gauge coupling to its lower-dimensional counterpart. In particular, in the limit where $\mathrm{Vol}(\Sigma) \rightarrow \infty$, the 4d bulk dynamics trivializes.

Changing perspective, we can also think of our localized matter as descending from a PW system on the three-manifold $\widetilde{M}_3 = \mathbb{R}_t \times \Sigma$, which results in a 4d $\mathcal{N} = 1$ chiral multiplet. By itself, this would produce a gauge anomaly if we had tried to compactify the $\mathbb{R}_t$ direction. In the context of our 4d bulk and 3d edge mode, this is just anomaly inflow from the bulk to the boundary \cite{Callan:1984sa}, as applied for example in \cite{Acharya:2001gy, Acharya:2002vs}. Note also that in the context of the PW system, we could localize additional matter fields at other points of $\widetilde{M}_3 = \mathbb{R}_t \times \Sigma$. This would amount to introducing additional domain wall fermions  which couple to each other via the bulk. Our local $Spin(7)$ system generalizes this further because there is no need for matter to be localized on a common $S^1$.

Instead of directly proceeding in terms of the equations of motion for localized zero modes, we could have instead phrased our analysis in terms of the bulk topological insulator. Indeed, as emphasized in \cite{Seiberg:2016rsg, Seiberg:2016gmd}, for example, some features of the bulk / boundary dynamics can be captured in purely topological terms. For example, we can speak of the jump in the $\theta$ angle as we pass from one side of the interface to the other. Now, in our context the $\theta$ angle of the 4d bulk descends from the period integral of the M-theory three-form potential $C_3$:
\begin{equation}
\theta = \underset{M_3}{\int} C_3
\end{equation}
where $M_3 = \Sigma \times S^1$. If we consider integrating out the 4d Dirac fermion, then we can view this as a system with no bulk matter but instead, a position dependent $\theta$ angle, and thus we can alternatively view this as a position dependent $C_3$. This effective $G_4$-flux would then have delta function support precisely at the location of our matter field:
\begin{equation}
G_4 = \frac{1}{2} \delta(t)dt \wedge \delta_{\{u=0\}} \wedge d \theta.
\end{equation}
Here $d \theta$ is the unit normalized volume form on the circle and $\delta_{\{u=0\}}=\frac{1}{2i}\delta(u)\delta(\overline{u})dud\overline{u}$. Note that the coefficient here is $1/2$. This is actually required to properly account for the localized matter field, which can be thought of as  Chern-Simons theory at ``level $1/2$'' (which only makes sense if we couple to a 4d bulk).

\section{Gluing Constructions \label{sec:FIRSTEXAMPLES}}

Up to now, our discussion of the local $Spin(7)$ system has involved working in a local patch where our 7d SYM theory is placed on a non-compact four-manifold $M_4$. Based on our previous analysis, we expect that the localized matter can be understood as 3d $\mathcal{N} = 2$ matter (at least locally), simply due to the fact that they typically fill out complex representations of the gauge group. Of course, since we are dealing with a local $Spin(7)$ compactification, we only expect to retain 3d $\mathcal{N} = 1$ supersymmetry, a feature which emerges upon working with a compact $M_4$. Indeed, the matter multiplets which are delocalized over $M_4$ fill out $\mathcal{N} = 1$ multiplets. Moreover, on a general $M_4$ we can write down more general interaction terms which would have been forbidden with $\mathcal{N} = 2$ supersymmetry.

With these considerations in mind, our aim in this section will be to develop a gluing construction for building more general solutions. The main idea will be to consider building blocks, either obtained from the local PW system or the local BHV system. In Appendix \ref{app:QUOT} we present a somewhat different method for building examples of local $Spin(7)$ systems based on a quotient construction of
BHV solution. This method should lift to compact geometries, but as it is somewhat orthogonal
to the other developments of the paper, we have placed it in an Appendix.

\subsection{Connected Sums of $G_2$ Local Models}\label{sec:ConnSum}

\begin{figure}[t!]
\centering
\includegraphics[width=0.5\textwidth, angle=0]{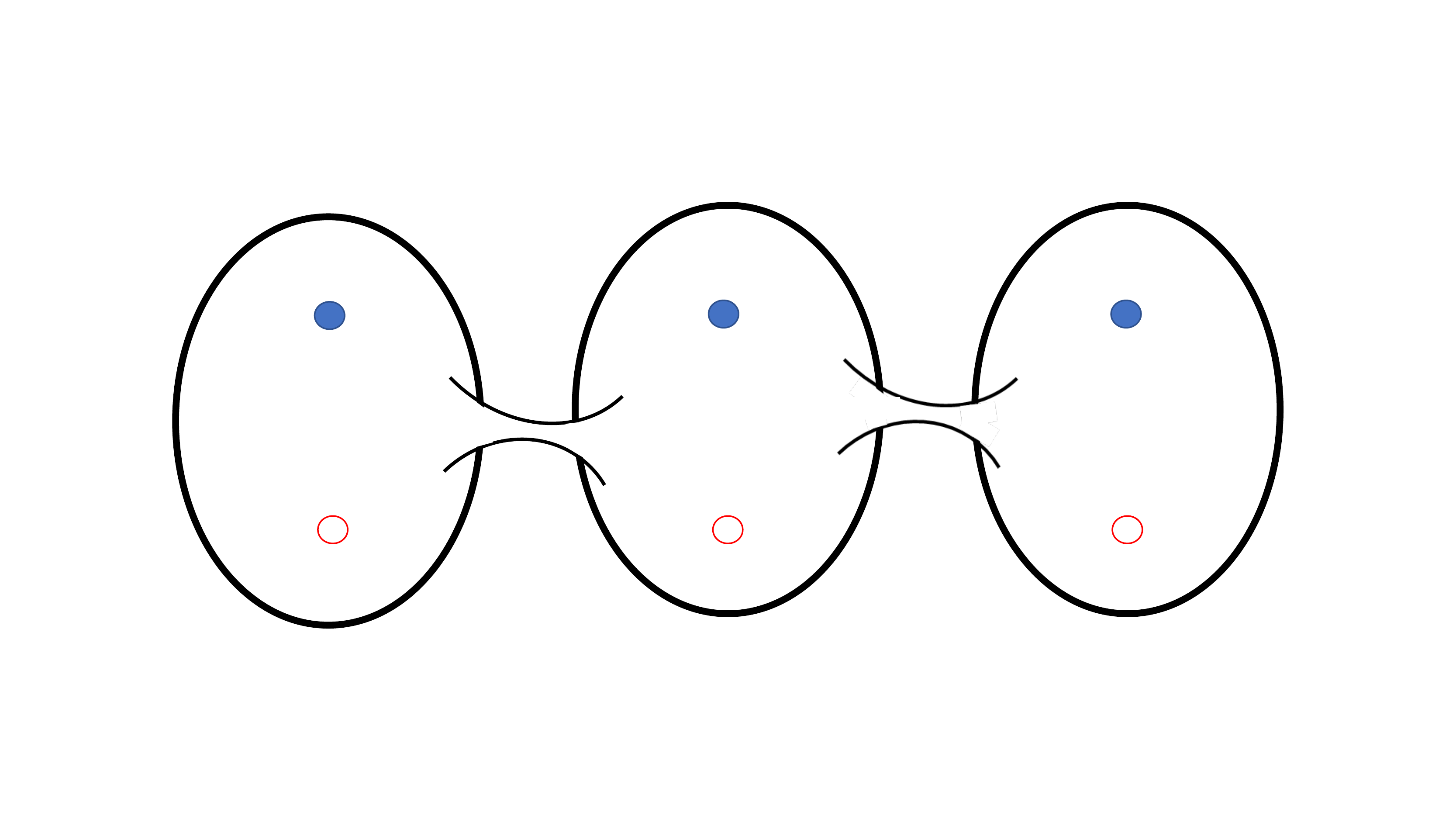}
\caption{Depiction of a connected sum construction of $M_4$ in terms of three summands of the form $S^1 \times M^{(i)}_3$. Each of the three blobs represents a four-manifold $M^{(i)}_3\times S^1$ for $M^{(i)}_{3}$ a three-manifold. The dots indicate localized matter on circles, where the solid blue dots represent localized chiral matter in $\mathbf{R}$ and the open red dots represent localized chiral matter in $\overline{\mathbf{R}}$ from the point of view of the PW system on the three-manifold $M^{(i)}_3$.}
\label{fig:connsum}
\end{figure}

We now construct a class of local $Spin(7)$ systems by starting with PW backgrounds on four-manifolds of the form $M_{4}^{(i)} = M^{(i)}_3 \times S^1$, i.e. we consider solutions to our local $Spin(7)$ equations which are ``trivial'' along the $S^1$ factor. Starting from $M_{4}^{(i)}$ and $M_{(4)}^{(j)}$ a pair of such building blocks, we can glue these four-manifolds together by cutting out a four-ball from each, and then gluing along a neck region. Topologically, this is just the standard connected sum $M_4^{(i)} \# M_{4}^{(j)}$. In the present context, however, we demand more, because each four-manifold is also equipped with a Higgs bundle, and we need to ensure that these solutions can \textit{also} be extended across the neck region. See figure \ref{fig:connsum}
for a depiction of this gluing procedure, where we also indicate possible zeros for the Higgs field.

In more detail, the connected sum construction on its own is a topological operation after which there may be one of several prescriptions to assign a metric to the resulting space. We can carry this out for any collection of for manifolds of the form $M_{3}^{(i)} \times S^1$ on each
block, but for ease of exposition, we discuss it in detail in the special case where we have $S^3 \times S^1$ summands.
If the building blocks have constant Ricci curvature one may be interested in requiring the composite space to also have constant Ricci curvature. In \cite{tmna/1479265331}, after assuming a certain non-degeneracy condition of a Poisson operator which $S^3\times S^1$ satisfies on the building blocks, there is a general perturbative procedure to prove the existence of such a metric on the total space. To illustrate, we can consider $S^3\times S^1$ blocks with round metric,
\begin{equation}
ds^2=d\phi^2+d\Omega^2_{S^3}=d\phi^2+d\psi^2+\sin^2\psi(d\theta^2+\sin^2\theta d\varphi^2) \; \; \; \; 0\leq \phi,\varphi\leq 2\pi \; \; \; 0\leq \theta,\psi \leq \pi
\end{equation}
by interpolating, with bump functions, to standard spatial wormholes connecting neighboring blocks. More specifically, to glue two neighboring $S^3\times S^1$ blocks, start by choosing the same point on each $p_0 \equiv (\phi_0,\varphi_0,\psi_0,\theta_0)$ such that it is far away from any zeros or poles of the Higgs field. We can define the radial coordinate for the four-ball neighborhoods $B^4_\epsilon(p_0)$ as
\begin{equation}
\rho^2 \equiv (\phi-\phi_0)^2+(\psi-\psi_0)^2+(\theta-\theta_0)^2 \sin^2(\psi-\psi_0) + (\varphi-\varphi_0)^2 \sin^2(\psi-\psi_0)\sin^2(\theta-\theta_0)
\end{equation}
where the positive and negative branches of $\rho$ each represent a copy of $S^3\times S^1$. Similarly we can also define a natural angular element $d\widetilde{\Omega}$ of the neighborhood as well so that the metric on $B^4_\epsilon(p_0)$ has the form
\begin{equation}
ds^2=d\rho^2+\rho^2 d\widetilde{\Omega}^2.
\end{equation}
This reproduces the statement that a neighborhood of a point in $S^3\times S^1$ is diffeomorphic to an open ball in $\mathbb{R}^4$. Let us define a bump function $b(\rho)$ such that $b(0)=b_0>0$ and $b(\pm \epsilon)=0$. Here $b_0<\epsilon$ is the inner radius of the wormhole. Then the metric:
\begin{equation}
ds^2=d\rho^2+ [ \rho^2 + b(\rho)^2 ] d\widetilde{\Omega}^2
\end{equation}
interpolates between the two branches of $\rho$ and hence the two copies of $S^3\times S^1$. This metric ansatz is symmetric enough such that the resulting four-manifold possess an orientation reversal: 1) Switch the two manifolds, sending $\rho\rightarrow -\rho$ and 2) apply an orientation-reversing map of $S^3$ to each factor.\footnote{A common such example that has two fixed points (the minimal number for $S^3$) is to send $(x_1,x_2,x_3,x_4)\rightarrow (x_1,-x_2,-x_3,-x_4)$ given $\sum_i x^2_i=R^2$. Note the antipodal map for $S^3$ is orientation-preserving.} This construction can then be generalized to any sequence of connected sums $\#_{i}(M^{(i)}_3\times S^1)$.

Giving an exact solution to the Higgs field would in principle be possible but non-trivial. Instead we can make the assumption that the gluing neck's diameter, $\epsilon$, is very small. By assumption $\Phi(p_0)\neq 0$, so the solution is well-approximated by simply setting $\Phi=\Phi(p_0)$ on the neck. Note that if the neck were large, we would have to calculate the corrections to the original building block Higgs field which alter (in pairs) our count of zeros, and there may be additional pairs of zeros of $\Phi$ on the neck. Also note that the existence of such a harmonic two-form (with singularities) is guaranteed by a (relative) Hodge theorem so we know there exists a solution arbitrarily close to the $\Phi|_{neck}=\Phi(p_0)$ approximation.

\subsubsection{Zero Mode Counting}

An advantage of this connected sum construction is that we can read off the matter content from these local building blocks. First of all, the codimension three matter in each individual summand of a connected sum construction is essentially unchanged from what we have in the
PW case \cite{Pantev:2009de, Braun:2018vhk}. There is a slight subtlety here because a priori, there could be additional ways for matter in conjugate representations to pair up, especially due to working in lower dimensions. One can visualize this in terms of Euclidean M2-branes which stretch across the gluing neck region. We revisit this issue in section \ref{sec:QUANTUM}.

\begin{figure}
\centering
\includegraphics[width=0.5\textwidth, angle=0]{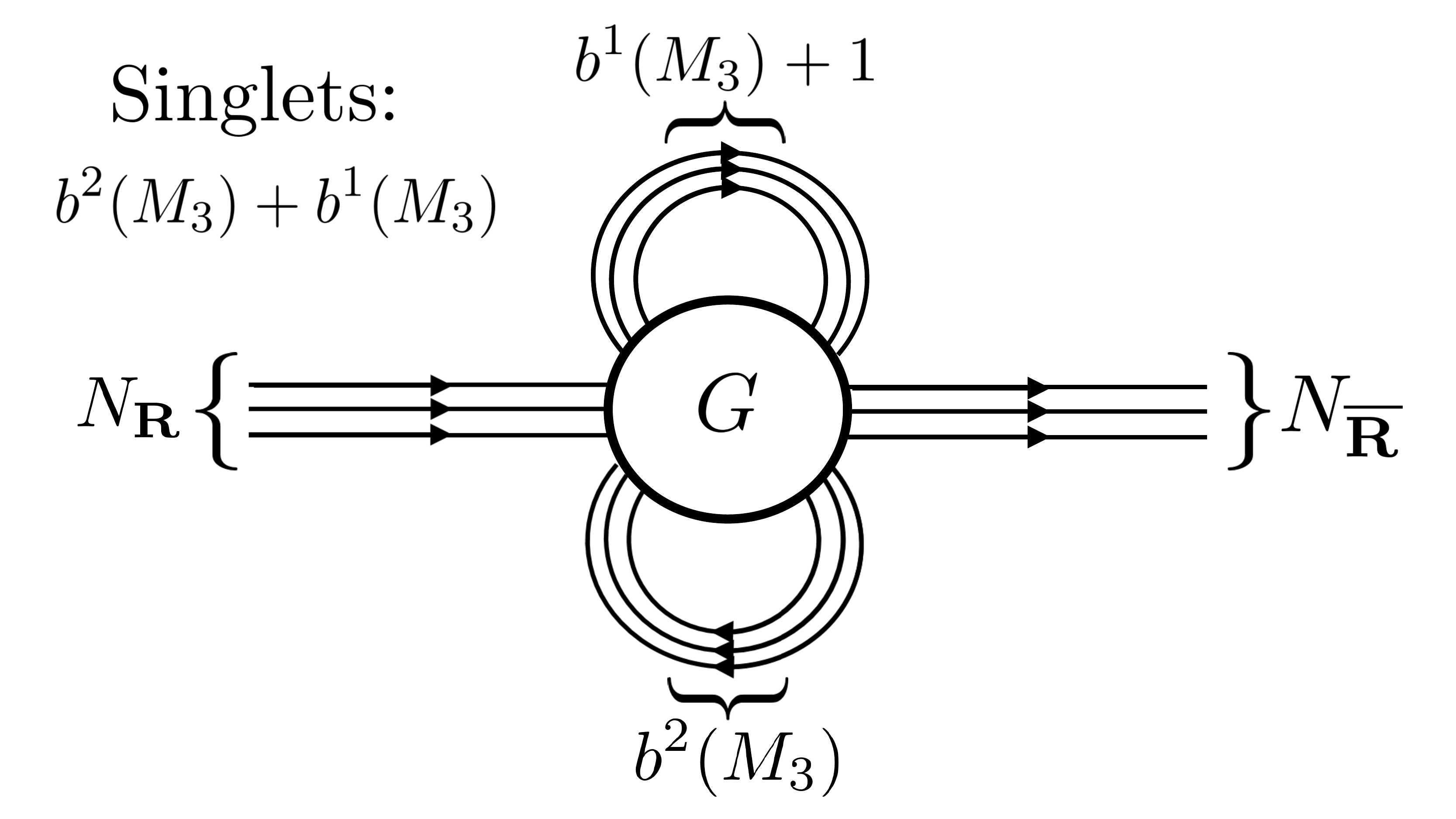}
\caption{Matter content of the 3d theory when $M_4=S^1 \times M_3$. The theory has $\mathcal{N}=2$ supersymmetry but the quiver is expressed in terms of $\mathcal{N}=1$ fields where double lines connected to the gauged node are adjoint fields, single lines denote fields in $\mathbf{R}$ and its conjugate, and no lines denote singlets.}
\label{fig:k=1}
\end{figure}

Putting this issue aside, we can determine the zero mode content of our newly constructed local $Spin(7)$ system just from analyzing the profile of the Higgs field. For localized matter on an individual building block $M_3 \times S^1$, there is not much difference from what we would get just from studying the PW system on the three-manifold $M_3$, as in references \cite{Pantev:2009de, Braun:2018vhk}. To quickly review these results, consider the simple Higgsing $\widetilde{G}\rightarrow G\times U(1)$ from turning on a one-form $\Phi_{\mathrm{PW}}$ on $M_3$. Equations for localized zero modes in the representation $\mathbf{R}_q$ are:
\begin{equation}
(d+q\Phi_{\mathrm{PW}})\psi_q=0
\end{equation}
\begin{equation}
(d^\dagger+q\iota_\Phi)\psi_q=0.
\end{equation}
By exponentiating the wavefuntions $\psi'=e^{-qf}\psi$ ($\Phi_{\mathrm{PW}}=df$)  these equations reduce to a local cohomology group. Indeed, we observe that $e^{-qf}de^{qf}=(d+q\Phi_{\mathrm{PW}})$. Note that this same step taken in solving for ground states in superquantum mechanics with a target space potential. To properly account for the $\Phi_{\mathrm{PW}}$ singularities, let us define the loci positive/negatively charged loci as $\Delta_\pm$ which consists of $n_{\pm}$ points on $M_3$ (formulas for more general singularity sources are given in the aforementioned references). We excise their neighborhoods as $\widetilde{M}_3 \equiv M_3 \backslash (\Delta_+\cup \Delta_-)$ and work in cohomology relative to negative sources. The relevant formulas for our localized zero modes are then given by the following Betti numbers:
\begin{equation}\label{PWmatter1}
b^1(\widetilde{M}_3,\Delta_-)=b^1(M_3)+n_--1= \; \text{\# of chiral modes in representation $\mathbf{R}$}
\end{equation}
\begin{equation}\label{PWmatter2}
b^2(\widetilde{M}_3,\Delta_-)=b^2(M_3)+n_+-1= \; \text{\# of chiral modes in representation $\overline{\mathbf{R}}$}
\end{equation}
See figure \ref{fig:k=1} for a depiction of the resulting quiver gauge theory in the case of compactification of the local $Spin(7)$ system on the four-manifold $M_4 = S^1 \times M_3$.

On the other hand, the bulk modes which are a spread across all of our new four-manifold $M_4$ will certainly be modified. First of all, given a breaking pattern such as $\widetilde{G} \rightarrow G \times U(1)^n $, we expect a 3d $\mathcal{N} = 1$ vector multiplet for the unbroken gauge group $G \times U(1)^n$.\footnote{Again, we are being cavalier with the global structure of the unbroken group. Additionally, some of these $U(1)$ factors may end up decoupling due to further interactions with a ``Green-Schwarz'' axion, but this is something we cannot address in the purely local context.} Additionally, we can expect bulk modes in 3d $\mathcal{N} = 1$ matter multiplets transforming in the adjoint representation of $G \times U(1)^n$ (i.e. they are neutral under the $U(1)$ factors). These are counted by $b^{1}(M_4)$ and $b^{2}_{\mathrm{SD}}(M_4)$, which respectively come from the internal vector potential, and the self-dual two-form of the 7d SYM theory. In terms of our connected sum building blocks we have:
\begin{align}
b^{1}(M_4) & = \underset{i}{\sum} b^{1}(M_4^{(i)}) = \underset{i}{\sum} (b^{1}(M_3^{(i)}) + 1) \\
b^{2}_{\mathrm{SD}}(M_4) & = \underset{i}{\sum} b^{2}_{\mathrm{SD}}(M_4^{(i)}) = \underset{i}{\sum} b^{2}(M_3^{(i)}) ,
\end{align}
in the obvious notation. By inspection, when we have more than one building block, the
zero modes do not automatically sort into ``complex'' 3d $\mathcal{N} = 2$ matter multiplets.
In fact, precisely because these modes transform in a real representation, and are spread over the entire four-manifold, we generically expect them to lift in pairs. From the structure of the form content, this involves a pairing between the one-forms and the self-dual two-forms. In subsequent sections we will revisit the precise remnant of these zero modes which can be detected by discrete anomalies in the 3d effective field theory.

A closely related comment is that the localized matter fills out 3d $\mathcal{N} = 2$ matter multiplets, which we can view as the dimensional reduction of 4d $\mathcal{N} = 1$ chiral matter on a circle. For bulk matter we expect the scalar degrees of freedom to split into two types, namely those which are even under a reflection of a spatial coordinate (i.e. $x^{i} \rightarrow - x^{i}$), and those which are odd under such a reflection. This in turn will impact how we count various contributions to the discrete anomalies. We defer a full treatment of this important issue to section \ref{sec:CRT} where we discuss reflections on the various kinds of matter fields in our system.

\subsection{Connected Sums of $CY_4$ Local Models}

In the previous subsection we illustrated how to start with a collection of 4d $\mathcal{N} = 1$ theories engineered in M-theory on local $G_2$ spaces, and, via a suitable gluing construction, build 3d $\mathcal{N} = 1$ systems. The main feature of all these constructions is that localized matter still fills out 3d $\mathcal{N} = 2$ supermultiplets, while bulk modes and vector multiplets only fill out 3d $\mathcal{N} = 1$ supermultiplets.

Now, another way to generate 4d $\mathcal{N} = 1$ vacua would be to start with F-theory on an elliptically fibered Calabi-Yau fourfold. Compactifying on a further circle would result in an M-theory background which retains 3d $\mathcal{N} = 2$ supersymmetry. In this case, chiral matter of the 4d system really results from two related effects. Geometrically, the enhancement in the singularity type along a curve would give us 4d $\mathcal{N} = 2$ hypermultiplets. Switching on a background flux from the 7-branes then results in 4d $\mathcal{N} = 1$ matter \cite{Beasley:2008dc, Donagi:2008ca}. From the perspective of an M-theory background, then, the geometrically localized matter will fill out 3d $\mathcal{N} = 4$ supermultiplets, but flux will then lead to 3d $\mathcal{N} = 2$ supermultiplets.

\begin{figure}[t!]
\centering
\includegraphics[width=0.5\textwidth, angle=0]{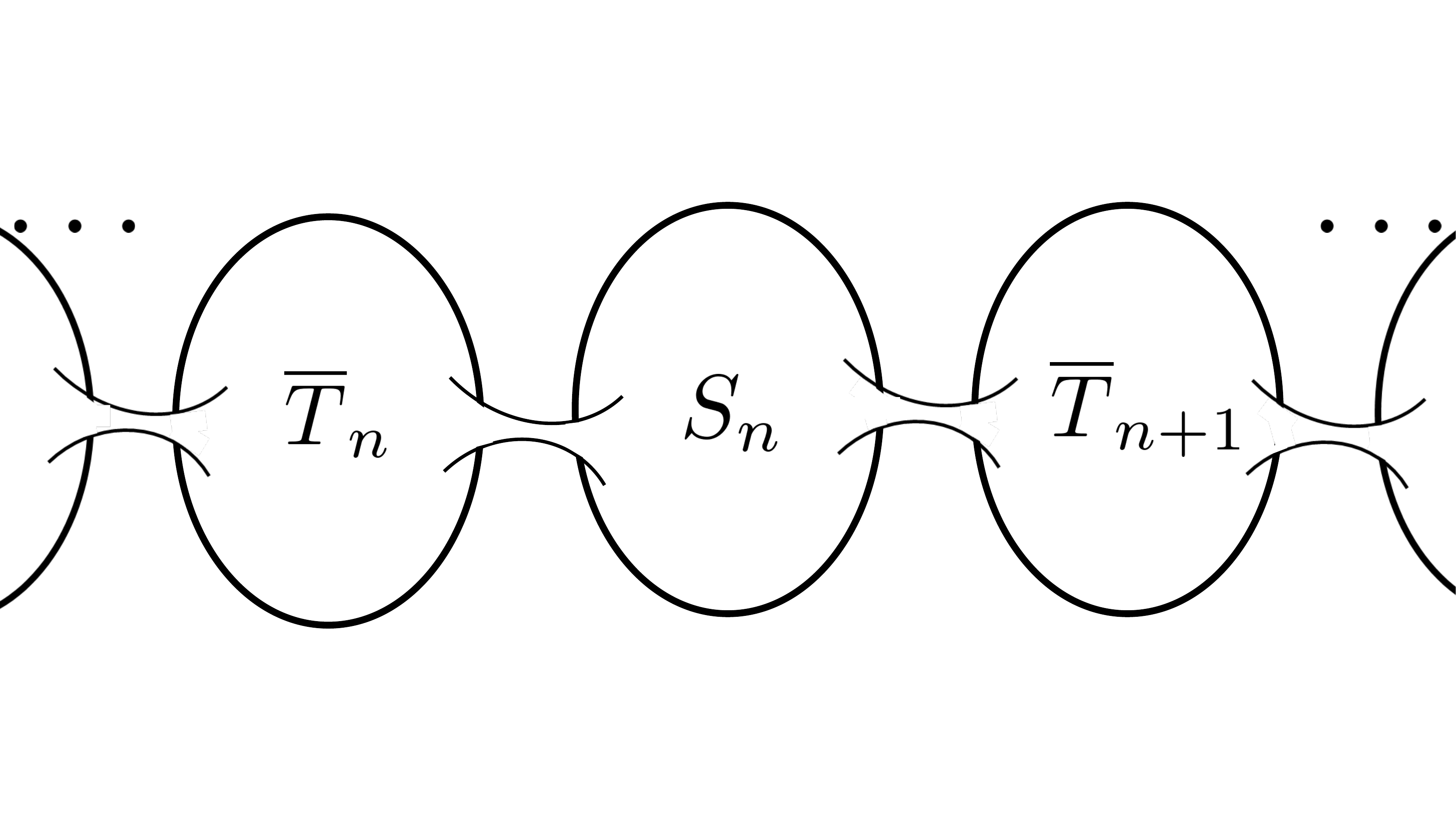}
\caption{An illustration of our gluing construction of K\"ahler manifolds.}
\label{fig:kahlergluing}
\end{figure}

As already mentioned, the local model associated with F-theory on a $CY_4$ is just the BHV system on a K\"ahler surface $S$,
as studied in \cite{Donagi:2008ca, Beasley:2008dc, Beasley:2008kw}. Of course, $S$ is also a four-manifold, so we can build up connected sums of such building blocks to arrive at more general four-manifolds. Since we are interested in 3d $\mathcal{N} = 1$ vacua, we actually require that the resulting four-manifold is not K\"ahler. One way to ensure this is to simply glue together K\"ahler surfaces with a suitable orientation reversal in the gluing process. For example, we can glue two copies of $\mathbb{CP}^2$ to arrive at $\mathbb{CP}^2 \# \overline{\mathbb{CP}^2}$, and this has signature zero,\footnote{Given a four-manifold $M$ with signature $\sigma(M)$, we have $\sigma(M \# \overline{\mathbb{CP}^2}) = \sigma(M) + \sigma(\overline{\mathbb{CP}^2}) = \sigma(M) - 1$.} and moreover, it is non-K\"ahler.

Compared with our discussion where we glued PW building blocks, we face an additional complication here in that the orientation reversal of $S$ a K\"ahler surface does not simply result in another K\"ahler surface (although $S$ and $\overline{S}$ are homeomorphic). If we must then deal with the local $Spin(7)$ system on $\overline{S}$ anyway, one might ask what has been gained by introducing a gluing construction at all?

The main point is that in our construction, we can assume that the Higgs field profile is only non-trivial on the K\"ahler surface summands, and remains trivial on the orientation reversed summands (which are non-K\"ahler). This sort of construction requires a specific profile for the Higgs field in the gluing region between a K\"ahler summand $S_i$ and a non-K\"ahler summand $\overline{S_j}$. In particular, we demand that the Higgs field tends to zero there. That this can be arranged was shown in reference \cite{Cvetic:2020piw}, although the price we pay is that we only get an approximate BHV solution in the K\"ahler surface region since all three components of $\Phi_{\mathrm{SD}}$ are necessarily switched on (although the contribution from the component parallel to the K\"ahler form on $S_i$ is exponentially suppressed). Proceeding in this way, we can engineer examples where the localized matter is still of the type found in a BHV system, but where the bulk modes are now of the more general kind found in local $Spin(7)$ systems. Observe also that nothing stops us from building up several such K\"ahler and non-K\"ahler summands.

With this in mind, we can also proceed to write down the classical zero mode content for our 3d effective field theory.
For the BHV system, we have the well-known contributions for matter localized on curves (possibly in the presence of gauge field fluxes), as discussed for example in \cite{Donagi:2008ca, Beasley:2008dc}. Let us therefore focus on the contributions from the ``bulk modes'' which fill out genuine 3d $\mathcal{N} = 1$ matter multiplets in the adjoint representation of $G \times U(1)^n \subset \widetilde{G} $, with notation as in subsection \ref{sec:ConnSum}. Writing our $M_4$ as a connected sum of $S_i$ K\"ahler summands and $\overline{T_k}$ summands with $T_k$ K\"ahler as in figure \ref{fig:kahlergluing}, we have:\footnote{The precise gluing does not affect these topological quantities.}
\begin{align}
b^{1}(M_4) & = \underset{i}{\sum} b^{1}(S_i) + \underset{k}{\sum} b^{1}(\overline{T_k})
= \underset{i}{\sum} 2 h^{1,0}(S_i) + \underset{k}{\sum} 2 h^{1,0}(T_k) \\
b^{2}_{\mathrm{SD}}(M_4) & =  \underset{i}{\sum} b^{2}_{\mathrm{SD}}(S_i) + \underset{k}{\sum} b^{2}_{ASD}(T_k)
= \underset{i}{\sum} (2 h^{2,0}(S_i) + 1) + \underset{k}{\sum} (h^{1,1}(T_k) - 1),
\end{align}
where in the above, we used the fact that orientation reversal interchanges the self-dual and anti-self-dual two-forms. Similar formulae hold for the Betti numbers of vector bundles built in this way.

As a final amusing comment, we note that Hopf surfaces are complex but not K\"ahler, and are diffeomorphic to
$S^{3} \times S^{1}$, which is quite similar to the PW building block discussed in subsection \ref{sec:ConnSum}.

\section{$\mathsf{C}$, $\mathsf{R}$, \& $\mathsf{T}$ Transformations \label{sec:CRT}}

Up to now, our discussion has basically focused on the \textit{classical} geometry of a local $Spin(7)$ compactification. By this, we simply mean that we have obtained our zero mode content from the linearized approximation to the Vafa-Witten equations our four-manifold $M_4$. Since we are dealing with 3d $\mathcal{N} = 1$ vacua, we can expect many strong coupling effects to enter the low energy effective field theory. On general grounds, the best we can hope for us to is to extract those features of the compactification which are robust against such strong coupling effects.

One such feature is the structure of global anomalies in 3d theories. These involve the study
of the 3d theory on a three-manifold $N_3$, and tracking the response of the partition
function $Z[N_3]$ under symmetry transformations. With an eye towards understanding
universal aspects of local $Spin(7)$ backgrounds, our aim here will be to extract constraints from
various kinds of discrete symmetries which are in some sense universal. The classic examples of this type include reflection of a
spatial coordinate $x^i \mapsto - x^i$, as well as a geometric time reversal operator $x^0 \mapsto - x^0$, and in cases where we also have complex representations of a gauge group, we can also speak of charge conjugation operations as well.\footnote{In some of the literature these reflection transformations are sometimes referred to as ``parity transformations''.}

Tracking these discrete symmetries from a top-down perspective turns out to be somewhat subtle,
because the parity assignments in the internal directions often end up impacting the parity assignments
in the 3d uncompactified directions!\footnote{From a bottom up perspective, one might be tempted
to assert that once the 3d QFT is \textit{defined} we can simply study the discrete symmetries of this system.
Part of the issue with this approach is that it pre-supposes that we know what 3d QFT we actually got in
the first place from compactifying on a local $Spin(7)$ geometry!}

To illustrate some of the subtleties, consider the compactification of a higher-dimensional
vector boson on a torus $T^n$. We can compose reflections of the 3d spacetime with internal
reflections on the torus. Since our vector boson is a one-form, there can be a non-trivial
mixture between these operations which will in turn impact whether we refer to the resulting
3d spin zero degrees of freedom as scalars or pseudo-scalars.

Our aim in this section and the next will be to give a top-down treatment of such 3d discrete transformations
by recasting them in terms of operations on the extra-dimensional geometry. This is important both
in terms of understanding the geometric origin of these transformations, as well as in terms of
understanding what geometric constraints are sufficient to ensure the existence of such symmetries.

Since we are primarily interested in local $Spin(7)$ systems, our focus will be on understanding
the discrete symmetries of 7d SYM theory. There are several canonical routes to realizing this
gauge theory, and it is helpful to consider different ways to engineer this system. In gauge theory terms,
one simple way to proceed is to start with 10d SYM theory, as obtained for example from heterotic strings
in flat space. Compactification on a $T^3$ then results in 7d SYM. A complementary starting point
is M-theory on an ADE singularity. From either starting point, we can ask how geometric reflection
operations on the respective 10d and 11d (Euclidean) spacetimes descend to our 7d theory.
Further compactification on a four-manifold then provides a general method for tracking the descent of these symmetries to a 3d system.

As a point of notation, we will often be working with an $n$-dimensional Euclidean signature
spacetime with local coordinates $(x^0 ,..., x^{n-1})$. We introduce the operation
$\mathsf{R}_i$ which acts as:
\begin{equation}
\mathsf{R}_i : x^i \rightarrow -x^i, \,\,\,\text{and} \,\,\, \mathsf{R}_i : x^{j} \rightarrow x^{j} \,\,\,\text{for} \, j \neq i.
\end{equation}
The case of $i = 0$ corresponds to a reflection of the Euclidean spacetime ``time coordinate''.
In continuing back to Lorentzian signature via $x^{0} \mapsto i x^{0} $, the action of $\mathsf{R}_0$
would act as the combination $\mathsf{CT}$, which can be specified even when charge conjugation $\mathsf{C}$ and time
reversal $\mathsf{T}$ do not make sense separately (see e.g. \cite{Witten:2015aba} for further discussion).
As nomenclature, we shall also refer to a scalar, vector and $p$-form as those which transform with their
standard geometric operations. We append the modifier ``pseudo'' or ``twisted'' whenever there is an additional
minus sign under geometric reflection operations in Euclidean signature.

\subsection{10d Origin of 7d Symmetries}

In this subsection we start from the discrete transformations
of 10d SYM theory with gauge group $G$ and study the descent
of these discrete transformations under compactification on
a $T^3$. This results in discrete symmetry operations for 7d SYM theory
in flat space.

To frame the discussion to follow, recall that the
field content of 10d SYM consists of a vector boson $A_M$ and a Majorana-Weyl spinor $\zeta$.
Our conventions for 10d spinors are summarized in Appendix \ref{app:SPINOR}.
For the vector boson $A_M$, the action of the various reflection symmetries (in Euclidean signature) is simply:
\begin{equation}
\mathsf{R}^{10d}_M : A_{N}(x) \mapsto (-1)^{\delta_{M,N}}A_{N}(\mathsf{R}^{10d}_M x),
\end{equation}
namely we flip the sign of the component of the gauge field which undergoes reflection. It is also convenient
to state this in terms of the transformation on the one-form $A = A_M dx^M$ which transforms as:
\begin{equation}
\mathsf{R}^{10d}_M : A(x) \mapsto A(\mathsf{R}^{10d}_M x).
\end{equation}
Although all fields transform in the adjoint representation, we can also speak of various
``charge conjugation operations'' which amount to automorphisms of the gauge group.

Turning next to the fermionic content of the theory, it is convenient
for our present purposes to work in terms of a Dirac spinor of $Spin(9,1)$
subject to various chirality and reality constraints. In terms of the 10d gamma matrices $\Gamma^{10d}_{M}$
acting on a Dirac spinor field
$\Psi(x)$, we have:
\begin{equation}\label{eqn:10dgaugino}
\mathsf{R}^{10d}_{M} : \Psi(x) \mapsto  \xi \Gamma^{10d}_{M} \Psi( \mathsf{R}^{10d}_{M} x).
\end{equation}
Here, we have introduced an arbitrary complex phase $\xi$, although physical considerations in any compactified
system restrict us to fourth roots of unity, i.e. $\xi^{4} = 1$. Of course, in 10d SYM, we have a Majorana-Weyl
spinor rather than a Dirac spinor. Consequently, there is no reflection symmetry per se in the 10d theory. We can, however,
still speak of symmetry transformations such as $\mathsf{R}^{10d}_{i} \mathsf{R}^{10d}_j$ which compose multiple reflections.
It is these composite operations which we expect to descend in various ways to compactified theories.

Let us now compactify on a $T^3$. We keep all field profiles trivial on the $T^3$, so we expect to get 7d SYM theory with gauge group $G$. Recall that the bosonic content consists of a vector boson and an R-symmetry triplet $\phi_a$, and the fermionic content consists of a
7d Dirac fermion which we can write as a pair of fermions $\Psi_I$ subject to a
symplectic-Majorana constraint. This is just inherited from the 10d Majorana-Weyl condition (see Appendix \ref{app:SPINOR}
for the precise form of this constraint). The flat space action is given by:
\al{S = \int dt\,d^{\,6}x\, \text{tr}\left[-\frac{1}{4} F_{\mu \nu} F^{\mu \nu}-\frac{1}{4} D_\mu \phi_a D^\mu \phi_a+ \frac{1}{4} [\phi_a,\phi_b][\phi_a,\phi_b]-\frac{i}{2} \overline{\Psi}^I \slashed{D}\Psi_I-\frac{i}{2} \sigma^a_{IJ}\overline \Psi^I [\phi_a,\Psi^J]\right]\,.
}

Let us now turn to the discrete reflection symmetries of the 7d system. We begin by working from a ``bottom up'' perspective,
emphasizing only the 7d geometric reflections. We will then look at how these descend from 10d.
We have the 7d reflections $\mathsf{R}^{7d}_i$,
as well as the ``analytic continuation'' of $\mathsf{R}^{7d}_0$ which we denote as $\mathsf{CT}^{7d}$.
To begin, we start with the action of $\mathsf{CT}^{7d}$ and $\mathsf{R}^{7d}_i$ on the fermions:
\al{ \mathsf{CT}^{7d}\,  \Psi_I(t, x_i) &= i\Gamma^{7d}_0 \Psi_I(-t,x_i)\,,\\
\mathsf{R}^{7d}_i\,  \Psi_I(t, x_j) &= \Gamma^{7d}_i \Psi_I(t,(-1)^{\delta_{ij}}x_j)\,.
}
Note that these actions are compatible with the symplectic-Majorana condition.
The action of $\mathsf{CT}^{7d}$ and $\mathsf R^{7d}_i$ leave the kinetic term invariant
if the action on the gauge field is\footnote{It is important to remember that $\mathsf{CT}^{7d} $
is anti-linear.}
\al{ \mathsf{CT}^{7d} A_\mu = (-1)^{\delta_{\mu 0}}A_\mu\,,\\
 \mathsf{R}^{7d}_i A_\mu = (-1)^{\delta_{\mu i}}A_\mu\,.
}
The last term that needs to be checked is the coupling between the fermions and the scalars.
One can show that:
\al{ \mathsf{CT}^{7d} \,\overline{\Psi}^I \Psi^J &= \overline{\Psi}^I \Psi^J\,,\\
\mathsf{R}^{7d}_i \,\overline{\Psi}^I \Psi^J &= -\overline{\Psi}^I \Psi^J\,.
}
Therefore in order for these transformations to be symmetries they need to act as
\al{ \mathsf{CT}^{7d}\phi_a = -\phi_a\,, \label{eqn:triplettransI} \\
 \mathsf{R}^{7d}_i \,\phi_a = -\phi_a\,. \label{eqn:triplettransII}
}

The fact that the R-symmetry triplet $\phi_a$ transforms as pseudo-scalars under the $\mathsf{R}^{7d}$ is
somewhat counterintuitive. At this point, we need to recognize that compactification on a $T^3$ can impact the reflection transformation
rules. In particular, the 7d \textit{physical} reflection symmetry is related to a composition of several 10d \textit{geometric} reflections.
For example, we have:
\begin{equation}
\mathsf{R}^{7d}_{i} =  \mathsf{R}^{10d}_{i}\mathsf{R}^{10d}_{7} \mathsf{R}^{10d}_{8} \mathsf{R}^{10d}_{9} ,
\end{equation}
and as already stated, these act on the fields as:
\begin{align}
\mathsf{R}^{7d}_i(\phi_a) & = -\phi_a, \; \; \; \; \; \mathsf{R}^{7d}_i(A)=A \\
\mathsf{R}^{7d}_i(\Psi_I) & =\mathsf{R}^{10d}_{789}\mathsf{R}^{10d}_i (\Psi^{10d}_{MW})=+ \Gamma^{7d}_i \Psi_I,
\end{align}
where we have written $A = A_i dx^i$ as a one-form. The overall sign in the fermion transformation rule follows if we assume we are reducing a positive chirality Majorana-Weyl fermion from 10d with phase $\xi = +1$

An implicit feature of our discussion so far has been a specific choice of reflection symmetry on the 7d fermions:
\begin{align}
\mathsf{CT}^{7d}\,  \Psi_I(\mathsf{CT}^{7d} x) &= i\Gamma^{7d}_0 \Psi_I(\mathsf{CT}^{7d} x)\,,\\
\mathsf{R}^{7d}_i\,  \Psi_I(\mathsf{R}^{7d}_i x) &= \Gamma^{7d}_i \Psi_I(\mathsf{R}^{7d}_i x)\,.
\end{align}
In particular, we see that this means $(\mathsf{CT}^{7d})^2 = (-1)^{F}$, and $(\mathsf{R}^{7d}_{i})^2 = 1$,
where $(-1)^F$ acts on a single fermion as $-1$. In other words, we have implicitly chosen to work on a 7d manifold
with $\mathsf{Pin}^{+}$ structure. We can alternatively ask whether we could have specified 7d SYM on a 7d manifold
with $\mathsf{Pin}^{-}$ structure. This alternate possibility would have occurred if we had demanded the transformation rules:
\begin{align}
\mathsf{CT}^{7d,-}\,  \Psi_I(\mathsf{CT}^{7d,-} x) &= \Gamma^{7d}_0 \Psi_I(\mathsf{CT}^{7d,-} x)\,,\\
\mathsf{R}^{7d,-}_i\,  \Psi_I(\mathsf{R}^{7d,-}_i x) &= i\Gamma^{7d}_i \Psi_I(\mathsf{R}^{7d,-}_i x)\,,
\end{align}
which would have resulted in $(\mathsf{CT}^{7d,-})^2 = 1$, and $(\mathsf{R}^{7d,-}_{i})^2 = (-1)^F$.

Indeed, depending on the number of coordinates in 10d that we reflect, we will obtain
different 7d structures, either $\mathsf{Pin}^+$ or $\mathsf{Pin}^-$.
The rule is the following one: a reflection of $2r$ coordinates in 10d gives a $\mathsf{Pin}^+$ structure
if $r$ is even or a $\mathsf{Pin}^-$ structure if $r$ is odd. The reason for this is that when
doing such a reflection twice, one is performing a $2 \pi$ rotation in $r$ two-dimensional planes
upon which spinors acquire a $(-1)^r$ factor. See for example \cite{Montero:2020icj} for more
details and examples in other dimensions. This explains the rule for reflections we obtained before:
the $\mathsf{R}_i$ transformation gives a $\mathsf{Pin}^+$ structure thus requiring the reflection
of all internal coordinates and therefore a flip in sign of the adjoint scalars.
A $\mathsf{Pin}^-$ structure can be obtained by modifying the transformation rules for the spinors,
for example taking
\al{\mathsf{R}^{7d,-}_i\,  \Psi_I(t, x_j) &= i (-1)^{I} \Gamma_i \Psi_I(t,(-1)^{\delta_{ij}}x_j)\,.
}
This would force the following transformation on the adjoint scalars
\al{ \mathsf{R}^{7d,-}_i \,\phi_1 &= \phi_1\,,\\
\mathsf{R}^{7d,-}_i \,\phi_2 &= \phi_2\,,\\
\mathsf{R}^{7d,-}_i\,\phi_3 &= -\phi_3\,,
}
meaning that only one internal coordinate is reflected. In this case, a topological twist of 7d SYM would necessarily
require us to simultaneously switch on a background $SU(2)$ R-symmetry gauge field as well as a discrete reflection symmetry gauge field.
While this would certainly be interesting to study further, in what follows we exclusively consider the case where our 7d theory is placed on a
$\mathsf{Pin}^+$ background.

\subsection{11d Origin of 7d Symmetries}

In the previous subsection we focused on reflections of 7d SYM, as generated from compactification of 10d SYM.
From an effective field theory standpoint, this is sufficient to understand many aspects of how reflections
will descend to 3d vacua of local $Spin(7)$ geometries.

On the other hand it is somewhat unsatisfactory because it deprecates the role of the original compactification
geometry. Since part of our aim is to understand how compactifications of singular $Spin(7)$ spaces
can give rise to various 3d $\mathcal{N} = 1$ theories, we clearly need to also track the geometric origin
of these discrete symmetry transformations from an M-theory perspective.
With this in mind, our plan in this subsection will be to study how the same sorts
of transformations descend from M-theory compactified on an
ADE singularity.

This already leads to a puzzle when considering the ``parity assignment'' for the adjoint-valued Higgs fields of 7d SYM.
Recall that metric moduli of an M-theory compactification descend to scalars, as opposed to pseudo-scalars.
On the other hand, we also know from the previous subsection that the adjoint-valued Higgs fields
of 7d SYM transform (on a $\mathsf{Pin}^+$ background) as pseudo-scalars. From the perspective of 10d SYM theory,
this comes about because a suitable combination of the 10d geometric reflections descends to the 7d reflection symmetry,
namely $\mathsf{R}_{i}^{7d} = \mathsf{R}_{i}^{10d} \mathsf{R}_{7}^{10d} \mathsf{R}_{8}^{10d} \mathsf{R}_{9}^{10d}$.
Here, we ask how all of these reflections instead descend from an 11d starting point.

To avoid conflating the various notions of reflection symmetries discussed earlier, we refer to the 11d Euclidean
reflection symmetries as $\mathcal{R}_{i}$, and the analytic continuation of $\mathcal{R}_0$ to Lorentzian signature
as $\mathcal{CT}$. Our aim will be to understand how these geometric reflections produce reflections on the fields of 7d SYM.

To frame the discussion to follow, it is actually helpful to first begin with the reflection transformations on the M-theory
fields. We expect to be able to consider M-theory on $\mathsf{Pin}^{+}$ backgrounds because the 11d Majorana condition for the supercharges is compatible with a $\mathsf{Pin}^{+}$ rather than $\mathsf{Pin}^{-}$ structure (see e.g. \cite{Berg:2000ne, Tachikawa:2018njr}). The field content of M-theory consists of a ``three-form'' potential $C_3$, as well as the metric and 11d gravitino.

Consider first the transformation rules for $C_3$. An important subtlety here is that this is actually a \textit{pseudo} three-form, or in more technical terms a twisted three-form,\footnote{Recall that a twisted differential form is one that is twisted by the orientation bundle. A twisted differential form $\rho$ on any manifold $M$ behaves as follows: given a map $f : M \rightarrow M$ that is orientation reversing then $f^* \rho = - \rho$.} as can be seen by analyzing
the reflection symmetry transformation properties of the Chern-Simons coupling $C \wedge G \wedge G$ (see e.g. \cite{Cremmer:1978km}).\footnote{Briefly, consider M-theory on an orientable background $Y_{11}$. The
reflection symmetries act as $\mathcal{R}_i:\int_{Y_{11}}CGG\rightarrow (-1)^3\int_{\overline{Y_{11}}} CGG=\int_{Y_{11}}CGG\,$, where $\overline{Y_{11}}$ denotes the orientation reversal of $Y_{11}$.} In components, the reflection transformation is:
\begin{equation} \label{eqn:REFLECT}
\mathcal{R}_S : C_{TUV}(X) \mapsto -(-1)^{\delta_{ST}}(-1)^{\delta_{SU}}(-1)^{\delta_{SV}} C_{TUV} (\mathcal{R}_S X),
\end{equation}
where the tensor indices run over the 11d spacetime directions. Again, the additional minus
sign compared with a geometric three-form tells us we are dealing with a twisted differential form.
The transformation rules for the metric are straightforward, and obey:
\begin{equation}
\mathcal{R}_S: G_{UV}(X) \mapsto  (-1)^{\delta_{SU}} (-1)^{\delta_{SV}} G_{UV}(\mathcal{R}_S X).
\end{equation}
Finally, we have the transformation rules for the gravitino:
\begin{equation}
\mathcal{R}_S: \Psi_{U}(X) \mapsto  (-1)^{\delta_{SU}} \Gamma^{11d}_{S} \Psi_{U}(\mathcal{R}_S X).
\end{equation}
We get bosonic degrees of freedom from the dimensional reduction of the three-form potential and the metric, and
their fermionic superpartners from the reduction of the gravitino degrees of freedom. In the case where we compactify on a space with singularities, we get additional degrees of freedom from branes wrapped on collapsing cycles.

Indeed, the case of primary interest to us here is where we compactify on an ADE singularity $\mathbb{C}^2 / \Gamma_{\mathfrak{g}}$ with $\Gamma_\mathfrak{g} \subset SU(2)$ a finite subgroup of ADE type, as indicated by the Lie algebra label $\mathfrak{g}$. Recall that this engineers 7d SYM. Resolving the singularity, the effective divisors intersect according to (minus) the Cartan matrix for the corresponding Lie algebra. The ``off-diagonal'' massive W-bosons of the gauge theory come from M2-branes wrapped over the simple roots. It is already instructive to consider the dimensional reduction of the M-theory three-form potential on the resolved space. Labelling the compactly supported basis of two-forms as $\omega_{I}$, we can decompose the three-form as:
\begin{equation}
C_{3} = \underset{I}{\sum} A^{(I)}_{1} \wedge \omega_{I},
\end{equation}
so we appear to get a collection of \textit{pseudo} one-forms in seven dimensions. Indeed, if we also track the dimensional reduction of the deformation moduli of the ADE singularity, we observe that these are just metric degrees of freedom which fill out an R-symmetry triplet.
From this perspective, it would appear that we get a 7d pseudo-vector multiplet rather than a vector multiplet for a $U(1)^r$ gauge theory.

By inspection, however, we see that if there happened to be a symmetry which acted on the $\omega_{I}$ as $\omega_{I} \mapsto - \omega_{I}$,
then we could compose this with the geometric reflections $\mathcal{R}_{i}^{11d}$ to again reach a 7d theory of a vector multiplet. What then is the geometric origin of these transformations? At some level, it is just the statement that our ADE singularity is building up the root space of a Lie algebra, and so we are free to consider the action of the automorphisms of this Lie algebra on the geometry. In some cases, the required automorphisms will just be inner automorphisms (namely they involve the adjoint action of the Lie algebra), and in other cases they will be outer automorphisms.

To illustrate, it is already instructive to return to 10d SYM theory compactified on a circle. According to our analysis of the previous subsection, we expect to get 9d vector multiplet, with bosonic field content consisting of a 9d vector and an adjoint-valued pseudo-scalar $\phi_{9d}$. Giving a vev to this pseudo-scalar in a direction of the Cartan subalgebra would break reflection symmetries. Observe, however, that we really have two symmetries, one is reflection $\mathsf{R}^{9d}$, and one is a ``charge conjugation'' operation $\mathsf{C}^{9d}: \langle \phi_{9d} \rangle \mapsto - \langle \phi_{9d} \rangle$, which descends from an automorphism of the Lie algebra. In this case, the combinations $\mathsf{C}^{9d} \mathsf{R}^{9d}$ are preserved on the moduli space. Of course, if we had started with the pseudo-vector multiplet theory, we would have instead retained $\mathsf{R}^{9d}$ and broken $\mathsf{C}^{9d} \mathsf{R}^{9d}$. Clearly, similar considerations descend to 7d SYM.

Returning to our discussion of M-theory backgrounds, we observe that the proposed transformation $\omega_{I} \mapsto - \omega_{I}$ is carrying out the desired operation, but now in geometric terms. Of course, in the full compactification we need to consider more than just the action of these conjugation operations on the Cartan subalgebra, and so we refer to this class of geometric operations as $\mathcal{A}_{\mathfrak{g}}$
to emphasize their connection to the automorphisms of the gauge algebra / singular geometry.

We summarize in table \ref{tab:AUTO} the action on the Lie algebra generators $T^A$ for
each automorphism associated with generalized charge conjugation $\mathcal{A}_{\mathfrak{g}}$.
\begin{table}[t!]
\begin{center}
\begin{tabular}{c|c|c|c|c|c}
 & $\mathfrak{u}(1)$ & $\mathfrak{su}(N)$ & $\mathfrak{so}(2N)$ & $\mathfrak{e}_6$ & $\mathfrak{e}_{7,8}$\\
 \hline
$\mathcal{A}_{\mathfrak{g}}(T^A)$ &$ - T^A$ &$-(T^a)^t$ & $\mathcal{O}^{-1}(T^A)\mathcal{O}$ & $\iota(F)$ & $\iota(F)$
.
\end{tabular}
\end{center}
\caption{Table illustrating the generalized charge conjugation operations generated by automorphisms $\mathcal{A}$ on the generators $T^A$ of a Lie algebra. In the $\mathfrak{so}(2N)$ case, $\mathcal{O}$ is a matrix of $O(2N)$ with determinant $-1$. In the case of $\mathfrak{e}_{6,7,8}$, $\iota$ refers to an involution, which is an outer automorphism for $\mathfrak{e}_6$ and an inner automorphism for $\mathfrak{e}_{7,8}$.}\label{tab:AUTO}
\end{table}
In the table, the matrix $T^A \in O(2N)$ has $\text{det}=-1$, and $\iota$ are different involutions on the $\mathfrak{e}_{6,7,8} $ Lie algebras that act as $-1$ on the maximal tori components of the fields (thereby extending the $U(1)$ definition of $\mathcal{A}_{\mathfrak{g}}$).
For $\mathfrak{e}_{6}$, this is an outer automorphism, while for
$\mathfrak{e}_7$ and $\mathfrak{e}_8$, it is an inner
automorphism.\footnote{Even though it may be undone by a gauge transformation for $\mathfrak{e}_{7,8}$,
it allows us to define the transformation on the moduli space of Higgsings
between various gauge groups (hyperk\"ahler moduli space of $ALE$ fibers from the M-theory viewpoint).}

Before composing $\mathcal{A}_{\mathfrak{g}}$ with $\mathcal{R}_i$, we take note of an interesting property of $\mathcal{R}_i$ on the non-abelian 7d vector multiplet. Since the M2-brane charge is reflection-odd, an M2-brane wrapping  $S^2$ cycles associated to simple roots of $\mathfrak{g}$ has its $U(1)^r$ charge vector transform as $\mathcal{R}_i: q_I\mapsto -q_I$ which means that in the singular limit, this amounts to a Lie algebra involution. So to complete the action of $\mathcal{R}_i $ to the off-diagonal field components, for say $\mathfrak{g}=\mathfrak{su}(N)$, requires
\begin{equation}
\mathcal R_i: A\rightarrow -(A)^t  \; \; \; \; \; \phi_a\rightarrow (\phi_a)^t
\end{equation}
where the overall sign is fixed by knowledge of the action of the maximal tori components.
Therefore, after composition with $\mathcal{A}_{\mathfrak{g}}$, we get back $\mathsf{R}^{7d}_i$ as before:
\begin{equation}
\mathsf{R}^{7d}_i=\mathcal{A}_{\mathfrak{g}} \circ \mathcal{R}_i
\end{equation}

It is now convenient to package all of this in terms of a table illustrating how the different symmetries of a higher-dimensional compactification descend to transformations. We collect in table \ref{tab:CRT} the transformation rules for 7d $\mathfrak{su}(N)$ gauge theory, illustrating the origins from M-theory on an A-type singularity. Similar tables can be constructed for the other Lie algebras using table \ref{tab:AUTO}.

\begin{table}[t!]
\begin{center}
\begin{tabular}{c|c|c|c|c}
 & $\mathcal R_i$ & $\mathcal {CT}$ & $\mathsf{R}^{7d}_i$ & $\mathsf{CT}^{7d}$ \\
 \hline
$A_\mu$ &$-(A)^t(-1)^{\delta \mu i}$ & $-(-A)^t(-1)^{\delta \mu 0}$ & $(-1)^{\delta \mu i}(A)$ & $(-1)^{\delta \mu 0}(A)$ \\
\hline
$\phi_a$ & $+(\phi_a)^t$& $+(\phi_a)^t$ &$-(\phi_a)$&$-(\phi_a)$\\
\hline
$\Psi_I$ & $-\Gamma_i (\Psi_I)^t$ & $-i\Gamma_0 (\Psi_I)^t$ &$\Gamma_i(\Psi_I)$ &$i\Gamma_0 (\Psi_I)$\\
\hline
\end{tabular}
\end{center}
\caption{Table of reflection transformation properties for the fields $A_\mu$, $\phi_a$ and $\Psi_I$ of 7d SYM with gauge algebra $\mathfrak{su}(N)$, with all fields presented as $N \times N$ matrices in the adjoint representation. We collect here both the
11d reflection transformations, as well as the 7d reflection transformations.}\label{tab:CRT}
\end{table}

The upshot of this analysis is that it is somewhat a matter of taste whether we refer to our 7d compactified theory as a pseudo-vector multiplet or as a vector multiplet. Said differently, different duality frames, be it heterotic on a $T^3$, or M-theory on a K3 surface might ``canonically favor'' one presentation or the other, but the further composition with an automorphism allows us to pass between the two presentations. In this sense, we are filling in an additional entry in heterotic / M-theory duality.

As an additional comment, our emphasis here has been on realizing 7d SYM theory with a simply laced gauge algebra. One can also ask about the non-simply laced case, which corresponds to quotienting the Lie algebra by a suitable outer automorphism. In M-theory terms, this amounts to activating a background discrete flux ``at infinity'' (see e.g. \cite{Witten:1997bs, deBoer:2001wca, Tachikawa:2015wka}), and in F-theory terms it can be obtained by starting with a non-compact elliptically fibered K3 surface, and upon compactifying on a further circle, twisting by this outer automorphism. Indeed, a general comment is that 7d SYM theory with a non-simply laced gauge theory requires the spin $1$ degrees of freedom to transform as a vector, simply because of the structure of the kinetic term $\mathrm{Tr}_{\mathfrak{g}} F^2$ under reflection symmetries.

\section{4d and 3d Reflections via 7d SYM \label{sec:4D3D}}

In the previous section we analyzed the reflection symmetries associated with 7d SYM theory.
In particular, we saw that compactification of 10d SYM (as obtained from heterotic on a $T^3$) and
M-theory compactification on an ADE singularity lead to different reflection parity assignments for the 7d fields, which are
related by a further automorphism action. Since we are ultimately interested in further compactification
of our system to 4d and especially 3d vacua, our aim here will be to understand the resulting reflection parity
assignments for zero modes generated by our local $Spin(7)$ system. With the analysis of the previous section in mind,
our starting point will be 7d SYM theory, with bosonic content
given by a 7d vector potential and a triplet of pseudo-scalars associated with the Higgs field. Our aim will be to track
the impact of further discrete transformations which come from compactifying on a three-manifold $M_3$ or four-manifold $M_4$.


This section is organized as follows. As a warmup, we first return to the case of the PW system compactified
on an orientable three-manifold $M_3$, and track how reflections act on
the resulting 4d fields. Then, we turn to the case of 3d $\mathcal{N} = 1$ vacua as obtained from our local $Spin(7)$ system
compactified on an orientable four-manifold $M_4$, and study the same question. To a certain extent, the reflection assignments
for local matter are dictated by effective field theory considerations, since we know that we
get 4d $\mathcal{N} = 1$ chiral multiplets from the PW system, and compactification on a circle produces 3d $\mathcal{N} = 2$ matter multiplets.
The situation is far less straightforward in the case of the bulk matter fields, since these fill out genuine 3d $\mathcal{N} = 1$
matter multiplets.

\subsection{4d Reflections}

As a warmup to the full problem, in this subsection we treat the case of 7d SYM compactified on an orientable three-manifold $M_3$.
Our task will be to understand the higher-dimensional origin of various reflection / parity assignments for modes of the resulting 4d system.

To a certain extent, we are just producing a 4d $\mathcal{N} = 1$ supersymmetric system, and we know how the
standard 4d operations referred to as $\mathsf{C}^{4d}$, $\mathsf{P}^{4d}$ and $\mathsf{T}^{4d}$ act on any 4d field theory, supersymmetric or not. In terms of our reflection operations, we of course have:
\begin{equation}
\mathsf{P}^{4d} = \mathsf{R}^{4d}_{1} \mathsf{R}^{4d}_{2} \mathsf{R}^{4d}_{3} \,\,\, \text{and}  \,\,\,  \mathsf{CT}^{4d} = \mathsf{R}^{4d}_{0},
\end{equation}
where in the last equation we are again analytically continuing from Euclidean to Lorentzian signature. Recall also that on a 4d $\mathcal{N} = 1$ chiral multiplet with scalar component $s_{4d} = s_1 + i s_2$, $s_1$ is a scalar, but $s_2$ is a pseudo-scalar (see e.g. reference \cite{Weinberg:2000cr}). Under parity, the chiral multiplet $S$ transforms as $S \rightarrow \overline{S}$, which also sends the left-handed
Weyl fermion of the supermultiplet to a right-handed Weyl fermion in the conjugate representation.

What complicates the story is that we are now asking how to lift these operations back to statements about a local $G_2$ system,
which as we have already remarked, can be analyzed by starting with 10d SYM theory compactified on $T^{\ast} M_3$. To determine the parity transformations for the 4d fields, we observe that although $T^{\ast} M_3$ is even-dimensional, we can, (since an oriented $M_3$ is automatically parallelizable) speak of an orientation reversal operation $\mathsf{O}_{T^{\ast}}$ on the cotangent directions as well, which locally looks like:
\begin{equation}
\mathsf{O}_{T^{\ast}} = \mathsf{R}^{10d}_{7} \mathsf{R}^{10d}_{8} \mathsf{R}^{10d}_{9},
\end{equation}
which we recognize as just the ``internal reflections'' associated with the compactification of 10d SYM theory on a $T^3$. Putting these comments together, we recognize that the complexified connection $\mathcal{A} = A + i \phi_{\mathrm{PW}}$ of the PW system breaks up into
a 4d scalar and a 4d pseudo-scalar.

Summarizing, we conclude that the appropriate reflection symmetries
of the 4d effective field theory descend as:
\begin{equation} \label{eqn:4d10dlift}
\mathsf{R}^{4d}_{i} = \mathsf{R}^{7d}_{i},
\end{equation}
where now $i = 0,1,2,3$ for the 4d Euclidean directions. Of course, a particular compactification need not preserve parity, but this depends
on the specific matter content and interactions of the resulting effective field theory.

\subsection{3d Reflections}\label{subsec:3dref}

Let us now turn to the reflection assignments for the zero modes generated from compactification of 7d SYM theory on a four-manifold $M_4$.
Here, it is helpful to start with the fermionic content, including the supercharges of the 7d theory.

Now, a 7d Dirac spinor can be decomposed into the fields $\lambda_\ell$ and $\lambda_r$ which transform as $(\textbf{2},\textbf{2},\textbf{1},\textbf{2})$ and $(\textbf{2},\textbf{1},\textbf{2},\textbf{2})$, respectively, under $Spin(1,2)\times SU(2)_\ell\times SU(2)_r \times SU(2)_R$, where here $SU(2)_\ell\times SU(2)_r$ denotes the Lorentz group on $M_4$ and $SU(2)_R$ refers to the R-symmetry. At the level of structure groups, we topologically twist by identifying $SU(2)_{r'} \equiv \textnormal{diag}(SU(2)_r \times SU(2)_R)$ so that in terms of the representations of $Spin(1,2)\times SU(2)_\ell\times SU(2)_{r'}$, the 3d fermions of interest are
\begin{equation}
\lambda_\ell\equiv \chi  \leftrightarrow (\textbf{2},\textbf{2},\textbf{2})\,,
\end{equation}
\begin{equation}
\lambda_{r'}\equiv (\psi,\lambda)  \leftrightarrow  (\textbf{2},\textbf{1},\textbf{3})\oplus (\textbf{2},\textbf{1},\textbf{1})\,.
\end{equation}
The twisted fields, $\chi$, $\psi$, and $\lambda$ are the 3d superpartners for the fluctuations $a$, $\varphi$, and $A_{3d}$, respectively.
From the 7d SYM action, we see under a parity transformation in a 3d spatial direction, $\mathsf{R}^{7d}_i$, that $a$ is even, while $\varphi$
is odd. Furthermore, since $\Gamma_i=\sigma_i\otimes \gamma^{(4d)}_5$, the 3d fermions transform as (we suppress the action on the spacetime coordinates):
\begin{equation}\label{eqn:parities}
\mathsf{R}^{7d}_i(\chi)=+\gamma_i \chi \; \; \;  \; \mathsf{R}^{7d}_i(\psi)=-\gamma_i \psi \;\; \; \; \mathsf{R}^{7d}_i(\lambda)=-\gamma_i \lambda\,.
\end{equation}
We note that the reflection transformations for the 3d gaugino $\lambda$ are the same as for the supercharge. Indeed, the supercharge transforms as $\mathsf{R}^{7d}_i(Q_\alpha)=-\gamma_iQ_\alpha$, and considering $Q_{\alpha}=\frac{\partial}{\partial \theta^\alpha}-i\theta^\beta \gamma_{\beta}^{\alpha \mu}\partial_\mu$, we see that the the $\mathcal{N}=1$ superspace coordinate, $\theta_\alpha$, transforms as $\mathsf{R}^{7d}_i(\theta_\alpha)=-\gamma_i\theta_\alpha$, $\mathsf{R}^{7d}_i(Q_\alpha)=-\gamma_iQ_\alpha$.

It is convenient to package the field content of the 7d SYM theory in terms of 3d $\mathcal{N} = 1$ superfields. In the case of $A_{3d}$ and the 3d gaugino, this is just the 3d $\mathcal{N} = 1$ vector multiplet, where $A_{3d}$ transforms as a vector (as opposed to a pseudo-vector) and the 3d gaugino is reflection odd (as in equation (\ref{eqn:parities})). As for the bulk modes associated with the internal fluctuations of the vector potential $a$ and the self-dual Higgs field $\varphi$, these split up into adjoint valued 3d $\mathcal{N} = 1$ pseudoscalar and scalar multiplets:
\begin{equation}
\textnormal{pseudoscalar multiplet:  } \boldsymbol{\varphi}=\varphi+i\theta^\alpha \psi_\alpha + \frac{i}{2}\theta^\alpha \theta_\alpha F_{\varphi}
\end{equation}
\begin{equation}
\textnormal{scalar multiplet:  } \boldsymbol{a}=a+i\theta^\alpha \chi_\alpha +\frac{i}{2} \theta^\alpha \theta_\alpha F_{a}.
\end{equation}
On a general $M_4$, there is no natural way to combine these fields into
``complexified combinations'' as would be appropriate for 3d $\mathcal{N} =2 $ matter multiplets.

Let us now turn to the localized matter fields. We treat the ``generic situation'' where matter is localized on a one-dimensional subspace of $M_4$. Recall that this matter can be treated locally as a zero mode of the PW system on a three-manifold which is then further compactified on a circle. At this level of analysis, it is therefore enough for us to track the reflection assignments of a chiral and anti-chiral superfield after reduction on circle. Of course, these zero modes descend from trapped fluctuations of the bulk modes, but we leave this point implicit in what follows.

Now, for a 4d $\mathcal{N} = 1$ chiral multiplet reduced on a circle, we start with a 4d real scalar and a pseudo-scalar, as well as a left-handed Weyl fermion. Reduction on a circle gives us a 3d real scalar and a pseudo-scalar, as well as a 3d Dirac fermion $\psi_{D}$. Since the 4d theory was chiral, it is a bit of a misnmomer to speak of reflection in just one direction. We can, however, compose two such reflections to generate a sensible 3d operation:
\begin{equation}\label{eq:4d3dperp}
\mathsf{R}_{i}^{3d} \psi_{D} = \mathsf{R}_{i}^{4d} \mathsf{R}_{\bot}^{4d} \psi_{D},
\end{equation}
where we treat $\psi_{D}$ as a 4d spinor, and $\mathsf{R}_{\bot}^{4d}$ denotes reflection along the circle. So, it does make sense to speak of a definite reflection assignment for this 3d Dirac fermion.

What then, is the reflection assignment for this zero mode? The first comment is that we expect opposite reflection assignments for reduction of a left-handed and right-handed Weyl fermion. The second comment is that we expect that circle reduction of a
4d $\mathcal{N} = 1$ gauge theory will lead to Weyl fermions all with the same reflection assignment. Since we have already fixed the reflection assignment for our 3d $\mathcal{N} = 1$ vector multiplet to be $-1$, we conclude that the 3d Dirac fermion $\psi_D$ obtained from reduction of a left-handed Weyl fermion will produce reflection assignment $-1$. Conversely, if we had started with a right-handed Weyl fermion,
the resulting 3d Dirac fermion $\widetilde{\psi}_D$, would have had a reflection assignment of $+1$. We can explicitly see this if we compactify a 4d Dirac fermion $\begin{pmatrix}
           \Psi_L \\
           \Psi_R
        \end{pmatrix}$ to two 3d Dirac fermions. Consider the standard Weyl basis of 4d gamma matrices:
\begin{equation}
\gamma_0=i\sigma_2\otimes \sigma_3, \; \gamma_1=\sigma_2\otimes \sigma_1, \; \gamma_2=\sigma_2\otimes \sigma_2, \; \gamma_\perp=\sigma_1\otimes \mathsf{1},
\end{equation}
then we have that $\mathsf{R}_1\mathsf{R}_\perp(\Psi_L)=-i
\gamma_1\gamma_\perp \begin{pmatrix}
           \Psi_L \\
           \Psi_R
        \end{pmatrix}=\gamma^{(3d)}_1 \begin{pmatrix}
           -\Psi_L \\
           \Psi_R
        \end{pmatrix}$ after dimensional reduction which returns our proposed 3d reflection phases.

Geometrically, recall that for the PW system on a three-manifold $M_3$, we get chiral versus anti-chiral matter depending on the signature of the Hessian $\ast_{M_3} d \ast_{M_3} \phi_{\mathrm{PW}}$. In particular, a Hessian with signature $(+,+,-)$ produces a left-handed Weyl fermion (a chiral mode) while a Hessian with signature $(-,-,+)$ produces a right-handed Weyl fermion (an anti-chiral mode). Proceeding now to the case of the local $Spin(7)$ system with matter localized on a circle, we see that these two choices of signature are related by an orientation reversal in the directions transverse to the matter circle.

Indeed, in a local neighborhood of the matter circle, we can define the operation $\mathsf{R}_{i}^{10d} \mathsf{R}^{10d}_{3456789}$. We observe that $\mathsf{R}^{10d}_{3456789}$ is a local orientation reversal operation on $\mathbb{R}^{7}$, but there is no generic expectation that on $\Lambda^{2}_{\mathrm{SD}} M_4$ that such an action is a symmetry of the geometry. For local matter, however, where the geometry really is just $S^1 \times \mathbb{R}^{6}$, we do have such a symmetry action, and this specifies a well-defined notion of how reflections descend to three dimensions.

Let us summarize the reflection assignments for the various bosonic and fermionic matter fields. Starting first
with the bulk fluctuations $a$, $\varphi$, as well as the localized ``complex'' scalars
$\tau_{+,+,-}$ and $\tau_{-,-,+}$, we have:\footnote{For the zero modes localized on a circle $S_{\bot}^{1}$,
our 3d reflection $\mathsf{R}_{i}^{3d} = \mathsf{R}_{i}^{4d} \mathsf{R}^{4d}_{\bot}$. This introduces an additional
minus sign in the reflection assignment for the pseudo-scalar of the 4d $\mathcal{N} = 1$ chiral multiplet, so that in 3d we wind up
with a pair of scalars or a pair of pseudo-scalars. We can also consider composition with a local charge conjugation operation $\mathsf{C}$,
and then it is a matter of convention what one wishes to call ``reflection'', i.e. $\mathsf{R}$ versus $\mathsf{CR}$.
Here, we have opted to use the same conventions for reflection on 3d Dirac fermions used in reference \cite{Witten:2015aba}.}
\begin{align}
a \rightarrow + a, \,\,\, \varphi \rightarrow - \varphi, \,\,\,
\tau_{+,+,-} \rightarrow - \tau_{+,+,-}, \,\,\, \tau_{-,-,+} \rightarrow + \tau_{-,-,+}.
\end{align}
Here, we remark that the signature of the internal Hessian flips sign under this 3d reflection.
For the fermionic fluctuations associated with each of these modes, we have:
\begin{equation}\label{eqn:finalreflections}
\chi^{(a)} \rightarrow +\gamma_{i} \chi^{(a)}, \,\,\, \psi^{(\varphi)} \rightarrow - \gamma_{i} \psi^{(\varphi)}, \,\,\,
\psi_{+,+,-} \rightarrow - \gamma_{i} \psi_{+,+,-}, \,\,\, \psi_{-,-,+} \rightarrow + \gamma_{i} \psi_{-,-,+},
\end{equation}
in the obvious notation.

An additional comment here is that if we neglect interaction terms, there are a large number of additional symmetries which act on our fields.
Indeed, in addition to the 3d reflection symmetry $\mathsf{R}_{i}^{3d}$, we can also speak of various charge conjugation actions which act on individual localized Dirac fermions as $\psi_{D} \rightarrow \psi_{D}^{\ast}$. From the perspective of compactifying 7d SYM, this provides us with a large set of candidate symmetries. Of course, in an actual compactification, we expect many of these symmetries to wind up being broken. The way to extract the ones which are preserved involves simply working out the 3d effective field theory and checking which symmetries are preserved at the level of the classical interaction terms. Our discussion here has primarily emphasized what a 3d field theorist would see, but a priori, there is no guarantee that a given compactification will actually preserve these symmetries. We return to this issue in section \ref{sec:QUANTUM}. For additional discussion of the various symmetries which appear in the free field limit, see Appendix \ref{app:CHARGEREF}.

\section{Parity Anomalies\label{sec:ANOMO}}

Having analyzed the reflection symmetry transformations for the matter content of 7d SYM coupled to various lower-dimensional defects,
we now turn to the anomalies generated by the matter content of this system. Again, our motivation for doing so is that
anomalies are robust against strong coupling effects, and so are particularly important to track in local $Spin(7)$ systems.
In Appendix \ref{app:BORDISM} we review some additional aspects of parity
anomalies of 3d gauge theories which we implicitly assume below. Again, we note that as discussed in \cite{Witten:2016cio}, it is
more appropriate to view these anomalies as associated with a reflection.

In general terms, we shall be interested in the resulting 3d effective field theory we get after compactification.
In this 3d theory, we can consider working on a Euclidean signature three manifold $N_3$. If we have a symmetry $\mathsf{S}$
of the classical action, we can ask about the response to the partition function $Z(N_3)$ under such a symmetry
transformation:\footnote{Here we assume the existence of a partition function. A common situation, especially in string compactification
on a non-compact background is that our 3d QFT is actually a relative QFT in the sense that we have a vector of partition functions. We ignore this subtlety in what follows. For further discussion, see e.g. \cite{Tachikawa:2013hya, DelZotto:2015isa, Witten_1997, Aharony_1998, Witten_1998, Moore_2005, Belov:2006jd, Freed_2007, witten2008conformal, Witten:2009at, Henningson:2010rc, freed2014relative, Monnier:2014txa, Monnier:2016jlo, Monnier:2017klz}.}
\begin{equation}
Z(N_3) \rightarrow \exp( 2\pi i \nu_{\mathsf{S}} / K_{\mathsf{S}}) Z(\mathsf{S} N_3),
\end{equation}
where here, we allow for the possibility that $\mathsf{S}$ is a spacetime transformation which acts on $N_3$.
The parameter $K_{\mathsf{S}}$ is a positive integer, and so we often refer to $\nu_{\mathsf{S}}$ as valued in the
finite group $\mathbb{Z}_{K_{\mathsf{S}}}$.
In our setting, we can consider the action of the 3d reflection symmetries $\mathsf{R}_{i}^{3d}$, of which there is a corresponding
``gravitational-parity anomaly'' $\nu_{\mathsf{R}}$. We remark that this phase is defined modulo $\mathbb{Z}_{16}$, as explained in reference \cite{Witten:2016cio} (see also \cite{Hsieh:2015xaa}). Some systems also have an independent charge conjugation symmetry, and so in principle we could also discuss $\nu_{\mathsf{CR}}$, which in Lorentzian signature is synonymous with $\nu_{\mathsf{T}}$, i.e. a time reversal anomaly.

Of course, we can also have mixed ``parity'' / gauge symmetry anomalies. For a theory with gauge group $H$, we can label these as $\nu_{\mathsf{R} H}$. In practical terms, one can compute this by tracking a possible one-loop shift to the Chern-Simons coupling from massless fermions.\footnote{One way to think about this has to do with introducing a Pauli-Villars regulator for
the loop integral. Picking a sign for the Pauli-Villars mass then produces a consequent shift from a single massless mode.}
For a massless fermion in a representation $\mathbf{R}$, we get a shift at one-loop given by
\al{ \delta k =  \alpha\, \mathbf{T}(\mathbf{R})\,,
}
where $\mathbf{T}(\mathbf{R})$ is the index of the representation $\mathbf{R}$, and we implicitly mean a 3d Dirac fermion with $\alpha=1$ if the representation is pseudo-real or complex, and a 3d Majorana fermion with $\alpha= 1/2 $ if the representation is real. The normalization is such that $\mathbf{T}(\mathbf{N})= 1/2 $ for the fundamental representation of $SU(N)$. As an example one can see that for a $\mathcal N=1$ $SU(N)$ theory with $N_f$ flavors the shift is
\al{\delta k = \frac{N_f}{2} + \frac{h^\lor}{2}\,,
}
where $h^\lor$ is the dual Coxeter number of the gauge group (for $SU(N)$ we have that $h^\lor = N$). Given that the bare Chern--Simons level $k$ is an integer in order to be able to have a zero Chern--Simons level at one-loop one must ensure the condition that $N_f = N \mod 2$ for these theories, matching with e.g. reference \cite{Gaiotto:2018yjh}. In table \ref{tab:group}, we collect our conventions for various indices of representations. Lastly, it is tempting to also consider the parity anomalies associated with placing 7d SYM on a general background which has some reflection symmetries, but we leave this to future work.

\begin{table}[!t]
\centering
\begin{tabular}{c|l|l|l}
$G$ &  &   &\\
\hline
\hline
$SU(N)$ & $\mathbf{T}(\mathbf{N}) = 1/2$ & $\mathbf{T}(\mathbf{adj}) = N$\\
\hline
$Spin(N)$ & $\mathbf{T}(\mathbf{N}) = 1$ & $\mathbf{T}(\mathbf{adj}) = N-2$\\
\hline
$E_6$ & $\mathbf{T}(\mathbf{27}) = 3$ & $\mathbf{T}(\mathbf{adj}) = 12$\\
\hline
$E_7$ & $\mathbf{T}(\mathbf{56}) = 6$ & $\mathbf{T}(\mathbf{adj}) = 18$\\
\hline
$E_8$ & $\mathbf{T}(\mathbf{248}) = 30$\\
\hline
\hline
$SU(5)$ & $\mathbf{T}(\mathbf{5}) = 1/2$ & $\mathbf{T}(\mathbf{10}) = 3/2$ &$\mathbf{T}(\mathbf{adj}) = 5$\\
\hline
$Spin(10)$ & $\mathbf{T}(\mathbf{10}) = 1$ & $\mathbf{T}(\mathbf{16}) = 2$ &$\mathbf{T}(\mathbf{adj}) = 8$\\
\hline
\end{tabular}
\caption{Indices for various representations of ADE groups}
\label{tab:group}
\end{table}

A general comment here is that in passing from our local $Spin(7)$ system to our 3d effective field theory,
we can expect some of the symmetries of the 3d theory to be inherited from ``top-down'' considerations, while some may only be emergent after some details of the compactification have decoupled. In this section we work in a specific idealized limit where we assume that the only interactions between zero modes are mediated by the 3d $\mathcal{N} = 1$ vector multiplet. In this case, the issue of parity assignments just follows from tracking the overall matter content of the system, an issue we just addressed in section \ref{sec:4D3D}. Of course, we expect that in a full compactification there may be various parity breaking effects. For example, integrating out the Kaluza-Klein (KK) modes of the four-manifold
can be visualized as starting with a collection of 3d Dirac fermions with mass terms added. Since the sign of these mass terms will shift the bare Chern-Simons coupling, the net impact from such zero modes can also produce an explicit parity breaking interaction (the Chern-Simons term itself). If, however, our $M_4$ enjoys an internal reflection symmetry, then these KK modes come in pairs, and there will be no overall shift
to the level.

With this simplified situation in mind, we just need to track the contributions
from the bulk zero modes and the localized zero modes
to the various parity anomalies. A general rule of thumb is that for a single
3d Majorana fermion $\chi$ with parity $\varepsilon_{\chi} = \pm 1$ under a reflection:
\begin{equation}
\mathsf{R}_{i}^{3d} \chi = \varepsilon_{\chi} \gamma_{i}^{3d} \chi,
\end{equation}
the contribution to the anomaly $\nu_{\mathsf{R}}$ is just $\nu_{\mathsf{R}}(\chi) = - \varepsilon_{\chi}$,
where our overall choice of minus sign convention has been introduced to avoid clutter in later equations.
By the same reasoning, if we have a 3d Dirac fermion $\psi$ with parity $\varepsilon_{\psi} = \pm 1$ which transforms in a complex representation $\mathbf{R}$ of the unbroken gauge group $H = G \times U(1)^n \subset \widetilde{G}$ (with notation as in section \ref{sec:Spin7Review}),
then the contribution to the anomaly is $\nu_{\mathsf{R}}(\psi) = \varepsilon_{\psi} \mathrm{dim}_{\mathbb{R}} \mathbf{R} $, i.e. we can just count the number of real degrees of freedom. Returning to our reflection assignments
from equation (\ref{eqn:finalreflections}) which we reproduce here:
\begin{equation}
\chi^{(a)} \rightarrow + \gamma_{i} \chi^{(a)}, \,\,\, \psi^{(\varphi)} \rightarrow - \gamma_{i} \psi^{(\varphi)}, \,\,\,
\psi_{+,+,-} \rightarrow - \gamma_{i} \psi_{+,+,-} \,\,\, \psi_{-,-,+} \rightarrow + \gamma_{i} \psi_{-,-,+}.
\end{equation}
There are bulk and local contributions to $\nu_{\mathsf{R}}$:
\begin{equation}
\nu_{\mathsf{R}} = \nu^{\mathrm{bulk}}_{\mathsf{R}} + \nu^{\mathrm{local}}_{\mathsf{R}}.
\end{equation}
The contribution from the bulk modes comes from the 3d gaugino of the vector multiplet,
and the superpartners of the internal one-form and self-dual two-form:
\begin{equation} \label{bulknu}
\nu^{\mathrm{bulk}}_{\mathsf{R}} = \mathrm{dim}_{\mathbb{R}} H \times (1 - b^{1} + b^{2}_{\mathrm{SD}}) \mod 16,
\end{equation}
where the overall factor of $\mathrm{dim}_{\mathbb{R}} H$ is due to the fact that
all zero modes in the bulk transform in the adjoint representation. Note that this also includes the
singlets associated with the ``adjoint'' of the $U(1)$ factors of $H = G \times U(1)^n \subset \widetilde{G}$.
An important feature of equation \ref{bulknu} is that we can express it in terms of the Euler character and signature of
$M_4$, namely we have:
\begin{equation} \label{eqn:bulknu}
\nu^{\mathrm{bulk}}_{\mathsf{R}} = \mathrm{dim}_{\mathbb{R}} H \times \frac{1}{2}( \chi(M_4) + \sigma(M_4)) \mod 16,
\end{equation}
where $\chi(M_4) = 2 - 2b^1 + b^2 $ and $\sigma(M_4) = b^{2}_{\mathrm{SD}} - b^{2}_{\mathrm{ASD}}$.\footnote{Note that we met this topological quantity previously in section \ref{sec:Spin7Review} in the context of determining the number of zero-circles of $\Phi_{\mathrm{SD}}$ modulo two (see \cite{perutz2006zero} for some discussion on the topological significance of this quantity).} So in other words, $\nu^{\mathrm{bulk}}_{\mathsf{R}}$ detects
a topological structure on $M_4$, at least modulo $\mathbb{Z}_{16}$. Another comment is
that this provides some a posteriori justification for the reflection assignments previously found in
section \ref{sec:4D3D}.

Turning next to the localized zero modes, we can have different complex representations contributing.
Additionally, there is an overall minus sign having to due with whether
we have a $(+,+,-)$ mode or a $(-,-,+)$ mode. Splitting up the zero modes into these two sets,
we get:
\begin{equation}
\nu^{\mathrm{local}}_{\mathsf{R}} =
\underset{i \in (+,+,-)}{\sum} \mathrm{dim}_{\mathbb{R}} \mathbf{R}_{i} - \underset{i^{\prime} \in (-,-,+)}{\sum} \mathrm{dim}_{\mathbb{R}} \mathbf{R}_{i^{\prime}} \mod 16.
\end{equation}

Consider next the mixed gauge-parity anomaly $\nu_{\mathsf{R} H}$. For $H$ a non-abelian simply connected gauge group, this is a mod $\mathbb{Z}_2$ phase, and for $H = U(1)^n$, this is a mod $\mathbb{Z}_4$ phase. For more general gauge group choices, this depends on the reduced $\mathsf{Pin}^{+}$ bordism groups, a point we discuss further in Appendix \ref{app:BORDISM}. In any event, we again just need to count the different contributions from the bulk and localized zero modes. In general terms, for a 3d Majorana fermion $\chi$ with reflection assignment $\varepsilon_{\chi}$ in a real representation $\mathbf{R}$ of $H$, the contribution to $\nu_{\mathsf{R} H}$ is
$\nu_{\mathsf{R} H}(\chi) = \varepsilon_{\chi} \mathbf{T}(\mathbf{R})$, while for a 3d Dirac fermion $\psi$ with reflection assignment
$\varepsilon_{\psi}$ in a complex or pseudo-real representation $\mathbf{R}$ of $H$,
the contribution to $\nu_{\mathsf{R} H}$ is
$\nu_{\mathsf{R} H}(\psi) = 2 \varepsilon_{\psi} \mathbf{T}(\mathbf{R})$.
Consequently, the contributions to $\nu_{\mathsf{R} H}$ from
the bulk and localized zero modes assemble in the expected way:
\begin{align}
& \nu_{\mathsf{R}H}  = \nu^{\mathrm{bulk}}_{\mathsf{R}H} + \nu^{\mathrm{local}}_{\mathsf{R}H} \mod 2, \\
& \nu^{\mathrm{bulk}}_{\mathsf{R}H} = h^{\lor}(H) \times \frac{1}{2}( \chi(M_4) + \sigma(M_4)) \mod 2, \\
& \nu^{\mathrm{local}}_{\mathsf{R}H} =
\underset{i \in (+,+,-)}{\sum} 2\mathbf{T}(\mathbf{R}_{i}) - \underset{i^{\prime} \in (-,-,+)}{\sum} 2\mathbf{T}(\mathbf{R}_{i^{\prime}}) \mod 2.
\end{align}
As a brief comment, we see that the contribution to the Chern-Simons
level (working modulo the integers) is just $2 \nu_{\mathbf{R} H}$.

Having spelled out the general contributions to anomalies from our
bulk and localized matter, we are now ready to use this to track
various quantum corrections which appear in local $Spin(7)$ systems.
We turn to this next.

\section{Quantum Corrections to $Spin(7)$ Matter \label{sec:QUANTUM}}

As already mentioned in previous sections, part of our goal in studying various reflection symmetries
of 7d SYM is to use this additional structure to potentially constrain the effects of quantum corrections.
Now, an important issue we face is that a priori, there is no reason to expect M-theory on a generic $Spin(7)$ background
to preserve reflection symmetries, and there is a remnant of this same issue in compactification of 7d SYM compactified on a
four-manifold $M_4$. The way this shows up in the resulting 3d effective field theory is that various \textit{candidate}
discrete symmetry transformations will end up being broken by interaction terms.

On the other hand, we can also consider tuning the moduli of the compactification so that a reflection symmetry is preserved. This is
actually more common than one might initially anticipate. For example, since M-theory also
includes objects such as orientifold M2-planes (i.e. OM2-planes), we expect that for a suitable choice of moduli, we have internal
reflection symmetries which act on the $Spin(7)$ space. Indeed, there are also well-known compact examples of $Spin(7)$ spaces which exhibit such symmetries (see e.g. \cite{Joyce:1999nk, JoyceBook}). In the context of a local model, this can also be arranged, for example, the local $G_2$ space $\Lambda^{2}_{\mathrm{SD}} S^4$ with a round metric for the four-sphere has many such symmetries.

We also expect that it is possible to compactify M-theory on various non-orientable backgrounds,
so long as they admit a $\mathsf{Pin}^{+}$ structure. From an effective field theory point of view, this involves
gauging some combination of reflection symmetries, but for this to be so, one must have a reflection symmetry in the first place!

With these considerations in mind, we adopt the following strategy. We begin by building a local model which has an internal reflection symmetry. From what we have worked out in the previous sections, we know that this will then descend to a reflection symmetry of the 3d effective field theory. This in turn puts strong constraints on possible quantum corrections to the matter content. In particular, we can then use 't Hooft
anomaly matching to track the value of $\nu_{\mathsf{S}}$ for any spacetime symmetry $\mathsf{S}$. So, even if a particular localized or bulk zero mode does not survive, it can still leave a remnant in its contribution to $\nu_{\mathsf{S}}$.

At a more generic point of the moduli space where compactification effects explicitly break a reflection symmetry, it is still in principle possible for the resulting 3d effective field theory to be reflection symmetric. For example, if we add an irrelevant reflection breaking operator to an effective field theory, then in the deep infrared one can still recover such a symmetry. Of course, relevant interaction terms will
impact the surviving symmetries.

Of particular significance is the Chern-Simons interaction term. This can receive a ``bare'' contribution from the ambient four-form
flux of an M-theory compactification, as well as from integrating out KK modes of the compactification. One of our aims in this section will be
to determine the overall value of this coupling. Our general claim is that we can use the zero mode content of the 3d effective field theory as a proxy for determining the presence (or absence) of such a contribution. We note that in some cases, the appearance of a background $G$-flux is sometimes required by a quantization condition \cite{Gukov:2001hf, Gukov:2002zg}, and we will see a sharp analog of this in the context of our local gauge theory analysis.

Additionally, we also must pay attention to non-perturbative instanton corrections. In the closely related context of 4d $\mathcal{N} = 1$ theories generated from local $G_2$ geometries, such effects are responsible for generating mass terms for vector-like pairs, as well as Yukawa interaction terms. The former is a parity invariant contribution, while the latter typically breaks parity. Since we have emphasized that our local $Spin(7)$ system can be built up from PW building blocks, we can expect similar effects to be present in our system as well.

In the process of compactifying, then, we encounter several different mass scales. First of all, we have the UV cutoff of the 7d SYM theory $\Lambda_{\mathrm{UV}}^{7d}$. Then, we have a generic radius of compactification for the four-manifold, which introduces a Kaluza-Klein mass scale $\Lambda_{KK}$. Below this, we expect to have a 3d effective field theory. The 3d gauge coupling will flow to strong coupling since the gauge field kinetic term is a relevant operator, and this occurs at a strong coupling scale $\Lambda_{\mathrm{strong}}^{3d} \sim \Lambda_{KK} (\Lambda_{KK}/\Lambda_{\mathrm{UV}}^{7d})^3 $. Euclidean M2-brane instanton effects can generate masses $\Lambda_{\mathrm{inst}}$
for vector-like pairs, but are exponentially suppressed by a volume factor. On an an isotropic $M_4$, this will be small relative to $\Lambda_{\mathrm{strong}}^{3d} $.
Finally, we have the ``deep infrared'' $\Lambda_{\mathrm{IR}}^{3d}$ where even these massive states have been integrated out.
All told, then, we have the hierarchy of mass scales:
\begin{equation}
\Lambda_{\mathrm{UV}}^{7d} \gg \Lambda_{\mathrm{KK}} \gg \Lambda_{\mathrm{strong}}^{3d} \gg \Lambda_{\mathrm{inst}} \gg \Lambda_{\mathrm{IR}}^{3d}.
\end{equation}

Our plan in the remainder of this section will be to track these compactification effects in more detail. We begin by analyzing
the impact of compactification effects on the Chern-Simons level of the 3d gauge theory. We then turn to an analysis of
quantum corrections. We then turn to some explicit geometric examples illustrating these general
considerations. Appendix \ref{app:SQM} contains additional details on instanton corrections to the local $Spin(7)$ system of equations.

\subsection{Chern-Simons Level Contributions}

In this subsection we turn to a more a detailed analysis of how various contributions to the
compactification can shift the Chern-Simons level. In particular, we discuss the conditions necessary for
such effects to cancel out so that a 3d reflection symmetry might still be present.

In a general compactification, there are various sources of contributions to the Chern-Simons level.
First of all, we can study the contribution from the zero modes, and as we discussed in section \ref{sec:ANOMO},
these contribute to various ``parity'' anomalies. Second, we have the contributions from the Kaluza-Klein states.
At first glance, it would appear that such modes cannot contribute, simply because they are heavy states, and we will integrate them out anyway.
There is a subtlety here, because in 3d, we also know that for a single Dirac fermion $\psi_{D}$ coupled to a background $U(1)$ gauge field,
adding a mass term $m \overline{\psi_{D}} \psi_{D}$ and integrating out the fermion produces a shift in the Chern-Simons level of the background $U(1)$, with the sign of the shift dictated by the sign of $m$. A priori, then, it might happen that a massive KK state could shift the background Chern-Simons level.

To better understand this issue, we return to the 7d SYM theory coupled to defects. As we have already mentioned, one way for us to study this
system of intersecting branes is to begin with a parent gauge theory with gauge group $\widetilde{G}$, and then consider adjoint
Higgsing to $G \times U(1)^n \subset \widetilde{G}$. In general then, we need to pay attention to mass terms generated by the internal components of the 7d gauge field, and the background Higgs field. Writing out
the 7d Dirac equation (and including the background Higgs fields as well) we have:
\begin{equation}
D_{3d} \Psi_{I} + \mathcal{D} \Psi_{I} = 0,
\end{equation}
where we have introduced the twisted differential operator:
\begin{equation}
\mathcal{D} = \begin{pmatrix}
\Phi_{\mathrm{SD}}\times & D_{sig.} \\
D_{sig.} & -\Phi_{\mathrm{SD}}\wedge
\end{pmatrix}.
\end{equation}
The point is that even though our original 7d SYM theory is reflection symmetric
(when the overall 7d Chern-Simons level is zero\footnote{At least in type IIA backgrounds,
the 7d Chern-Simons term requires a Romans mass background, and we are assuming this is absent.}),
activating vevs for the Higgs fields introduces an explicit parity breaking term
precisely because these modes transform as pseudo-scalars. So, from a 3d perspective, we can think
of $D_{sig}$ as generating a parity preserving mass term for Kaluza-Klein modes,
while the action of $\Phi_{\mathrm{SD}}$ will produce a parity breaking mass term. A related comment is
that there is a coupling between localized modes and the bulk Higgs field given by the superpotential term
\cite{Heckman:2018mxl}:
\begin{equation}
W_{3d} \supset \underset{L}{\int} \Sigma^{\dag} \Phi_{\mathrm{SD}} \Sigma,
\end{equation}
for $\mathcal{N} = 2$ matter $\Sigma$ trapped on a line $L$.

In the presence of a background position dependent profile for $\Phi$, then, we have an explicit parity breaking contribution
to the background. Integrating out the corresponding massive states generated by this profile,
we see that the background value of $\Phi_{\mathrm{SD}}$ can impact the Chern-Simons level of the 3d effective field theory.
To a certain extent, we have already seen how $\Phi_{\mathrm{SD}}$ impacts reflection assignments in the 3d effective field theory, because
each localized zero mode contributes to $\nu_{\mathsf{R}}$, i.e. we get a contribution to the gravitational-parity anomaly. So, generically we expect Kaluza-Klein states which pick up a mass from $\Phi_{\mathrm{SD}}$ to induce a shift in the 3d Chern-Simons level. On the other hand, we have already explained that a background profile for $\Phi_{\mathrm{SD}}$ can produce trapped zero modes, which also induces a
contribution to the 3d parity anomaly, i.e. we can view it as producing a shift in the background Chern-Simons level.

Our main claim is that these two contributions always cancel each other out. The main idea is to consider an initially
aligned stack of six-branes and to gradually switch on a background value of $\Phi_{\mathrm{SD}}$. Initially,
we can pair up positive and negative parity KK modes through mass terms such as $\overline{\psi_{+}} \psi_{-}$.
Once we tilt the 6-branes, however, one of these modes becomes trapped (i.e. it is either a $\tau_{+,+,-}$ or $\tau_{-,-,+}$ zero mode). This also shifts the mass matrix for the KK modes. Since this is a small perturbation, we expect the value of $\nu_{\mathsf{R}}$ to be the same before and after, i.e., we can locally split up the contributions to the effective Chern-Simons level as:
\begin{equation}\label{eqn:KKLoc}
k(\mathrm{KK}) + k(\mathrm{Loc.})  = 0,
\end{equation}
where the second term corresponds to all the contributions from ``local matter.''
More precisely, we have two sources for local contributions. First of all,
we have the contributions from the local matter.
Second, we have contributions which are associated with singularities
for the Higgs field which are generically localized on one-dimensional subspaces.
Recall, however, that such singularities are associated with giving vevs to the
scalars of matter localized on a line of some bigger parent gauge theory \cite{Heckman:2018mxl, Beasley:2008dc}.
So, we break this up further as a schematic contribution of the form:
\begin{equation}
k(\mathrm{KK}) + k(\mathrm{Loc. \, Matt.}) +  k(\mathrm{Loc. \, Sing.}) = 0.
\end{equation}
So, at least from the perspective of the 3d effective field theory,
integrating the KK states generates an explicit parity breaking contribution,
but this is \textit{exactly opposite} to the contribution from local matter.
From this perspective, the classical action (after integrating out the KK modes)
breaks parity, but the one loop contribution from the local zero modes restores it.

One might view this as slightly undesirable, since the classical action now breaks parity.
Of course, we can also restore classical parity by considering a broader class of M-theory backgrounds.
Indeed, in a general M-theory background on an eight-manifold $X_8$,
it is well-known that the four-form flux $G_4$
can induce further shifts in the effective Chern-Simons level through the
7d term (see e.g. \cite{Gukov:1999ya, Gukov:2001hf}):
\begin{equation}
\underset{7d}{\int} \frac{1}{4 \pi} CS_{3}(A) \wedge \frac{i^{\ast}G_4}{2 \pi},
\end{equation}
where $i^{\ast} G_4$ is the pullback of the four-form flux onto $M_4 \subset X_8$.

In fact, sometimes a four-form flux \textit{must} be switched on
since the quantization condition is \cite{Witten:1996md}:
\begin{equation}
\bigg[\frac{G_4}{2\pi}\bigg]-\frac{p_1(X_8)}{4} \in H^4(X_8,\mathbb{Z}).
\end{equation}
We would like to understand whether a half-quantized
flux needs to be switched on when we have
$M_4$ a four-manifold of ADE singularities. Rather than attempt
to directly define and compute the Pontryagin class on the singular space $X_8$,\footnote{In the smooth case, at least,
one can proceed as follows (see also \cite{Gukov:2001hf}.
Consider $M_4$ a calibrated four-cycle inside $X_8$ such that the normal bundle is smooth.
In this case, the local geometry is just a Bryant-Salamon space \cite{bryant1989}, as given by the right-handed spinor bundle over $M_4$, which
we write as $\Sigma_R M_4$. The restriction of $p_1(X_8)$ to $M_4$ is then (up to 2-torsion)
\begin{equation}
p_1(X_8)|_{M_4}= p_1(\Sigma_R M_4) + p_1(M_4),
\end{equation}
where we have used an adjunction formula. On the righthand side, we are computing the
first Pontryagin class of the right-handed spinor bundle of $M_4$, i.e.
we get an element of $H^{4}(M_4, \mathbb{Z})$. Using the relations:
\begin{equation}\label{eqn:X8X7}
p_1(\Sigma_R M_4)=\frac{1}{2}p_1(\Lambda^2_{+}M_4e)+2c_1^2(\Sigma_R M_4)=\frac{1}{2}(p_1(M_4)+2 e(M_4))+2c_1^2(\Sigma_R M_4),
\end{equation}
where $e(M_4)$ is the Euler class, we have in total that
\begin{equation}
p_1(X_8)|_{M_4}=\frac{3}{2}p_1(M_4)+e(M_4)+2c_1^2(\Sigma_R M_4).
\end{equation}
Since the signature satisfies $3 \sigma(M_4) = p_{1}(M_4)$, we get that the integral over our characteristic classes are:
\begin{equation}
\underset{M_4}{\int} p_1(X_8)|_{M_4} = \frac{9}{2} \sigma(M_4) + \chi(M_4)+2c_1^2(\Sigma_R M_4),
\end{equation}
with $\sigma(M_4)$, $\chi(M_4)$, and $c_1^2(\Sigma_R M_4)$ the signature, Euler character, and squared first Chern number of the spinor bundle of $M_4$ (note we use the same symbol for the Chern number and Chern class). We are interested in $p_1(X_8)/4$, so since we assume $M_4$ is spin, the signature is divisible by 16 (by Rokhlin's theorem) and $c^2_1(\Sigma_R M_4)$ is even so working modulo 4 this becomes:
\begin{equation}
p_1(X_8)|_{M_4}\equiv e(M) \mod 4.
\end{equation}
So, we see that a half-integer $G_4$ flux is absent when $\chi(M_4)=0 \mod 4 $, and is present when
$\chi(M_4) \neq 0 \mod 4 $.}
we can instead contemplate what happens with pure 7d SYM theory with gauge group $H$
compactified on $M_4$. From our previous analysis, we know that the bulk zero modes
generate a contribution to the mixed gauge-parity anomaly given by (treated as an integer):
\begin{equation}
\nu_{\mathsf{R} H} = h^{\lor}(H) \times \frac{1}{2}( \chi(M_4) + \sigma(M_4)).
\end{equation}
Observe that when this is an odd number, we get a corresponding half-integer shift in the Chern-Simons level. Since $1/2( \chi(M_4) + \sigma(M_4))$ is an integer, we conclude that this happens precisely when $h^{\lor}(H)$
is an odd number, as can happen for $SU(N)$ gauge theory with $N$ odd.\footnote{We can also
get $h^{\lor}(H)$ odd if $H$ is not simply laced.} In this case, we see that much as we had in equation (\ref{eqn:KKLoc}),
the mixed gauge-parity anomaly allows us to detect the presence of a half-integer $G_4$-flux.
In particular, we have:
\begin{equation}
k(\mathrm{G-flux}) + k(\mathrm{Bulk\,Zero\,Modes}) \in \mathbb{Z},
\end{equation}
in the obvious notation.

\subsection{Quantum Corrections \label{ssec:QUANTCORR}}

Let us now turn to a more explicit analysis of quantum corrections. To a certain extent, we have already detailed
the robust observables as captured by anomalies. Indeed, supersymmetry provides
only mild protection against such effects because the system we have engineered has only 3d $\mathcal{N} = 1$ supersymmetry.
That said, we note that the localized matter fields really fill out 3d $\mathcal{N} = 2 $ multiplets while the bulk modes
are in 3d $\mathcal{N} = 1$ multiplets. In passing from 7d SYM down to 3d, we can ask about the impact of various loop corrections, as generated
both by the zero modes and by the KK modes. For ease of exposition, we primarily focus on Yukawa-like interactions of the form:
\begin{equation}
c \phi \overline{\psi_{-}} \psi_{+} + h.c,
\end{equation}
where $\phi$ is a pseudo-scalar, $\psi_{-}$ is a reflection odd Dirac fermion and $\psi_+$ is a reflection even fermion. Here,
$c$ is a dimensionless parameter, and we have taken our scalars to have scaling dimension $1$. This is natural in our setting, because the kinetic term for such scalars directly descends from compactification of 7d SYM, and so they come with a kinetic term of the form:
\begin{equation}\label{eqn:normalization}
\frac{\mathrm{Vol}(M_4)}{g^2_{7d}} (\partial \phi)^2 \sim \frac{1}{g^2_{3d}} (\partial \phi)^2.
\end{equation}
Suppose now that $\phi$ and $\psi_{-}$ are actually KK modes. This occurs, for example,
in considering a bulk mode coupling to the modes localized on a circle
$L$ via the term:
\begin{equation}
W_{3d} \supset \underset{L}{\int} \Sigma_{\mathrm{KK}}^{\dag} \Phi^{\mathrm{KK}}_{\mathrm{SD}} \Sigma_{(0)},
\end{equation}
where $\Sigma_{(0)}$ is a zero mode, $\Sigma_{KK}$ is a KK mode localized on the line $L$,
and $\Phi^{KK}_{\mathrm{SD}}$ is a bulk KK Mode. If we consider loop diagrams with internal KK modes, we see that generically,
we might pick up an additional mass term for some of our candidate zero modes, provided such interaction terms
are compatible with the reflection (and other discrete) symmetries retained by the compactification. We observe that the overall
mass scale generated in this way is:
\begin{equation}
m_{3d} \sim g^{2}_{3d}(\Lambda_{\mathrm{KK}}) \sim \Lambda_{\mathrm{KK}} \times \left( \frac{\Lambda_{\mathrm{KK}}}{\Lambda_{\mathrm{UV}}^{7d}} \right)^3 \ll \Lambda_{\mathrm{KK}},
\end{equation}
i.e. we can generate mass terms, but these only become prominent as we approach the scale of strong coupling in the 3d theory.

In addition to effects from integrating out KK states, we also expect contributions from Euclidean M2-branes and M5-branes.
Let us begin with the Euclidean M2-brane contributions. Recall that in the PW system,
these instanton effects, can, for example, lift vector-like pairs of matter fields, as follows from a direct analysis of the associated supersymmetric quantum mechanics for the 4d theory compactified further on a $T^3$. Similar considerations apply
in local $Spin(7)$ systems, where we ``smear'' the M2-brane over the matter one-cycles. We comment on this
in further detail in Appendix \ref{app:SQM}. The general qualitative point
is that we can suppress the size of mass terms for local matter by keeping the corresponding zero mode circles geometrically sequestered.
This produces a mass scale $\Lambda_{\mathrm{inst}} \ll \Lambda^{3d}_{\mathrm{strong}}$. Additionally, M2-brane instanton corrections can generate Yukawa-like interactions for our 3d $\mathcal{N} = 2$ matter. This is essentially the same mechanism for Yukawa interactions
generated in 4d $\mathcal{N} = 1$ vacua of local $G_2$ models.\footnote{This is another distinction with expectations
from the BHV system, where Yukawas can be generated by classical geometry with textures generated by further intersections and / or fluxes and instantons (see e.g. \cite{Beasley:2008dc, Beasley:2008kw, Heckman:2008qa, Bouchard:2009bu, Cecotti:2009zf, Marchesano:2009rz, Marchesano:2015dfa, Cvetic:2019sgs}).}

We can also have M5-brane instanton effects. In the related context of M-theory on a Calabi-Yau fourfold,
it is well-known that M5-branes wrapped on holomorphic divisors $D$ with arithmetic genus $\chi(D, \mathcal{O}_D) = 1$
can contribute to the resulting superpotential \cite{Witten:1996bn}, and in the presence of localized singularities, there is a natural
extension of these considerations which can be interpreted as a strong coupling contribution in 3d gauge theories \cite{Katz:1996th}. In our setting where we have a four-manifold of ADE singularities generating a local $Spin(7)$ space, we can clearly identify six-cycles coming from one of the collapsing $S^{2}$'s in the fiber, over $M_4$. Note also that there can be further enhancements in the singularity type when we have local matter, and this in turn means that these instantons can generate additional non-perturbatively induced interaction terms for the matter.

An important comment here is that we have also emphasized that at least in some cases, we can equivalently study the resulting 3d effective field theory by working with type IIA string theory on the local $G_2$ space $\Lambda_{\mathrm{SD}}^{2} M_4$, with spacetime filling D6-branes wrapping $M_4$. Here, we expect that M2-brane instantons descend to worldsheet instantons and that M5-brane instanton corrections descend to Euclidean D4-branes. In the D4-brane case, it wraps all of the $M_4$, as well as a finite interval stretched between the different sheets of the associated spectral cover. This is compatible with the standard way in which M-theory on an ALE space reduces to type IIA when we reduce along the circle direction of the ALE space.\footnote{See for example references \cite{Blumenhagen:2006xt, Blumenhagen:2009qh} for further discussion on
such instanton corrections in the context of intersecting D6-brane models in type IIA compactified on a Calabi-Yau threefold.}

\section{Examples} \label{sec:EXAMPLES}

In this section we turn to some explicit examples. Our operating theme will
be to construct specific four-manifolds and Higgs field profiles
which offer some level of protection against various quantum corrections.

\subsection{Topological Insulator Revisited} \label{ssec:TOPOAGAIN}

As a first example, let us begin by revisiting the case of a topological insulator considered in section \ref{sec:TOPOINS}. Recall that in this case, we took a PW system on the three-manifold $M_3 = S^1 \times \Sigma$ with a single 4d $\mathcal{N} = 2$ hypermultiplet localized at a point of $\Sigma$, and spread over $S^1$. Introducing a position dependent profile for the Higgs field in a transverse direction $\mathbb{R}_{\bot}$, we have a single 3d $\mathcal{N} = 2$ matter field trapped at the point $x_\bot = 0$ of $\mathbb{R}_{\bot}$. The reflection assignment for the associated 3d Dirac fermion is then dictated by the relative change in sign for this Higgs field, and it of course transforms in some representation $\mathbf{R}$ of the unbroken gauge group $H$.

In the present setup where our $M_4$ is non-compact, the unbroken gauge group $H$ is actually just a flavor symmetry of the 3d field theory.
We can calculate the anomalies associated with reflections, as well as mixed reflection-gauge transformation anomalies. Indeed, this 3d mode
will generate a shift in the background Chern-Simons level for $H$ given by (with one choice of reflection assignment):
\begin{equation}
\delta k= \mathbf{T}(\mathbf{R}),
\end{equation}
so for example, in the case where the Lie algebra of $H$ is $\mathfrak{su}(N)$, we get matter in fundamental representation from the unfolding $\mathfrak{su}(N+1) \rightarrow \mathfrak{su}(N) \times \mathfrak{u}(1)$, and the two-index anti-symmetric from $\mathfrak{so}(2N) \rightarrow \mathfrak{su}(N) \times \mathfrak{u}(1)$ (see e.g. \cite{Katz:1996xe}). The contribution to $\delta k$ is a half-integer shift for the fundamental representation $\mathbf{N}$s, and for a two-index anti-symmetric representation $\mathbf{\Lambda^{2} N}$ we get $(N-2)/2$, which is also a half-integer when $N$ is odd. The corresponding gravitational-parity anomalies are just $\nu_{\mathsf{R}}(\mathbf{N}) = 2N$ and $\nu_{\mathsf{R}}(\mathbf{\Lambda^{2} N}) = N(N-1)$.

Another interesting case to consider is matter transforming in a representation of $\mathfrak{so}(10)$.
We get matter in the spinor representation $\mathbf{16}$ from unfolding $\mathfrak{e}_6 \rightarrow \mathfrak{so}(10) \times \mathfrak{u}(1)$.
We get matter in the vector representation $\mathbf{10}$ from unfolding $\mathfrak{so}(12) \rightarrow \mathfrak{so}(10) \times \mathfrak{u}(1)$. A general comment here is that even though the $\mathbf{10}$ is a real representation of $\mathfrak{so}(10)$, we are still dealing with a 3d Dirac fermion, i.e. it contributes as ten complex fermions.\footnote{From the perspective of the unfolding breaking pattern, observe that these modes are still charged under a $\mathfrak{u}(1)$, so in that sense they are automatically still complex representations.} For the spinor, we note that $\mathbf{T}(\mathbf{16}) = 2$, i.e. there is no mixed parity-$Spin(10)$ anomaly. Note also that $\nu_{\mathsf{R}}(\mathbf{16}) = 0 \mod 16$. For the vector representation, we have $\mathbf{T}(\mathbf{10}) = 1$, so again we do not get a mixed gauge-parity anomaly. In contrast to the spinor
representation, now we have $\nu_{\mathsf{R}}(\mathbf{10}) = 20 = 4 \mod 16$.

In the case where we just have a localized mode on a non-compact $M_4$,
we also see that the KK modes of the 7d SYM theory have actually decoupled from the 3d boundary mode. This just follows from the
fact that the corresponding Yukawa interactions discussed around equation (\ref{eqn:normalization})
are set by $g^2_{3d}$, but this is zero in the case where $H$ is a flavor symmetry.

It is also natural to ask about what happens if we compactify $\mathbb{R}_{\bot}$ to a circle, which we write as $S^1_{\bot}$.
In this case, we expect the Higgs field to generically have additional zeros. One way to argue for this is to
observe that on the three-manifold specified by $\widetilde{M_3} = \Sigma \times S^{1}_{\bot}$, having a single zero mode would have
generated a 4d $\mathcal{N} = 1$ field theory with a gauge anomaly. The simplest resolution is the assumption that there is an additional
zero mode somewhere else on $S^{1}_{\bot} \times \Sigma$, which in 4d terms would give rise to a ``vector-like pair''. More precisely, a Gauss' law type constraint on the PW system tells us that in this case we should expect a single $(+,+,-)$ mode and a single $(-,-,+)$ mode in the
same representation $\mathbf{R}$ (see e.g. \cite{Braun:2018vhk}). From a 3d perspective, we get two 3d Dirac fermions, one with reflection assignment $+1$, and the other with reflection assignment $-1$. We can again switch off all mass term contributions to this pair of localized modes by decompactifying $\Sigma$. Of course, in this case we really just have two decoupled 3d fields, but we can reintroduce interactions by taking $\Sigma$ large, but of finite volume.

\subsection{Vector-Like Models}\label{ssec:veclike}

Let us now turn to some constructions with ``vector-like'' matter spectra,
i.e. constructions in which the local matter is reflection symmetric. Our plan will be to take
a 3d $\mathcal{N} = 2$ $\mathrm{QFT}$ and its ``reflection conjugate'' $\mathsf{R}(\mathrm{QFT})$,
and then geometrically glue them together to build a new 3d theory, i.e. $\mathrm{QFT} \# \mathsf{R}(\mathrm{QFT})$,
through the operation discussed in section \ref{sec:ConnSum}.

To frame the discussion to follow, suppose we have successfully engineered a 4d $\mathcal{N} = 1$ \textit{chiral} model
using the PW system on a three-manifold $M_3$. In this case, we have various matter fields localized at points of
$M_3$ transforming in complex representations $\mathbf{R}_i$ of the unbroken gauge group $H$. Denote by $\varepsilon_i$ the
parity assignment for each such local matter field, i.e. $\varepsilon = -1$ for $(+,+,-)$ matter and $\varepsilon = +1$ for $(-,-,+)$
matter. If we compactify this chiral theory on a circle, we just get a 3d $\mathcal{N} = 2$ theory. One way for us to
build more elaborate examples is to use our gluing construction from section \ref{sec:ConnSum}.
This will produce 3d $\mathcal{N} = 2$ matter coupled to the reduction of 3d $\mathcal{N} = 1$ ``bulk matter'' as well as the 3d $\mathcal{N}  = 1$ vector multiplet.

Now, since we are dealing with 4d chiral theories, there is no a priori guarantee that
a reflection symmetry will survive in the resulting 3d effective field theory.
One way to ensure this, however, is to take an individual building block
$M_{4} \simeq M_{3} \times S^{1}$ and its associated Higgs bundle, as well as its
orientation reversed counterpart $\overline{M_{4}} $. We can accomplish this by performing an orientation reversal
on just the $M_{3}$ factor, while leaving the $S^1$ factor untouched. When we do this, all of the $(+,+,-)$ matter
will now become $(-,-,+)$ matter, i.e. we get exactly the opposite reflection assignments for our 3d $\mathcal{N} = 2$ matter.
Gluing together $M_{4}$ and $\overline{M_{4}}$, we then get a 3d $\mathcal{N} = 1$ theory where all the localized matter is reflection symmetric, i.e. there is no net contribution to the various parity anomalies.
\begin{figure}[t!]
\begin{center}
\includegraphics[scale = 0.5, trim = {1.0cm 4.0cm 0cm 5.0cm}]{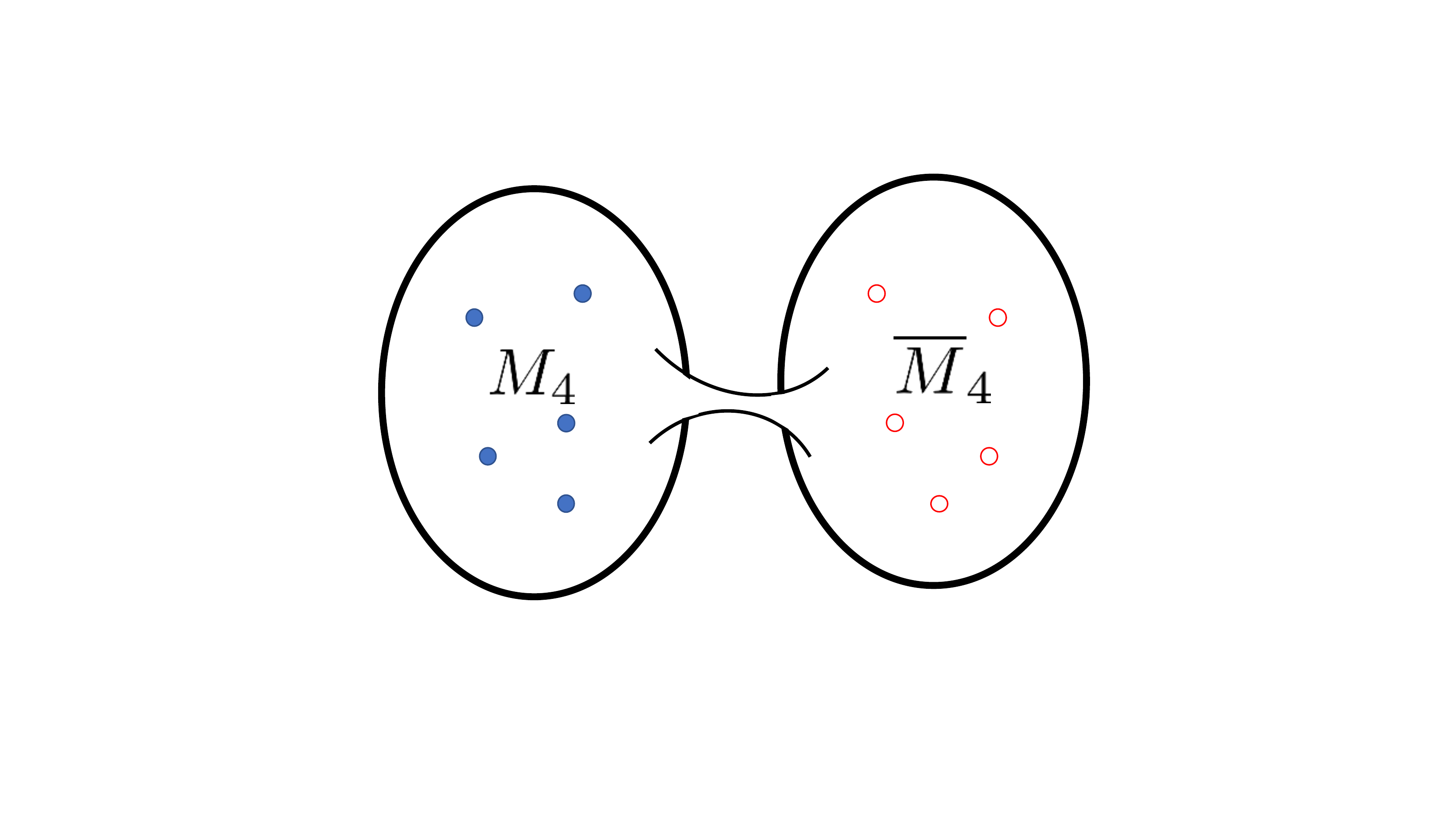}
\caption{Depiction of the connected sum construction used in the section. The separate blue and red dots indicate localized modes (on circles) with opposite chiralities/Hessians.}
\label{fig:newconnsum}
\end{center}
\end{figure}
Consider next the bulk matter of the glued theory. Recall that in general, we have:
\begin{equation}
\chi(M_4 \# \overline{M_4}) = 2\chi(M_4) -2 \,\,\, \text{and} \,\,\, \sigma(M_4 \# \overline{M_4}) = 0,
\end{equation}
so, for example, the contribution to the gravitational-parity anomaly written in equation (\ref{eqn:bulknu})
$M_4$, namely we have:
\begin{equation}\label{eqn:bulknuagain}
\nu^{\mathrm{bulk}}_{\mathsf{R}}(M_4 \# \overline{M_4}) = \mathrm{dim}_{\mathbb{R}} H \times  (\chi(M_4) - 1).
\end{equation}
Using the fact that our building block was just $M_{4} = M_{3} \times S^1$, we also have $\chi(M_4) = 0$, so we can also write:
\begin{equation}\label{eqn:bulknuagain}
\nu^{\mathrm{bulk}}_{\mathsf{R}}(M_4 \# \overline{M_4}) = - \mathrm{dim}_{\mathbb{R}} H.
\end{equation}

More generally, if we had $\ell$ such chiral building blocks, we can build more general backgrounds.
Depending on the details of this, we can engineer additional discrete symmetries just from permuting these different chiral
factors. The further gluing to a orientation reversed $\overline{M_4}$ will then produce another reflection symmetric spectrum, but with additional discrete symmetries. Note that independent of the details of these gluing operations, the contribution to the gravitational-parity anomaly obeys equation (\ref{eqn:bulknuagain}). Indeed, if we can decompose $M_4 = (Q_4)^{\# \ell},$ for some four-manifold $Q_4$,
then we can also write:\footnote{This is just because $\chi((Q_4)^{\# \ell}) = \ell \chi(Q_4) - 2 (\ell - 1)$.}
\begin{equation}
\nu^{\mathrm{bulk}}_{\mathsf{R}}(M_4 \# \overline{M_4}) = \mathrm{dim}_{\mathbb{R}} H \times (\ell \chi(Q_4) - 2\ell + 1),
\end{equation}
and again, if we further specialize to the case $Q_4 = Q_3 \times S^1$ for some three-manifold $Q_3$, we get:
\begin{equation}
\nu^{\mathrm{bulk}}_{\mathsf{R}}(M_4 \# \overline{M_4}) = \mathrm{dim}_{\mathbb{R}} H \times (- 2\ell + 1),
\end{equation}

\subsection{GUT-Like Models}

In the previous section we discussed a general method for building examples with vector-like matter.
In particular, we could start with a PW background which would realize a 4d $\mathcal{N} = 1$ theory
with matter spectrum a Grand Unified Theory (GUT), and then we could add to it the corresponding mirror reflection. On the other hand,
it is natural to consider more general possibilities where we may not have a purely geometric reflection symmetry, but which may nevertheless have vanishing parity anomalies.

To build an example of this sort, suppose we have already engineered a PW system on a three-manifold $M_3$ which
has a bulk $SO(10)$ gauge group and local matter in the $\mathbf{16}$ and $\mathbf{10}$ (see Appendix \ref{app:SO10PW} for
some additional details). For ease of exposition, we assume all the local matter is of $(+,+,-)$ type, and so we also require singularities in the Higgs field to satisfy the corresponding Gauss' law constraint for the spectral cover construction.
Now, we can simply take this model and compactify on a circle to arrive at a 3d $\mathcal{N} = 2$ theory,
as obtained from compactifying on $M_3 \times S^1$. To get a 3d $\mathcal{N} = 1$ theory, we can glue our $M_3 \times S^1$ to another four-manifold $N_4$ equipped with a trivial Higgs bundle (i.e. no local matter present)  e.g.,
$M_4 = (M_3 \times S^1) \# N_4$. This results in bulk matter fields in 3d $\mathcal{N} = 1$ supermultiplets. A general comment here is that we can engineer a model with $\mathbf{10}$'s and $\mathbf{16}$'s starting from either the parent gauge group $E_8$, or from $E_7$. In the latter case,
we have a $\mathbb{Z}_2$ center, and this descends to non-trivial higher-form symmetries in the 3d effective field theory. If we build a model with just $\mathbf{16}$'s, we could in principle start with an $E_6$ gauge theory, which has a $\mathbb{Z}_3$ center.

Recall from subsection \ref{ssec:TOPOAGAIN} that a single $\mathbf{16}$ of
$\mathfrak{so}(10)$ produces no anomaly, while a single $\mathbf{10}$ contributes
$\nu_{\mathsf{R} SO(10)} = 0 \mod 2$.
From the bulk modes in the adjoint of $\mathfrak{so}(10)$, the contributions to the
gauge-parity anomaly and gravitational-parity anomalies are:
\begin{align}
& \nu_{\mathsf{R} SO(10)}^{\mathrm{bulk}} = h^{\lor}(SO(10)) \times \frac{1}{2}(\chi(M_4) + \sigma(M_4)) \equiv 0 \mod 2 \\
& \nu_{\mathsf{R}}^{\mathrm{bulk}} = \mathrm{dim}_{\mathbb{R}}(SO(10)) \times \frac{1}{2}(\chi(M_4) + \sigma(M_4)) \equiv
- \frac{3}{2}(\chi(M_4) + \sigma(M_4)) \mod 16.
\end{align}
All told, then, the contribution to $\nu_{\mathsf{R}}$ just depends on the number of $\mathbf{10}$'s
and the adjoint-valued bulk modes:
\begin{equation}
\nu_{\mathsf{R}} \equiv 4 N_{\mathbf{10}} - \frac{3}{2}(\chi(M_4) + \sigma(M_4)) \mod 16.
\end{equation}

As we already mentioned previously, the anomalies provide robust information which survives into the deep infrared.
In particular, even if we classically localize a matter field, it might happen that strong quantum corrections still
gap the system, leaving only some topological order behind. As an interesting special case, suppose we have engineered a model
with local matter in the $\mathbf{16}$ of $SO(10)$, and where the only bulk matter comes from the 3d $\mathcal{N} = 1$
vector multiplet. There is no gauge-parity anomaly in this system, and the contribution to the gravitational-parity anomaly can be
set to zero by taking $M_4 = (S^{3} \times S^{1})^{\# \ell}$ with $\ell \equiv 1 \mod 16$. In this case, the 3d theory in the deep infrared
has trivial topological order. At a general level, we expect this to occur due to strong coupling effects in the 3d effective field theory, as we already discussed in subsection \ref{ssec:QUANTCORR}.

We can also engineer other examples of GUT-like modes by again simply borrowing construction techniques from the PW system on $M_3$.
Consider, for example, an $E_6$ gauge theory with $N_{\mathbf{27}}$ $(+,+,-)$ matter fields in the $\mathbf{27}$.
We can get this starting from unfolding $E_7$, which has a $\mathbb{Z}_2$ center. In this case, the $\mathbb{Z}_2$ descends to the 3d
effective field theory as various higher-form symmetries.
A single $\mathbf{27}$ contributes $\nu_{\mathsf{R}E_6}(\mathbf{27}) = 0 \mod 2$, i.e. there is no gauge-parity anomaly from the local matter. The contribution to the gravitational-parity anomaly is $\nu_{\mathsf{R}}(\mathbf{27}) = 6 \mod 16$.
From the bulk modes in the adjoint of $\mathfrak{e}_6$, the contributions to the
gauge-parity anomaly and gravitational-parity anomalies are:
\begin{align}
& \nu_{\mathsf{R} E_6}^{\mathrm{bulk}} = h^{\lor}(E_6) \times \frac{1}{2}(\chi(M_4) + \sigma(M_4)) \equiv 0 \mod 2 \\
& \nu_{\mathsf{R}}^{\mathrm{bulk}} = \mathrm{dim}_{\mathbb{R}}(E_6) \times \frac{1}{2}(\chi(M_4) + \sigma(M_4)) \equiv 7 \chi(M_4) \mod 16,
\end{align}
where in the second equation we dropped the contribution from $\sigma(M_4)$, since it is divisible by $16$.
All told, then, the contribution to $\nu_{\mathsf{R}}$ just depends on the number of $\mathbf{27}$'s
and the adjoint-valued bulk modes:
\begin{equation}
\nu_{\mathsf{R}} \equiv 6 N_{\mathbf{27}} + 7 \chi(M_4) \mod 16.
\end{equation}

\section{Conclusions \label{sec:CONC}}

In this paper we have studied topologically robust quantities associated with M-theory compactified on a local $Spin(7)$ space.
In particular, we have analyzed the local matter present in configurations of intersecting six-branes as generated by the unfolding of
singularities associated with $M_4$ a four-manifold of ADE singularities. Our primary tool in carrying out this analysis has been
the study of 7d SYM theory coupled to defects, and the implications for the resulting 3d effective field theory. At the classical
level, we showed that matter in such systems is generically localized along codimension three subspaces of the four-manifold.
We also explained how to calculate various observables which are robust against quantum corrections. In particular,
we have shown how reflections and automorphisms of 10d SYM theory on a $T^3$ and 11d M-theory on
an ADE singularity descend to reflections in 7d SYM, and moreover, how these symmetries further descend to the 3d effective field theory.
We used this to extract the corresponding contributions to anomalies which are robust against quantum corrections.
We also presented various examples illustrating these general themes.
We also took some preliminary steps in determining the spectrum of extended objects in these theories
and their transformations under higher-form symmetries.
In the remainder of this section we discuss some avenues for future investigation.

Our primary emphasis has been on extracting some robust calculable features of local $Spin(7)$ systems.
It is natural to ask whether we can say more about the resulting 3d effective field theories.
For example, depending on the specific matter content, Chern-Simons level, and presence (or absence) of a higher-form symmetry,
we might expect to generate a rich class of possible 3d field theories. It would clearly be instructive to analyze these possibilities,
and also use this as a starting point for a geometrization of proposed field theoretic dualities. Along these lines, we anticipate
that flop transitions in the associated local $Spin(7)$ space will provide important topological insights on these issues.

Our emphasis in this paper has been on the study of 3d effective field theories generated by working with M-theory on a
non-compact $Spin(7)$ space. It is also natural to consider the resulting theories obtained from taking perturbative superstring theory on the
same backgrounds. For type IIB and IIA, this would give rise to 2d $\mathcal{N} = (0,2)$ and $\mathcal{N} = (1,1)$ theories, respectively, while for heterotic strings, this would generate 2d $\mathcal{N} = (0,1)$ theories. The case of type IIB backgrounds is particularly interesting, because here we recover a notion of holomorphy.

The main tool of analysis we have used in the study of local $Spin(7)$ systems is 7d SYM theory, and in particular,
the associated partial topological twist of $\mathcal{N} = 4 $ SYM theory. We also used some general gluing construction techniques
to build various examples of 3d systems. It is natural to ask how these gluing operations equipped with Higgs bundles
lift to a local $Spin(7)$ geometry, and more globally, to possibly compact $Spin(7)$ spaces.
As noted in \cite{Cvetic:2020piw}, these sorts of operations can often
be interpreted as a generalization of the standard connected sums
construction of references \cite{Kovalev:2001zr, haskins2014asymptotically}
(see also \cite{Braun:2018joh} for a proposed extension to $Spin(7)$ spaces).
Related to this, it would be interesting to better understand
the possible constraints which come from coupling our 3d theories
to gravity.

A more long-term goal in this direction would be to use the results obtained here to start building more explicit F-theory backgrounds on a
$Spin(7)$ space. Indeed, we anticipate that the analysis of anomalies and topological structures found here will provide additional insight into the ``$\mathcal{N} = 1/2$'' backgrounds considered in \cite{Heckman:2018mxl, Heckman:2019dsj}
(see also \cite{Bonetti:2013fma, Bonetti:2013nka}).

\newpage

\section*{Acknowledgments}

We thank T.B. Rochais for collaboration at an early stage of this project.
We thank A. Debray, M. Del Zotto, H. Gluck, M. Montero and W. Ziller for helpful discussions.
The work of MC, JJH and ET is supported in part by the DOE (HEP) Award
DE-SC0013528. The work of MC and ET was supported by the
Simons Foundation Collaboration grant \#724069 on ``Special Holonomy in Geometry, Analysis and Physics''.
The work of MC is also supported by the Fay R. and Eugene L. Langberg Endowed Chair (MC) and the Slovenian
Research Agency (ARRS No. P1-0306).
The work of JJH and GZ was supported by a University Research Foundation
grant at the University of Pennsylvania and DOE (HEP) Award DE-SC0021484.


\appendix

\section{More Details on Classical Zero Mode Localization\label{app:MOREZERO}}

In this Appendix, we provide further details on the zero mode localization analysis of section \ref{sec:NOLOCO},
with particular emphasis on the case where a flux turned is turned on. First off all, the
matrix $M$ used in the localization argument is
\al{M = \left(
\begin{array}{cccc}\label{eq:M}
 \frac{(2 \lambda +\mu ) \sqrt{\lambda ^2+\frac{4 N ^2 (\lambda +\mu )^2}{(3 \lambda
   +2 \mu )^2}+2 \lambda  \mu +\mu ^2}}{\lambda +\mu } & -\frac{i \lambda  N }{3
   \lambda +2 \mu } & 0 & 0 \\
 -\frac{i \lambda  N }{3 \lambda +2 \mu } & \sqrt{\lambda ^2+\frac{4 N ^2 (\lambda
   +\mu )^2}{(3 \lambda +2 \mu )^2}+2 \lambda  \mu +\mu ^2} & 0 & 0 \\
 0 & 0 & \sqrt{\lambda ^2+4 N ^2} & i N  \\
 0 & 0 & i N  & 0 \\
\end{array}
\right)\,.
}
Now, to expedite the analysis of the equations we can write them collectively as $\mathbf{D }\Psi = 0$ where $\Psi$ is the column vector
\al{\Psi = \left[\begin{array}{c}a_1\\a_2\\a_3\\a_4\\\varphi_1 \\ \varphi_2 \\ \varphi_3 \end{array}\right]
}
and $\mathbf D$ is the matrix operator
\al{\mathbf D = \left[\begin{array}{ccccccc}0 & - \Phi_3  & \Phi_2 & -\Phi_1 & D_4 & -D_3 & D_2\\ \Phi_3 & 0 & -\Phi_1 & -\Phi_2 & D_3 & D_4 & -D_1 \\ -\Phi_2 & \Phi_1 & 0 & -\Phi_3 & -D_2 & D_1 & D_4 \\ \Phi_1 & \Phi_2 & \Phi_3 & 0 & -D_1 & -D_2 & -D_3\\ -D_4 & -D_3 & D_2 & D_1 & 0 & \Phi_3 & - \Phi_2 \\ D_3 & -D_4 & -D_1 & D_2  & -\Phi_3 & 0 & \Phi_1 \\ -D_2 & D_1 & -D_4 & D_3 & \Phi_2 & -\Phi_1 & 0\end{array}\right]\,.
}
Here when writing $\Phi_i$ we mean $[\Phi_i, \cdot]$. To get a better understanding of the equations we can take the square of $\mathbf{D}$ and focus on its anti-symmetric part
\al{& \mathbf D^2|_{\text{asym}} = i \mathbf F \\&= \left(
\begin{array}{cccc}
 0 & -F_{\text{12}}+2 F_{\text{34}}-2 F_{\text{56}} &
   -F_{\text{13}}-2 \left(F_{\text{24}}+F_{\text{57}}\right) &
   -F_{\text{14}}+2 F_{\text{23}}-2 F_{\text{67}} \\
 F_{\text{12}}-2 F_{\text{34}}+2 F_{\text{56}} & 0 & 2
   F_{\text{14}}-F_{\text{23}}-2 F_{\text{67}} & -2
   F_{\text{13}}-F_{\text{24}}+2 F_{\text{57}} \\
 F_{\text{13}}+2 \left(F_{\text{24}}+F_{\text{57}}\right) & -2
   F_{\text{14}}+F_{\text{23}}+2 F_{\text{67}} & 0 & 2
   F_{\text{12}}-F_{\text{34}}-2 F_{\text{56}} \\
 F_{\text{14}}-2 F_{\text{23}}+2 F_{\text{67}} & 2
   F_{\text{13}}+F_{\text{24}}-2 F_{\text{57}} & -2
   F_{\text{12}}+F_{\text{34}}+2 F_{\text{56}} & 0 \\
 F_{\text{15}}-2 \left(F_{\text{26}}+F_{\text{37}}\right) & 2
   F_{\text{16}}+F_{\text{25}}+2 F_{\text{47}} & 2
   F_{\text{17}}+F_{\text{35}}-2 F_{\text{46}} & -2
   F_{\text{27}}+2 F_{\text{36}}+F_{\text{45}} \\
 F_{\text{16}}+2 F_{\text{25}}-2 F_{\text{47}} & -2
   F_{\text{15}}+F_{\text{26}}-2 F_{\text{37}} & 2
   F_{\text{27}}+F_{\text{36}}+2 F_{\text{45}} & 2
   F_{\text{17}}-2 F_{\text{35}}+F_{\text{46}} \\
 F_{\text{17}}+2 \left(F_{\text{35}}+F_{\text{46}}\right) &
   F_{\text{27}}+2 F_{\text{36}}-2 F_{\text{45}} & -2
   F_{\text{15}}-2 F_{\text{26}}+F_{\text{37}} & -2
   F_{\text{16}}+2 F_{\text{25}}+F_{\text{47}} \\
\end{array}
\right.\nonumber\\
&\left.
\begin{array}{ccc}
 2 \left(F_{\text{26}}+F_{\text{37}}\right)-F_{\text{15}} &
   -F_{\text{16}}-2 F_{\text{25}}+2 F_{\text{47}} &
   -F_{\text{17}}-2 \left(F_{\text{35}}+F_{\text{46}}\right) \\
 -2 F_{\text{16}}-F_{\text{25}}-2 F_{\text{47}} & 2
   F_{\text{15}}-F_{\text{26}}+2 F_{\text{37}} &
   -F_{\text{27}}-2 F_{\text{36}}+2 F_{\text{45}} \\
 -2 F_{\text{17}}-F_{\text{35}}+2 F_{\text{46}} & -2
   F_{\text{27}}-F_{\text{36}}-2 F_{\text{45}} & 2
   F_{\text{15}}+2 F_{\text{26}}-F_{\text{37}} \\
 2 F_{\text{27}}-2 F_{\text{36}}-F_{\text{45}} & -2
   F_{\text{17}}+2 F_{\text{35}}-F_{\text{46}} & 2
   F_{\text{16}}-2 F_{\text{25}}-F_{\text{47}} \\
 0 & -2 F_{\text{12}}-2 F_{\text{34}}-F_{\text{56}} & -2
   F_{\text{13}}+2 F_{\text{24}}-F_{\text{57}} \\
 2 F_{\text{12}}+2 F_{\text{34}}+F_{\text{56}} & 0 & -2
   F_{\text{14}}-2 F_{\text{23}}-F_{\text{67}} \\
 2 F_{\text{13}}-2 F_{\text{24}}+F_{\text{57}} & 2
   F_{\text{14}}+2 F_{\text{23}}+F_{\text{67}} & 0 \\
\end{array}
\right)\,.\nonumber
}
Here we wrote everything in terms of the curvature of the gauge bundle. The notation $F_{ij}$ is clear for $1\leq i,j \leq 4$. For $i\leq 4$ and $5\leq j\leq 7$ we have that $F_{ij} = D_i \phi_{j-3}$. For $5\leq i,j\leq 7$ we have that $F_{ij} = [\phi_{i-3},\phi_{j-3}]$. The matrix can be further simplified using the equations of motion that relate the various components of the curvature, however we will not do it for the moment. The reason we introduced this flux matrix is that if fluxes are constant it is possible to build solutions efficiently by just considering the eigenvectors of $\mathbf F$.

As an example we can take the first Spin(7)  background written above. One can show that one eigenvector has the form
\al{\left[
\begin{array}{c}\frac{\sqrt{\left(\lambda ^2+4 N ^2\right) \left((3 \lambda +2 \mu )^2+4 N
   ^2\right)}+3 \lambda ^2+2 \lambda  \mu +4 N ^2}{4 N
   }\\-i\frac{\sqrt{\sqrt{\left(\lambda ^2+4 N ^2\right) \left((3 \lambda +2 \mu )^2+4
   N ^2\right)}+5 \lambda ^2+6 \lambda  \mu +2 \mu ^2+4 N
   ^2}}{\sqrt{2}}\\0\\0\\-i\frac{2 \sqrt{2} N  (2 \lambda +\mu )
   \sqrt{\sqrt{\left(\lambda ^2+4 N ^2\right) \left((3 \lambda +2 \mu )^2+4 N
   ^2\right)}+5 \lambda ^2+6 \lambda  \mu +2 \mu ^2+4 N ^2}}{\sqrt{\left(\lambda ^2+4
   N ^2\right) \left((3 \lambda +2 \mu )^2+4 N ^2\right)}-3 \lambda ^2-2 \lambda
   \mu +4 N ^2}\\ \lambda +\mu \\0\end{array}\right]\,.
}
To find a solution, we write:
\al{\Psi = \left[
\begin{array}{c}\frac{\sqrt{\left(\lambda ^2+4 N ^2\right) \left((3 \lambda +2 \mu )^2+4 N
   ^2\right)}+3 \lambda ^2+2 \lambda  \mu +4 N ^2}{4 N
   }\\-i\frac{\sqrt{\sqrt{\left(\lambda ^2+4 N ^2\right) \left((3 \lambda +2 \mu )^2+4
   N ^2\right)}+5 \lambda ^2+6 \lambda  \mu +2 \mu ^2+4 N
   ^2}}{\sqrt{2}}\\0\\0\\-i\frac{2 \sqrt{2} N  (2 \lambda +\mu )
   \sqrt{\sqrt{\left(\lambda ^2+4 N ^2\right) \left((3 \lambda +2 \mu )^2+4 N
   ^2\right)}+5 \lambda ^2+6 \lambda  \mu +2 \mu ^2+4 N ^2}}{\sqrt{\left(\lambda ^2+4
   N ^2\right) \left((3 \lambda +2 \mu )^2+4 N ^2\right)}-3 \lambda ^2-2 \lambda
   \mu +4 N ^2}\\ \lambda +\mu \\0\end{array}\right] e^{-\frac{1}{2} \mathbf x.  M.\mathbf x}\,.
}
Plugging this back in the equations of motion the problem becomes a linear system in the matrix $M$ and the solution can be found rather quickly (we will not copy it here for the sake of brevity).

\section{Quotient Construction} \label{app:QUOT}

In this Appendix we present a quotient construction for generating local $Spin(7)$ systems. Our starting point are BHV solutions with a suitable $\mathbb{Z}_2$ action which can actually be lifted to a compact Calabi-Yau fourfold. This provides a way to generate examples of compact spaces. Since it is somewhat orthogonal to the main developments of the text, we have placed it here in an Appendix.

The main idea here is similar in spirit to Joyce's construction method \cite{Joyce:1999nk} which involves starting with a Calabi-Yau fourfold with an anti-holomorphic $\mathbb{Z}_2$ action. After quotienting by this action, one obtains an eight-manifold with possible fixed points, and it can be shown that in favorable circumstances this results in a $Spin(7)$ space. Our emphasis will be somewhat different since we focus on the local Higgs bundle construction, but which can in principle lift to a compact geometry.

At the level of eight-manifolds, the idea will be to start from a
Calabi-Yau fourfold and implement a $\mathbb{Z}_{2}$ action which leaves the K\"ahler
form $J$ invariant, but which sends the holomorphic four-form to its complex
conjugate, possibly twisted by a phase:%
\begin{equation}
\Omega\mapsto e^{-i\theta}\overline{\Omega}.
\end{equation}
We can then consider a Cayley four-form for a candidate $Spin(7)$ geometry, as
given by:%
\begin{equation}
\Lambda=\operatorname{Re}\left(  e^{i\theta}\Omega\right)  + \frac{1}{2} J\wedge J.
\end{equation}

Since we are interested in building local models, we begin by studying possible quotients of four-manifolds, and then ask how these
can be lifted to suitable eight-manifolds. Our plan will be to construct examples using two basic geometric operations.
The first operation is an antipodal map on a $\mathbb{CP}^{1}$. In homogeneous
coordinates $[u_{i},v_{i}]$, it acts as:%
\begin{equation}
\tau_{i}:[u_{i},v_{i}]\mapsto\lbrack-\overline{v}_{i},\overline{u}_{i}],
\end{equation}
or, in terms of a local affine coordinate $z_{i}=u_{i}/v_{i}$, we have
$z_{i}\mapsto-1/\overline{z}_{i}$.

The second operation is a permutation of
coordinates on a product of $\mathbb{CP}^{1}$'s. Given local coordinates
$(z_{1},...,z_{n})$, this acts as:%
\begin{equation}
\sigma:(z_{1},...,z_{n})\mapsto(z_{\sigma(1)},...,z_{\sigma(n)}),
\end{equation}
where $\sigma$ is some permutation on $n$ letters.

Let us now turn to some local model considerations. Our starting point is the
standard one in local F-theory constructions, namely we begin with a K\"ahler surface
$S$ wrapped by a stack of seven-branes. In the local model, the moduli space
of the spectral cover makes reference to the non-compact Calabi-Yau threefold
$K_{S}\rightarrow S$, where $K_{S}$ is the canonical bundle.

A particularly important special case is given by $S=\mathbb{CP}%
^{1}\times\mathbb{CP}^{1}$. We will be interested in generating new local
models by taking various quotients, using the procedure of Massey (see e.g.
\cite{Massey, Lawson}). In this case, we have two anti-podal maps which we
denote as $\tau_{i}$, and $\langle \sigma \rangle$ is isomorphic to $\mathbb{Z}_{2}$. We also
introduce the combination $\tau \equiv \tau_{1}\tau_{2}$. Observe that the $\tau_{i}$
and $\sigma$'s generate the dihedral group $D_{8}$, the symmetries of a
square. Letting $J=\left\langle \sigma\right\rangle $, $K=\left\langle
\tau_{1},\tau_{2}\right\rangle $, $H=\left\langle \sigma,\tau\right\rangle $,
and $G=\left\langle \sigma,\tau_{1},\tau_{2}\right\rangle $ we have:%
\begin{align}
\mathbb{CP}^{1}\times\mathbb{CP}^{1}/J  &  =\mathbb{CP}^{2}\\
\mathbb{CP}^{1}\times\mathbb{CP}^{1}/K  &  =\mathbb{RP}^{2} \times \mathbb{RP}^{2}\\
\mathbb{CP}^{1}\times\mathbb{CP}^{1}/H  &  =S^{4}\\
\mathbb{CP}^{1}\times\mathbb{CP}^{1}/G  &  =\mathbb{RP}^{4}\text{.}%
\end{align}
We remark that some of these group actions have fixed point loci,
but this simply serves to define a branched covering \cite{Andreotti1958}.

These quotients are ``topological'' in nature since quantities
such as the curvature invariants of the underlying four-manifold change.
As such, one must exercise some caution in using this to construct backgrounds
which solve the corresponding supergravity equations of motion. On the other
hand, generating a candidate Cayley four-form provides evidence that the
quotient does make sense.

To proceed further, we write down the local presentation of the holomorphic
three-form on $X=K_{S}\rightarrow S$. Introduce local coordinates
$[u_{i},v_{i}]$ for each $\mathbb{CP}^{1}$ factor, and let $n$ denote the
normal coordinate direction, as given by a section of $K_{S}^{-1}$. Then, the
holomorphic three-form of the Calabi-Yau is:%
\begin{equation}
\Omega_{X}=dz_{1}\wedge dz_{2}\wedge dn.
\end{equation}
The K\"ahler form is given in these local coordinates by:%
\begin{equation}
J_{X}=i\frac{dz_{1}\wedge d\overline{z_{1}}}{(1+z_{1}\overline{z_{1}})^{2}%
}+i\frac{dz_{2}\wedge d\overline{z_{2}}}{(1+z_{2}\overline{z_{2}})^{2}}%
+i\frac{dn\wedge d\overline{n}}{(1+z_{1}\overline{z_{1}})^{-2}(1+z_{2}%
\overline{z_{2}})^{-2}}+...,\label{JKahler}%
\end{equation}
where we have only displayed the \textquotedblleft diagonal
elements\textquotedblright.

Let us consider the action under the various symmetry group actions, where we
do not change the normal coordinate. We denote the corresponding symmetry
generators as $\sigma_{S}$ and $\tau_{S}^{(i)}$, in the obvious notation. The
action of $\sigma_{S}$ is:
\begin{equation}
\sigma_{S}:dz_{1}\wedge dz_{2}\wedge dn\mapsto dz_{2}\wedge dz_{1}\wedge
dn=-dz_{1}\wedge dz_{2}\wedge dn,
\end{equation}
which we summarize as:%
\begin{equation}
\sigma_{S}:\Omega_{X}\mapsto-\Omega_{X}.
\end{equation}
By similar considerations, we have:%
\begin{equation}
\sigma_{S}:J_{X}\mapsto J_{X}.
\end{equation}

Consider next the action of $\tau_{S}^{(i)}$. In this case, we must remember
that since $n$ is a section of $K_{S}^{-1}=O(2h_{1}+2h_{2})$ with $h_{1}$ and
$h_{2}$ the hyperplane classes of $\mathbb{CP}^{1}\times\mathbb{CP}^{1}$, $dn$
will also transform under a coordinate change such as $z_{1}\mapsto1/z_{1}$ as
$dn\mapsto(z_{1})^{2}dn$. We then determine:%
\begin{equation}
\tau_{S}^{(1)}:dz_{1}\wedge dz_{2}\wedge dn\mapsto-\left(  \frac{z_{1}%
}{\overline{z_{1}}}\right)  ^{2}d\overline{z_{1}}\wedge dz_{2}\wedge dn,
\end{equation}
and similar considerations apply for $\tau_{S}^{(2)}$. Clearly, if we want to
extend our group actions to the Calabi-Yau, we must consider a more general
transformation law for the normal coordinate. Another issue is that our
coordinate transformation picks out only one of the coordinates, leaving us
with a mixed holomorphic / anti-holomorphic structure.

To solve both problems, we propose to primarily focus on the subgroup $H$
generated by $\sigma$ and $\tau\equiv\tau_{1}\tau_{2}$. In this case, the
corresponding action on the holomorphic three-form is:%
\begin{equation}
\tau_{S}:dz_{1}\wedge dz_{2}\wedge dn\mapsto\left(  \frac{z_{1}}%
{\overline{z_{1}}}\right)  ^{2}\left(  \frac{z_{2}}{\overline{z_{2}}}\right)
^{2}d\overline{z_{1}}\wedge d\overline{z_{2}}\wedge dn.
\end{equation}
To get a sensible extension of the $\tau$ action, we propose to extend by
demanding:%
\begin{equation}
\tau_{X}:n\mapsto\left(  \overline{z_{1}}\right)  ^{2}\left(  \overline{z_{2}%
}\right)  ^{2}\overline{n},\text{ \ \ and \ \ }dn\mapsto\left(  \overline
{z_{1}}\right)  ^{2}\left(  \overline{z_{2}}\right)  ^{2}d\overline{n}.
\end{equation}
that is, we consider complex conjugation on the normal coordinate as well. In
the case of the extension of $\sigma$, we also introduce a sign flip:%
\begin{equation}
\sigma_{X}:n\mapsto-n.
\end{equation}
This produces a sensible action on the holomorphic three-form for all
generators of the subgroup $H$. Indeed, we have:%
\begin{align}
\sigma_{X}  &  :\Omega_{X}\mapsto\Omega_{X}\\
\tau_{X}  &  :\Omega_{X}\mapsto\overline{\Omega_{X}}\text{.}%
\end{align}

Consider next the group action on the K\"ahler form. For $\sigma_{X}$, we
clearly have $\sigma_{X}(J_{X})=J_{X}$. Consider next the action of $\tau_{X}%
$. Plugging in our definitions, we have:%
\begin{equation}
\tau_{X}(J_{X})=i\frac{d\overline{z_{1}}\wedge dz_{1}}{(1+z_{1}\overline
{z_{1}})^{2}}+i\frac{d\overline{z_{2}}\wedge dz_{2}}{(1+z_{2}\overline{z_{2}%
})^{2}}+i\frac{d\overline{n}\wedge dn}{(1+z_{1}\overline{z_{1}})^{-2}%
(1+z_{2}\overline{z_{2}})^{-2}}+...=-J_{X}%
\end{equation}
so we see that the K\"ahler form $J_{X}$ flips sign. Summarizing the action on
the K\"ahler form of the local model, we have:%
\begin{align}
\sigma_{X} &  :J_{X}\mapsto J_{X}\\
\tau_{X} &  :J_{X}\mapsto-J_{X}\text{.}%
\end{align}

Based on this, we conclude that a quotient by the group $G$ will project out
the K\"ahler form. Not all is lost, however, because we can append to $X$ an
additional factor of $%
\mathbb{R}
$. Denote the corresponding space as $Y=X\times%
\mathbb{R}
$, where we let $n^{\prime}$ denote this direction in $%
\mathbb{R}
$. There is a natural extension of the aforementioned group actions, given by:%
\begin{align}
\sigma_{Y} &  :n^{\prime}\rightarrow n^{\prime}\\
\tau_{X} &  :n^{\prime}\rightarrow-n^{\prime}.
\end{align}
We now observe that the combination:%
\begin{equation}
\Phi_{(3)}=\operatorname{Re}(\Omega_{X})+J_{X}\wedge dn^{\prime},
\end{equation}
is invariant under both $\sigma_{Y}$ and $\sigma_{X}$. Here, the order one
coefficients are fixed by the demand that $\Phi_{(3)}$ is a calibration three-form.

As such, we can take the quotient, and arrive at a local model with reduced
supersmmetry. The holonomy group is not quite $G_{2}$, but for physical
applications it is \textquotedblleft close enough\textquotedblright%
.\footnote{As an additional comment, we remark that we are exploiting the fact
that we are dealing with a non-compact space, and the fact that we allow for
discrete group factors. For example, nothing stops us from starting with $X$ a
Calabi-Yau $(2m+1)$-fold, and from this starting point forming a
$(4m+3)$-dimensional manifold with a distinguished $2m+1$ form as given by
$\operatorname{Re}(\Omega)+J^{m}\wedge dn^{\prime}$, in the obvious notation.}
The upshot of this analysis is that we can start from a local
BHV\ construction, and then pass to a local $Spin(7)$ model.

Note that after performing such a quotient, our $\mathbb{CP}^1 \times \mathbb{CP}^1$ has become an
$S^4$. In cases where we have additional matter fields, we can expect that some matter curves of the original
$\mathbb{CP}^1 \times \mathbb{CP}^1$ model will also be identified.

Let us now ask whether these local considerations extend to actual
compactification geometries. Along these lines, we consider lifting $X$ to an
elliptically fibered Calabi-Yau fourfold. To be concrete, we take our elliptic
Calabi-Yau fourfold $Z\rightarrow B$ to have base $B=\mathbb{CP}^{1}%
\times\mathbb{CP}^{1}\times\mathbb{CP}^{1}$, with Weierstrass model:%
\begin{equation}
y^{2}=x^{3}+f_{8,8,8}x+g_{12,12,12}\text{ \ \ ,}%
\end{equation}
in the obvious notation. Since we are aiming for a quotient which produces a
$Spin(7)$ space, we seek out a quotient which preserves a candidate Cayley
four-form constructed from the holomorphic four-form and K\"ahler form of
the Calabi-Yau fourfold.

Based on the symmetries of the system, our plan will be to consider a single
combined antipodal action given by $\tau=\tau_{1}\tau_{2}\tau_{3}$. To
maintain contact with our previous considerations, we opt to consider pairwise
permutations of the $\mathbb{CP}^{1}$ factors, as denoted by $\sigma_{ij}$, which ends up generating
all of the symmetric group on three letters. Observe that quotienting $(\mathbb{CP}^{1})^3$ by the symmetric group on three
letters results in $\mathbb{CP}^3$. The further action by $\tau$ (which commutes with these pairwise permutations) is then a six-manifold
$\mathbb{CP}^3 / \langle \tau \rangle$, with $\mathbb{CP}^3$ specifying a two-sheeted branched cover over it.

We would now like to understand whether we can extend this quotient to a Calabi-Yau fourfold.
At least locally, there is no issue. To see why, let (by abuse of notation)
$n$ denote the normal coordinate for $B$ inside $Z$. Then, the proposed action
on all of the holomorphic coordinates is as follows:%
\begin{align}
\tau & :(z_{1},z_{2},z_{3},n)\mapsto(-1/\overline{z_{1}},-1/\overline{z_{2}%
},-1/\overline{z_{3}},-\overline{z_{1}}^{2}\overline{z_{1}}^{2}\overline
{z_{1}}^{2}\overline{n})\\
\sigma_{12}  & :(z_{1},z_{2},z_{3},n)\mapsto(z_{2},z_{1},z_{3},-n)\\
\sigma_{23}  & :(z_{1},z_{2},z_{3},n)\mapsto(z_{1},z_{3},z_{2},-n)\\
\sigma_{13}  & :(z_{1},z_{2},z_{3},n)\mapsto(z_{3},z_{2},z_{1},-n).
\end{align}
To extend this in a way compatible with the local geometry, we then require
that $dx /y$, the meromorphic one-form of the Weierstrass model, transforms as:
\begin{align}
\tau & :dx/y\mapsto-\overline{z_{1}}^{2}\overline{z_{2}}^{2}\overline{z_{3}%
}^{2}d\overline{x}/d\overline{y}\\
\sigma_{ij}  & :dx/y\mapsto-dx/y.
\end{align}

For now, we assume that an appropriate $f$ and $g$ in the Weierstrass model
have been chosen so that such a symmetry is available. We now ask about
the action on the holomorphic four-form and K\"ahler form. The
holomorphic four-form is given in local coordinates as:%
\begin{equation}
\Omega_{Z}=dz_{1}\wedge dz_{2}\wedge dz_{3}\wedge\frac{dx}{y}%
\end{equation}
which transforms as:%
\begin{align}
\tau(\Omega_{Z})  & =\overline{\Omega_{Z}}\\
\sigma_{ij}(\Omega_{Z})  & =\Omega_{Z},
\end{align}
much as we had in the case of the local model.

The transformation on the K\"ahler form is a bit more challenging to track since
we do not know the explicit K\"ahler metric of the model. Nevertheless, at least
in a local patch we expect to get a fair approximation by considering the
non-compact geometry $\mathcal{O}(2h_{1}+2h_{2}+2h_{3})\rightarrow B$. The K\"ahler form
is then of Fubini-Study type:%
\begin{equation}
J_{Z}=i\frac{dz_{1}\wedge d\overline{z_{1}}}{(1+z_{1}\overline{z_{1}})^{2}%
}+i\frac{dz_{2}\wedge d\overline{z_{2}}}{(1+z_{2}\overline{z_{2}})^{2}}%
+i\frac{dz_{3}\wedge d\overline{z_{3}}}{(1+z_{3}\overline{z_{3}})^{2}}%
+i\frac{dn\wedge d\overline{n}}{(1+z_{1}\overline{z_{1}})^{-2}(1+z_{2}%
\overline{z_{2}})^{-2}(1+z_{3}\overline{z_{3}})^{-2}}+...,
\end{equation}
where again, we have suppressed the display of the \textquotedblleft
off-diagonal\textquotedblright\ elements, and we observe that much as we had
in our local Calabi-Yau threefold example, we have:%
\begin{equation}
\tau(J_{Z}) = - J_{Z} \,\,\, \text{and} \,\,\, \sigma_{ij}(J_{Z})  = + J_{Z}.
\end{equation}
We now observe that the following four-form is invariant under these combined
actions:%
\begin{equation}
\Lambda_{(4)}=\operatorname{Re}\left(  \Omega_{Z}\right)  +\frac{1}{2}%
J_{Z}\wedge J_{Z},
\end{equation}
where the order one coefficients are fixed by the demand that $\Lambda_{(4)}$
is a calibration four-form. So, provided we can specify suitable Weierstrass coefficients compatible with
our quotient, we can obtain a reasonable construction of 3d $\mathcal{N}=1$
backgrounds.

One way to ensure that the group action is a symmetry of the
geometry is to impose additional conditions on the Weierstrass coefficients.
Explicitly, writing out $f_{8,8,8}$ as a polynomial in the homogeneous
coordinates, we have:%
\begin{equation}
f_{8,8,8}=\underset{0\leq i,j,k\leq8}{\sum}f_{ijk}(u_{1})^{i}(v_{1}%
)^{8-i}(u_{2})^{j}(v_{2})^{8-j}(u_{3})^{k}(v_{3})^{8-k}.
\end{equation}
Under the proposed transformations, we have:%
\begin{align}
\tau(f_{8,8,8})  & =\underset{0\leq i,j,k\leq8}{\sum}\overline{f_{ijk}%
}(-\overline{v_{1}})^{i}(\overline{u_{1}})^{8-i}(-\overline{v_{2}}%
)^{j}(\overline{u_{2}})^{8-j}(-\overline{v_{3}})^{k}(\overline{u_{3}})^{8-k}\\
\sigma_{12}(f_{\text{symm}})  & =\underset{0\leq i,j,k\leq8}{\sum}%
f_{ijk}(u_{2})^{i}(v_{2})^{8-i}(u_{1})^{j}(v_{1})^{8-j}(u_{3})^{k}%
(v_{3})^{8-k}\\
\sigma_{23}(f_{\text{symm}})  & =\underset{0\leq i,j,k\leq8}{\sum}%
f_{ijk}(u_{1})^{i}(v_{1})^{8-i}(u_{3})^{j}(v_{3})^{8-j}(u_{2})^{k}%
(v_{2})^{8-k}\\
\sigma_{13}(f_{\text{symm}})  & =\underset{0\leq i,j,k\leq8}{\sum}%
f_{ijk}(u_{3})^{i}(v_{3})^{8-i}(u_{2})^{j}(v_{2})^{8-j}(u_{1})^{k}%
(v_{1})^{8-k}.
\end{align}

As an example of how we can enforce a symmetry in this system, consider the
special case:%
\begin{equation}
f_{\text{special}}=f_{4}\left(  \left(  u_{1}^{2}v_{1}^{6}+v_{1}^{2}u_{1}%
^{6}\right)  \left(  u_{2}^{2}v_{2}^{6}+v_{2}^{2}u_{2}^{6}\right)  \left(
u_{3}^{2}v_{3}^{6}+v_{3}^{2}u_{3}^{6}\right)  \right)  ,
\end{equation}
with $f_{4}$ real. This leads to a collection of $SO(8)$ 7-branes on various
loci. There are also additional collisions of these 7-branes which lead to
conformal matter and \textquotedblleft conformal Yukawas\textquotedblright
\cite{DelZotto:2014hpa, Heckman:2014qba, Apruzzi:2018oge},
but these singularities can all be blown up. The
corresponding blown up geometries admit a natural extension of the group
action, so we conclude that these Calabi-Yau fourfolds provide a simple class
of examples exhibiting the main features. As another comment, we note that
enforcing the condition that the various coefficients in the original model
are real does not exclude any possible choice of singular fiber, see e.g.
\cite{Dierigl:2020wen} for further discussion on this point.

Clearly, there are many choices available, even with such symmetry
constraints.

\section{10d and 7d Spinor Conventions} \label{app:SPINOR}

In this Appendix we summarize our conventions for 10d spinors. We work in a mostly $+'s$ Lorentzian signature spacetime.
Our conventions follow those of Appendix B of reference \cite{Polchinski:1998rr}. Our 10d and 7d gamma matrices are denoted $\Gamma^{10d}_M$ and $\Gamma^{7d}_i$, where $M=0,\dots, 9$ and $\mu=0,\dots, 6$. These gamma matrices satisfy the usual anticommutation relations:
\begin{equation}
\{\Gamma^{10d}_{M}, \Gamma^{10d}_{N} \} = 2 \eta_{M N} \; \; \; \; \; \{\Gamma^{7d}_{\mu}, \Gamma^{7d}_{\nu} \} = 2 \eta_{\mu \nu}.
\end{equation}
We choose a 10d basis that decomposes in terms of the 7d one as
\al{  \Gamma^{10d}_{M=\mu} &= \sigma_1 \otimes \Gamma^{7d}_\mu \otimes \mathbf{1}\,,\\
\Gamma^{10d}_{M=7,8,9} &= \sigma_2 \otimes \mathbf{1} \otimes \sigma_{1,2,3}\,.
}
where we define the 7d gamma matrices as
\begin{equation}
\Gamma_0=i\sigma_2 \otimes \sigma_3 \otimes \mathbf{1}\; \; \; \; \; \Gamma_1=\sigma_1 \otimes \sigma_3 \otimes \mathbf{1} \; \; \; \; \; \Gamma_2=\sigma_3 \otimes \sigma_3 \otimes \mathbf{1}
\end{equation}
\begin{equation}
\Gamma_3=\mathbf{1} \otimes \sigma_1 \otimes \sigma_1 \; \; \; \; \; \Gamma_4=\mathbf{1} \otimes \sigma_1 \otimes \sigma_2 \; \; \; \; \; \Gamma_5=\mathbf{1} \otimes \sigma_1 \otimes \sigma_3 \; \; \; \; \; \Gamma_6=\mathbf{1} \otimes \sigma_2 \otimes \mathbf{1}.
\end{equation}
These conventions then determine that the 10d chirality matrix is $\Gamma^{10d}_{11}=\sigma_3\otimes \mathbf{1}\otimes \mathbf{1}$. A 10d Weyl spinor $\zeta$ of positive chirality that can be thus be written under the decomposition $\mathbb{R}^{9,1}=\mathbb{R}^{6,1}\times \mathbb{R}^3$ as
\al{ \zeta = \left(\begin{array}{c} 1 \\ 0\end{array}\right) \otimes \varepsilon_{7d} \otimes \eta_{3d}\,.
}.

To study the Majorana condition on this spinor, and thus make contact with the gaugino of 10 SYM, we need to understand the action of charge conjugation on these spinors. We will also see how this condition will lead to the symplectic Majorana condtion on 7d spinors after dimensional reduction. We start by introducing the 10d charge conjugation matrix, $B$, which has the property
\al{ B \, \Gamma^{10d}_M B^{-1} = (\Gamma^{10d}_M)^*\,.
}
Checking with the explicit representation given in \cite{Polchinski:1998rr} one can see that $B = B^\dag$ and $B B^* = 1$. The Majorana condition on $\zeta$ is
\al{ \zeta^* = B\, \zeta\,.
}
It is also convenient to define the matrix $C$ that satisfies
\al{ C \,\Gamma^{10d}_M C^{-1} = - (\Gamma^{10d}_M)^{T}\,.
}
Specifically $C = B \Gamma^{10d}_0$. One can choose the matrix $C$ as
\al{ C = \sigma_1 \otimes C_7 \otimes \sigma_2\,,
}
where $C_7$ satisfies $C_7\, \Gamma_\mu C_7^{-1} = - \Gamma_\mu^T$. This means that
\al{ B = -C \Gamma^{10d}_0 = - \mathbf{1} \otimes C_7 \Gamma_0 \otimes \sigma_2 = - \mathbf{1} \otimes B_7 \otimes \sigma_2\,.
}
and to see how this acts on 7d spinors, it is convenient to write
\al{ \varepsilon_{7d} \otimes \eta_{3d} = \left(\begin{array}{c}\varepsilon_1 \\ \varepsilon_2 \end{array}\right)\,.
}
Then the 10d Majorana condition reduces to
\al{ \varepsilon_1^* &= i B_7 \varepsilon_2\,,\\
\varepsilon_2^* &= -i B_7 \varepsilon_1\,,
}
which is nothing but the symplectic-Majorana condition in 7d. Note that $B_7 B_7^* = -1$.

%
%

\section{Quantum Mechanics of Euclidean M2-Branes}\label{app:SQM}
In this Appendix, we motivate the appearance of the twisted differential operator in the $Spin(7)$ superpotential and 7d Dirac equation twisted on $M_4$ by formulating the supersymmetric quantum mechanics (SQM) of the Euclidean M2-Branes in the $Spin(7)$ space is similar in spirit to references \cite{Pantev:2009de, Braun:2018vhk, Hubner:2020yde} for local $G_2$ models. We mainly focus on the case where we have the most concrete results, i.e. where $\Phi_{\mathrm{SD}}=0$. That being said, we believe this analysis is also helpful in understanding the zero modes of the operator:
\begin{equation}
\mathcal{D}_{\Phi_{\mathrm{SD}}} = \begin{pmatrix}
\Phi_{\mathrm{SD}}\times & D_{sig.} \\
D_{sig.} & -\Phi_{\mathrm{SD}}\wedge
\end{pmatrix},
\end{equation}
which we will refer to it as the $Spin(7)$ operator. Our aim will be to show that SQM picture determines the zero mode count when $\Phi_{\mathrm{SD}}=0$ (i.e. it will count the bulk zero modes).

Before diving into the details, we address the conceptual puzzle: what does it mean for these M2-branes to wrap calibrated 3-cycles in a $Spin(7)$ manifold, $X_8$, when there canonically is no such notion? The point is that their contributions from wrapping volume-minimizing 3-cycles can be still be given (at a formal level) by the spectrum of the above operator even though for $\Phi_{\mathrm{SD}}\neq 0$ on a general $M_4$, there is  not enough worldline supersymmetry to be able to extract much information.

 We now study the case of a trivial background, $\Phi_{\mathrm{SD}}=0$, starting with the standard Witten SQM on a target ($M_4$,$g$) with supercharges $Q=d$ and $Q=d^\dagger$ (i.e. no superpotential)\cite{10.4310/jdg/1214437492}, the Hilbert space is $\mathcal{H}=\oplus^{4}_{i=0}\Omega^i(M_4,\mathbb{C})$ and the space of supersymmetric ground states is simply $\oplus^{4}_{i=0}H^i(M_4,\mathbb{C})$. In a given $M_4$ coordinate patch, the theory has four bosons, $X^i$, four complex fermions $\psi^i$, and in particular has a $\mathbb{Z}_2$ symmetry we call $\tau$:\;$\psi^i\rightarrow \overline{\psi}^i$, $\overline{\psi}^i\rightarrow -\psi^i$. In terms of differential forms, $\tau$ is the signature operator in the above Hilbert space basis (recall $\overline{\psi}^i\leftrightarrow dx^i\wedge$) which acts as $\tau(\omega)=-i^{p(p-1)}*\omega$ on a $p$-form $\omega$, and note also that $\tau^2=(-1)^{F}$. Since $[\tau,\hat{H}]=0$, we can project this SQM onto the $+1$-eigenspace of $\tau$ leaving us with a Hilbert space $\mathcal{H}=\Omega^{0|4}_+\oplus \Omega^{1|3}_+\oplus \Omega^{2}_+$, where $\Omega_+^{1|3} \equiv \{ a-*a; a\in \Omega^1 \}$ and $\Omega^{0|4}_+ \equiv \{f+f\textnormal{Vol}_{M_4}; f\in \Omega^0\}$. This projected SQM is now the SQM of the M2-branes in a trivial background. From now on, we will often refer to $\Omega^{1|3}_+$ and $\Omega^{0|4}$ as $\Omega^1$ and $\Omega^0$ respectively.

Since, $[Q-\overline{Q},\tau]=0$ and $\{Q+\overline{Q},\tau\}=0$, we seem to have one remaining real supercharge $\mathcal{Q}_1 \equiv i\sqrt{2}(Q-\overline{Q})=i\sqrt{2}(d-d^{\dagger})$ with the Hamiltonian $\hat{H}=\frac{1}{2}\{ \mathcal{Q}_1,\mathcal{Q}_1^\dagger\}=\mathcal{Q}_1^2=\frac{1}{2}\Delta$. However, now the operator $\mathcal{Q}_2 \equiv (-1)^F\sqrt{2}(d-d^{\dagger})$ is an independent supercharge that commutes with $\tau$ so we have the same amount of supersymmetry as Witten's as SQM. Note also that $\mathcal{Q}_2^2=\frac{1}{2}\Delta$ and $\{\mathcal{Q}_1,\mathcal{Q}_2 \}=0$.

Consider next the ground states. These are specified by the equation $\mathcal{Q}_1 |\psi \rangle=0 $ for the $Spin(7)$ bulk zero modes. For example, given a state $(a-*a)\in\Omega^1$, we have:
\begin{equation}
\mathcal{Q}_1(a-*a)=0 \; \Leftrightarrow (d-d^\dagger)(a-*a)=0 \implies (da)^+=0, \; \; d^\dagger a=0
\end{equation}.
The $\mathbb{Z}$-grading associated to fermion number in Witten's SQM is broken to $\mathbb{Z}_2$ fermion parity in our SQM as evidenced by the fact that $(da)_{\mathrm{SD}}\in \Omega^2$ and $d^\dagger a\in \Omega^0$ do not have the same fermion number but do have the same fermion parity.

The action of the Spin(7) SQM will be the same as that of Witten's SQM on $(M_4,g_{M_4})$ where it is understood that the boundary conditions of the path integral will restrict to $\tau=+1$ states, and the allowed operators in correlation functions commute with $\tau$.
The Lagrangian is (see e.g. reference \cite{Hori:2003ic}):
\begin{equation}
L=\frac{1}{2}g_{ij}\dot{X}^i\dot{X}^j+\frac{i}{2}g_{ij}(\overline{\psi}^iD_t\psi^j-D_t\overline{\psi}^i\psi^j)-\frac{1}{2}R_{ijkl}\psi^i\overline{\psi}^j\psi^k\overline{\psi}^l
\end{equation}
where $D_t\psi^i \equiv \dot{\psi}^i+\Gamma^i_{jk}\dot{X}^j\psi^k$. The field transformations under $\mathcal{Q}_1$ are:
\begin{align}
& \delta X^i  =\epsilon(\psi^i+\overline{\psi}^i) \\
& \delta \psi^i =\epsilon(i\dot{X}^i-\Gamma^i_{jk}\overline{\psi}^j\psi^k) \\
& \delta \overline{\psi}^i=\epsilon(i\dot{X}^i+\Gamma^i_{jk}\overline{\psi}^j\psi^k).
\end{align}
We now briefly comment on the case when $\Phi_{\mathrm{SD}}\neq0$. In typically sigma-model SQM, we can add a superpotential $W$ in the usual way by deforming the supercharge to $-\frac{i}{\sqrt{2}}\mathcal{Q}_1=d+dW\wedge-d^{\dagger}-\iota_{dW}$ but this of course is not the same as turning on $\Phi_{\mathrm{SD}}$ in our $Spin(7)$ background since $dW$ is a one-form. In particular, if we were to rescale $\Phi_{\mathrm{SD}}\rightarrow t\Phi_{\mathrm{SD}}$ we expect perturbative (in $1/t$) ground states that are labeled by circles. Implementing this at the level of SQM seems challenging since $\mathcal{Q}$ has odd-fermion parity ($(-1)^F=-1$) but $\Phi_{\mathrm{SD}}$ is parity-even indicating that we have only one real supersymmetry generator. We leave a careful study of this case to future work.

\section{4d and 3d Reflections and Their Geometric Origins}\label{app:CHARGEREF}
Having already discussed the transformation $\mathsf{R}_i$ in 7d SYM in section \ref{sec:CRT}, our goal in this Appendix will be
to fill in some of the details of section \ref{sec:4D3D} in tracking how this reflection
(possibly composed with internal geometric symmetries) reduces to 4d and 3d reflection symmetries.

Starting with the 4d case, we first briefly review the topological twisting of 7d SYM compactified on a three-manifold $M_3$. Observe that the 7d Dirac spinor, decomposes as
\begin{equation}
\psi^{7d}_{\mathrm{Dirac}}=\psi^{4d}_{L}\otimes \psi^{3d}_{\mathrm{Dirac}} \; \; \mathrm{or} \; \; \psi^{4d}_{\mathrm{Maj}}\otimes \psi^{3d}_{\mathrm{Dirac}}
\end{equation}
where $\psi^{4d}_{L}$ is a left-handed Weyl fermion and $\psi^{4d}_{\mathrm{Maj.}}$ is a Majorana fermion. Note that in three Euclidean dimensions, the minimal spinor is Dirac due to the lack of a reality condition. As in 7d, we can think of this 3d Dirac spinor as a symplectic-Majorana fermion, and intuitively as filling out the representation $(\mathbf{2},\mathbf{2})$ of $SU(2)_{M_3}\times SU(2)_R$. After the topological twist, this becomes $\mathbf{3}\oplus \mathbf{1}$. The 7d spinor then decomposes into two pieces in 4d, the piece in the trivial representation of the twisted $M_3$ structure group, $\mathbf{1}$ is the 4d gaugino, and the piece that is a one-form along $M_3$ is the fermionic component of chiral multiplets. Note that the 4d scalars $a$ and $\varphi$ naturally arrange themselves into a the complex scalar components of chiral multiplets $S \equiv a+i\varphi$. The supercharge clearly takes values in $\mathbf{1}$ internally. Since we expect $\mathsf{R}^{7d}_1$ to be preserved in the compactification, we can appeal to the fact that  a reflection symmetry in 4d acts on the supercharge as (see reference \cite{Weinberg:2000cr})
$\mathsf{R}^{7d}_1 (Q^{4d}_L) \propto (Q^{4d}_L)^*$ and on chiral superfields as $\mathsf{R}^{7d}_1 (\boldsymbol{S})=\overline{\boldsymbol{S}}$. In terms of on-shell components $S$ and $\psi^{(S)}_L$ we have
\begin{align}
\mathsf{R}^{7d}_1(\psi^{(S)}_L)=i\sigma_1 C_{4d} (\psi^{(S)}_L)^* \\
\mathsf{R}^{7d}_1(S)=S^*.
\end{align}
The transformation of the scalar immediately agrees with expectations since $\varphi$ is a pseudoscalar while $a$ is a scalar in 7d SYM. The transformation of the Weyl spinor component follows from 7d as well but we leave the details to a footnote\footnote{Let's define a custom 7d gamma matrix basis as follows: $\Gamma_0=i\sigma_1\otimes \mathbf{1}\otimes \mathbf{1}$, $\Gamma_{1\leq i \leq 3}=\sigma_2\otimes \sigma_i\otimes \mathbf{1}$, $\Gamma_{4\leq i \leq 6}=\sigma_3\otimes \mathbf{1}\otimes \sigma_i$. Since the 7d gaugino is in the adjoint, we can employ the decomposition $\psi^{4d}_{\mathrm{Maj}}\otimes \psi^{3d}_{\mathrm{Dirac}}$ and simply act by $\Gamma_1$. Writing the 4d Majorana in terms of left-handed Weyl fermions $\psi^{4d}_{\mathrm{Maj}}=\begin{pmatrix}
           \psi_L \\
           C_{4d}\psi^*_L
        \end{pmatrix}$, acting by $\Gamma_1$ induces $\psi_L\rightarrow i\sigma_1 C_{4d}\psi^*_L$. After Higging, the adjoint decomposes into various (possibly complex) representations, where this transformation still stands. }

In the 3d case, we have already performed the topological twisting in subsection \ref{subsec:3dref}, as well as the action of $\mathsf{R}^{7d}_1$ on the 3d fields (see equation (\ref{eqn:parities})). What remains to be spelled out is the precise geometric origin of equation (\ref{eq:4d3dperp}), which we (essentially) reproduce here
\begin{equation}
\mathsf{R}_{i}^{3d} = \mathsf{R}_{i}^{4d} \mathsf{R}_{\bot}^{4d},
\end{equation}
along a compact $M_4$ that is more general that a product $M_3\times S^1$. Recall that the motivation for such a definition was so that the localized matter fermions transform under reflection as $\psi\rightarrow \pm \gamma_i \psi$, i.e. without a complex conjugation, which is more commonly met in 3d field theory literature. Our proposal is that we can define such a symmetry geometrically if $M_4$ possess an orientation reversing isotropy $\sigma_{M_4}$ so that
\begin{equation}\label{3drefnocc}
\mathsf{R}^{3d}_i \equiv \mathsf{R}^{7d}_i \circ \sigma_{M_4}.
\end{equation}
Looking first at what happens to the scalars, we see that $a$ and $\varphi$ are untouched leaving the combination that appears for localized matter $a+i\varphi$ without the unwanted conjugation. As for their fermionic partners, first recall that the 7d Dirac gaugino decomposes as
\begin{equation}
\psi^{7d}_{\mathrm{Dirac}}=\psi^{3d}_{\mathrm{Maj}}\otimes (\psi^{4d}_{L}+\psi^{4d}_{R})
\end{equation}
where, in the notation of section \ref{sec:4D3D}, the 3d Majorana fermions $\chi$ is a left-handed Weyl fermion on the internal $M_4$, while $\psi$ is right-handed. We can quickly find the transformation (up to a phase) by noting that $\sigma_{M_4}$ exchanges left-/right-handed fermions and recalling $\Gamma^{7d}_1\otimes \gamma^{M_4}_5$ which implies $\psi^{3d}_{\mathrm{Dirac}}=\chi+i\psi\rightarrow \gamma_1(\psi-i\chi)=-i \gamma_1\psi^{3d}_{\mathrm{Dirac}}$. The overall phase can be fixed to $+1$ a $\mathsf{Pin}^+$ action by working out the gamma matrix algebra in flat 7d space. Note that deriving the action on bulk modes involves knowing the action of $\sigma_{M_4}$ on various cohomology classes.

We have thankfully seen that there is no complex conjugation in the reflection transformations of 3d matter fields if we use the definition of Eq. (\ref{3drefnocc}). The puzzle remains however on how to derive the fact that localized modes with Hessians $(+,+,-)$ and $(-,-,+)$ should transform as $\psi^{3d}_{\mathrm{Dirac}}\rightarrow - \gamma_i \psi^{3d}_{\mathrm{Dirac}}$ and $\psi^{3d}_{\mathrm{Dirac}}\rightarrow + \gamma_i \psi^{3d}_{\mathrm{Dirac}}$ respectively. This was motivated in sections \ref{sec:4D3D} and \ref{sec:ANOMO} by several means, one being that a vector-like pair in 4d should not contribute to parity anomalies because one can write a parity conserving mass term. The resolution to this puzzle can be seen at the level of the superpotential. Under $\sigma_{M_4}$, the operator $D_{q\Phi_{SD}}\rightarrow -D_{q\Phi_{SD}}$ because $\Phi_{SD}$ is a pseudoscalar and the signature operator flips sign under orientation-reversal. This eigenvalue of this operator is the mass $M$ that couples vector-like pairs in the $Spin(7)$ superpotential
\begin{equation}
M\Phi^{(a)}_{\overline{\mathbf{R}}} \Phi^{(b)}_\mathbf{R}
\end{equation}
where $\Phi^{(a)}$ and $\Phi^{(b)}$ are complex multiplets in conjugate representations. The previous paragraph would suggest that these superfields both transform under $\mathsf{R}^{3d}_1$ as $\Phi^{(a,b)}\rightarrow +\Phi^{(a,b)}$. The important detail to note is that the above term is only parity-odd (i.e. conserves parity) if $M\rightarrow -M$ which as we have just shown is natural from the $Spin(7)$ perspective although quite strange from a bottom-up 3d perspective. The trick is to simply redefine $\Phi^{(a)}_{\mathbf{R}}$ and $M$ such that the former is parity-odd and the latter is parity-even\footnote{Said more pedantically, smooth orientable manifolds possess a notion of a constant global section of their determinant line bundles which for example when multiplying scalars turns them to pseudoscalars and vice-versa. In the product $M\Phi^{(a)}_{\mathbf{R}}$ we are simply redefining who that constant global section belongs to.} A similar statement can be made for the parity assignments of bulk mode states, as neutral bulk modes (before expanding in terms of KK-eigenstates)enter into the 3d $\mathcal{N}=1$ superpotential as
\begin{equation}
  \int_{M_4} \varphi \wedge da
\end{equation}
which again produced ``odd'' 3d masses due to the transformation $\sigma_{M_4}:\int_{M_4}\rightarrow \int_{\overline{M_4}}=-\int_{M_4}$. By transferring this assignment to the superfield $\varphi$, this reproduces the charge assignments used in the main text (see equation (\ref{eqn:finalreflections})).

\section{Pantev-Wijnholt system with $\mathfrak{so}(10)$ GUT matter}\label{app:SO10PW}

Here we look at how to engineer various (possibly chiral) matter representations in the PW system with gauge algebra $\mathfrak{so}(10)$ via an $E_8$ unfolding:
\begin{equation}
\mathfrak{e}_8  \supset \mathfrak{so}(10) \oplus \mathfrak{su}(4) \supset \mathfrak{so}(10)\oplus \mathfrak{u}(1)^3.
\end{equation}
The Higgs bundle $\Phi_{i=1,2,3}$ lives on a three-manifold $M_3$ satisfies $d\Phi_i=0$ and $d^\dagger\Phi_i=\sum_I v_{i,I} *\delta_{p_I}$ (provided $\sum_Iv_{i,I}=0$). We consider the solution that is defined by the three essentially unique harmonic functions (with singularities) $f_i$ that are solutions to the electrostatic problem $\Delta f=\sum_I v_{i,I} *\delta_{p_I}$ and define the Higgs fields as $df_i=\Phi_i$.

Now taking a look at the adjoint breaking
\begin{equation}
\textbf{248}\rightarrow \textbf{45}_{\vec{0}}+\sum^4_{i=1} \textbf{16}_i+\sum^4_{i=1}\overline{\textbf{16}}_i+\sum_{i<j}\textbf{10}_{ij}+\sum_{i< j}\textbf{1}_{ij}+3\times \textbf{1}_{\vec{0}}
\end{equation}
we see that we might have several species of localized matter in addition to the usual bulk modes $\textbf{24}_{\vec{0}}$ and $\textbf{1}_{\vec{0}}$. The subscript notation for the localized representations is a shorthand for the $U(1)$ charge vectors, let $\vec{q}_1=(1,0,0)$, $\vec{q}_{2,3}$ be defined accordingly, and $\vec{q}_4=(-1,-1,-1)$. Then our abelian charge assignments for the various $\mathfrak{so}(10)$ multiplets are $\textbf{16}_i=\textbf{16}_{\vec{q}_i}$, $\overline{\textbf{16}}_i=\overline{\textbf{16}}_{-\vec{q}_i}$,$\textbf{10}_{ij} \equiv \textbf{10}_{(\vec{q}_i+\vec{q}_j)}$ and $\textbf{1} \equiv \textbf{1}_{\pm(\vec{q}_i-\vec{q}_j)}$ . Based on the $U(1)$ charges, these various representations are localized around the following loci
\begin{equation}
\{\Phi_i=0\}\leftrightarrow \textbf{16}_i \; \text{or} \; \overline{\textbf{16}}_i
\end{equation}
\begin{equation}
\{\Phi_i+\Phi_j=0\} \leftrightarrow \textbf{10}_{ij}
\end{equation}
\begin{equation}
\{\Phi_i-\Phi_j=0\} \leftrightarrow \textbf{1}_{ij}
\end{equation}
where the representations on the left/right are localized at Morse index $\pm1$ zeros. Note that this has a nice spectral cover description in terms of intersections of various combinations of sheets. We can now generalize equations (\ref{PWmatter1}) and (\ref{PWmatter2})
to count the number of charged $SO(10)$ modes on the PW building block
\begin{align}
N_{\textbf{16}}=\sum_i n_i^-+b^1(M_3)-1, \; \; \; \; \;  N_{\overline{\textbf{16}}}=\sum_i n_i^++b^2(M_3)-1 \\
N_{\textbf{10}}=\sum_{i<j} (n^-_i+n^-_j)+\sum_{i<j} (n^+_i+n^+_j)+2b^1(M_3)-2
\end{align}
where $n^{\pm}_i$ denote the number of singular loci for the $i$th Higgs field with positive/negative residue with respect to the $U(1)_i$-charge.

\section{Parity Anomalies of 3d Gauge Theories}\label{app:BORDISM}

In this Appendix, we provide further details on the obstructions to placing 3d theories that are classically invariant under reflections and/or time-reversal, on a non-orientable manifold by reviewing the conditions that were spelled out in \cite{Witten:2016cio}. Namely, we will give a detailed characterization of the anomalies $\nu_{\mathsf{R}}$ (the gravitational-parity anomaly) and $\nu_{\mathsf{R}G}$ (mixed gauge-parity anomaly) discussed in the main text. In the following we will always focus on the case where $\mathsf{R}^2=+1$ (or alternatively $\mathsf T^2 = (-1)^F$ where $F$ is the fermion number), that is we consider the case of whether we are able to define the theory on a non-orientable manifold with a $\mathsf{Pin}^+$ structure.\footnote{The condition for a manifold $X$ to admit a $\mathsf {Pin}^+$ is that $w_2(TX)= 0$. To have a $\mathsf{Pin}^-$ structure the condition is that $w_2(TX)+w_1^2(TX) =0$.}

\subsection{Warmup: Gauge-Parity Anomaly}

As a warmup for our more technical discussion of anomalies via bordism groups, we discuss a convenient way of understanding gauge-parity anomalies. This is to simply start with the 3d theory of $N$ Majorana fermions, whose global symmetry group is $SO(N)$, from which various gauge-parity anomalies may be derived (at least when the gauge group is a subgroup of $SO(N)$),\footnote{To be precise the global symmetry group is $O(N)$ but in the following it will not make a difference.} and to check whether there is a gauge-parity anomaly according to \cite{Witten:2016cio} one should check whether any gauge bundle $V_{\mathcal R}$ on a four-manifold is `stably trivial', a condition that is equivalent to the vanishing of the Stiefel--Whitney classes
\al{w_1(V_\mathcal R) = w_2(V_\mathcal R)= w_4(V_\mathcal R) = 0\,.
}
Here $\mathcal R$ is the representation of the fermions under the gauge group $G$. The vanishing of the Stiefel--Whitney classes depends on the representation $\mathcal R$ and so the anomaly can be present or absent depending on the matter content.

In the following we will take an $\mathcal N=1$ gauge theory with $G = SU(N)$ and $N_f$ multiplets in the fundamental. In this case the matter representation $\mathcal R = \textbf{fund}^{\oplus N_f} \oplus \textbf{adj}$. We would like to take an $SU(N)$ bundle as a particular case of an $SO(2N)$ bundle and compute the Stiefel--Whitney classes. Luckily for any complex vector bundle $V$ one can show that (see for example exercise 14-B of \cite{milnor1974characteristic})
\al{ c_i (V) \text{ mod } 2 = w_{2i}(V)\,.
}
All odd degree Stiefel--Whitney classes vanish. Clearly for an $SU(N)$ bundle $c_1(V) = 0$ meaning that $w_2(V) = 0$. This means that we need to compute $c_2(V_{\mathcal R})$ for our example and take its mod 2 reduction to obtain $\nu_{\mathsf{R}SU(N)}$. In the following it is important to take the correct normalisation and we will follow the conventions of \cite{Ohmori:2014kda}. For an $SU(N_c)$ bundle $V_\rho$ we have that\footnote{As in \cite{Ohmori:2014kda} we absorbed a $2 \pi$ factor in the definition of the Yang--Mills field strength.}
\al{ c_2(V_\rho) = \frac{1}{2} \text{tr}_{\rho} \left(F \wedge F\right)\,.
}
In general it is convenient to convert the trace in a representation $\rho$ to a normalized trace $\text{Tr}$ with the property
\al{ \frac{1}{4}\int \text{Tr} \left(F \wedge F\right) \in \mathbb Z\,.
}
For $SU(N_c)$ the trace identities are
\al{ \text{tr}_{\mathbf{fund}} \left(F \wedge F \right) =  \frac{1}{2}\text{Tr} \left(F \wedge F \right)\,, \quad \text{tr}_{\mathbf{adj}} \left(F \wedge F \right) =  N_c\text{Tr} \left(F \wedge F \right)\,.
}
Therefore for our case
\al{ c_2(V_\mathcal R) = \frac{1}{4} \text{Tr}\left(F \wedge F \right) \left[N_f + 2 N_c\right]\,.
}
This means that the fourth Stiefel--Whitney class is
\al{ \int w_4(V_\mathcal R) = \left[N_f + 2N_c\right] m \text{ mod } 2\,,
}
where $m \in \mathbb Z$ is the instanton number. Clearly if we want the fourth Stiefel--Whitney class to be zero for any $SU(N_c)$ bundle we need to take $N_f + 2N_c = 0 \text{ mod } 2 $ or equivalently $N_f = 0 \text{ mod } 2$.

It is possible to generalize the previous discussion to other simply connected gauge groups for generic matter spectra. Given a theory with simply connected gauge group $G$ and fermions in the (possibly reducible) representation $\mathcal R$ absence of an anomaly can be written as
\al{ \nu_{\mathsf{R}G}\frac{1}{2}\int \text{tr}_{\mathcal R} (F \wedge F) = 0 \mod 2\,,
}
for any $G$-bundle. Using the index of the representation $\mathcal R$ it is possible to rephrase the condition as $2\mathbf T(\mathcal R) = 0 \mod 2$. The condition is equivalent to asking whether the number of fermionic zero modes in the background is even or odd. For example, taking $G= E_6$, and since $2 \mathbf T(\mathbf {27}) = 6$, there are no possible anomalies coming from a gauge bundle.
In the next section we will discuss a different method to detect possible anomalies which includes purely geometrical contributions and generalizes to other spacetime dimensions.

\subsection{Anomalies and Bordisms}

A different way to understand the presence of an anomaly when trying to define a theory on an unorientable manifold $X$ is to use the Dai--Freed theorem \cite{Dai:1994kq} to define the partition function of a spinor on such a manifold. The proper definition of the phase of the partition function of a spinor is $\exp(2 \pi i \eta_Y)$ where $Y$ is a manifold that satisfies $\partial Y = X$ and extends all structures (like Pin/Spin structures and gauge bundle) defined on $X$. The quantity $\eta_Y$ is the eta-invariant of the Dirac operator on $Y$. While this provides a sensible definition of the phase of the partition function of a spinor, this definition may depend on the choice of $Y$ or the extension chosen on $Y$ of structures of $X$. This is a reflection of an anomaly in the definition of the phase of the partition function. In order to detect such an anomaly, one can compare the phase of the partition function with two different extensions $Y_1$ and $Y_2$. Geometrically this can be accomplished as follows: since $Y_1$ and $Y_2$ have the same boundary $X$ it is possible to glue them along the boundary if one reverses the orientation of one of them, say $Y_2$. The gluing produces a closed manifold $\hat Y = Y_1 \sqcup \overline{Y_2} $ where $\overline{Y_2}$ is the orientation reversed $Y_2$. Given the properties of the eta-invariant $\eta_{Y_1 \sqcup Y_2} = \eta_{Y_1} +\eta_{Y_2}$ and that $\eta_{\overline Y} = - \eta_Y$ one finds that
\al{ \eta_{\hat Y} = \eta_{Y_1} - \eta_{Y_2}\,.
}
In order to ensure that the partition function is sensible it is necessary to require that $\eta_{Y} \in \mathbb Z$ for any closed four-dimensional manifold with a $\mathsf {Pin}^+$ structure. Another property of the eta invariant significantly simplifies the computations: it is invariant under bordisms between manifolds. Given two $d$-dimensional manifolds $X_1$ and $X_2$ we say that there is a bordism between them if there exists a $d+1$-dimensional manifold $Z$ such that $\partial Z = X_1 \sqcup \overline{X}_2$. If the manifolds $X_i$ carry additional structures (like (s)pin structures or gauge bundles) we require that they extend to $Z$. Therefore in the following we will need to check what are the allowed values of the eta invariant for different gauge groups to see if any anomaly for time-reversal is present when we define our 3d theories on a $\mathsf{Pin}^+$ manifold. In particular we will consider the bordism groups $\Omega_4^{\mathsf{Pin}^+}(M)$ where four-dimensional manifolds come equipped with a function to some space $M$. We will take $M = BG$ where $BG$ is the classifying space for a group $G$. The classifying space of a group $G$ is an infinite-dimensional space that has a principal $G$-bundle with total space $EG$, the so-called universal bundle, with the following property: any principal $G$-bundle over any manifold $X$ is the pull-back of $EG$ via some continuous function $f: X \rightarrow BG$. Therefore by considering manifolds equipped with maps to the classifying space of some group $G$ we end up considering all possible principal $G$-bundles over such manifolds. We shall call $\Omega_d^{\mathsf{Pin}^+}(\text{pt})$ the case without gauge group, that is the case where the function goes to a single point. This case classifies the possible anomalies that come only from geometry. In general there exists a map $\Phi : \Omega_d^{\mathsf{Pin}^+}(M)\rightarrow \Omega_d^{\mathsf{Pin}^+}(\text{pt})$. The map $\Phi$ deletes the details of the gauge bundle from equivalence classes in the bordism group $\Omega_d^{\mathsf{Pin}^+}(M)$. Given that this map is surjective it is possible to form a short exact sequence
\begin{equation}
\begin{tikzcd}
 0 \arrow[r]& \text{ker } \Phi \arrow[r]& \Omega^{\mathsf{Pin}^+}_d(M) \arrow[r,"\Phi"]& \Omega^{\mathsf{Pin}^+}_d(\text{pt}) \arrow[r]& 0\,.
 \end{tikzcd}
\end{equation}
We will call $\widetilde \Omega^{\mathsf{Pin}^+}_d(M) \equiv \text{ker } \Phi$ the reduced bordism group. Due to the existence of a map $\Psi : \Omega^{\mathsf{Pin}^+}_d(\text{pt}) \rightarrow \Omega^{\mathsf{Pin}^+}_d(M) $ which satisfies $\Phi \cdot \Psi = \mathbb I$ the above short exact sequence splits implying that $\Omega_d^{\mathsf{Pin}^+}(M)\simeq \Omega_d^{\mathsf{Pin}^+}(\text{pt})\oplus \widetilde \Omega_d^{\mathsf{Pin}^+}(M)$. The splitting of the bordism group into a purely geometric and a reduced part means that anomalies due to geometry can not be cured by introducing suitable gauge bundles and vice versa. For more details on this map and a proof of the splitting see for example Appendix A of \cite{Garcia-Etxebarria:2018ajm}. For the case at hand $ \Omega^{\mathsf{Pin}^+}_4(\text{pt}) = \mathbb Z_{16}$ with the generator being realized by a single Majorana fermion on $\mathbb{RP}^4$. This matches the computation of the eta invariant for this system done in \cite{Witten:2015aba}. This group classifies all possible gravitational-parity anomalies a 3d theory may have, i.e. $\nu_{\mathsf{R}}\in\Omega^{\mathsf{Pin}^+}_4(\text{pt})$.

We will discuss now the result of the computation of the reduced bordism groups for some classes of gauge groups which is what $\nu_{\mathsf{R}G}$ take values in. The computation uses the Atiyah--Hirzebruch spectral sequence which we choose not to review here, see for example \cite{Garcia-Etxebarria:2018ajm} for the necessary mathematical background to perform such computations.

For the case of $G=SU(N)$ the computation is rather easy and one obtains
\al{\widetilde \Omega^{\mathsf{Pin}^+}_4 (BSU(N)) &= \mathbb Z_2\,.
}
This confirms the computation done in the previous section where the anomaly is a mod 2 condition. The result for $N=3$ has already been obtained in \cite{Guo:2017xex} using the Adams spectral sequence. The
generator for this reduced bordism group is $S^4$ with one instanton.

The case of $G= U(1)$ is a bit more tricky and we obtain
\al{\widetilde \Omega^{\mathsf{Pin}^+}_4 (BU(1)) &= e(\mathbb Z_2,\mathbb Z_2)\,,
}
where $e(G,H)$ is an extension of $G$ by $H$. Specifically we have the short exact sequence
\begin{equation}
\begin{tikzcd}
 0 \arrow[r]& H \arrow[r]& e(G,H) \arrow[r]& G \arrow[r]& 0\,.
 \end{tikzcd}
\end{equation}
Such extensions are classified by the group $\text{Ext}(G,H)$ and for our purposes $\text{Ext}(\mathbb Z_m,\mathbb Z_n) = \mathbb Z_{\text{gcd}(m,n)}$. This implies that the bordism group can be either $\mathbb Z_2 \oplus \mathbb Z_2$ if the extension is trivial or $\mathbb Z_4$ if the extension is not trivial. To gain more information it would be necessary to compute the $\eta$-invariant for a fermion. Luckily for the case of $\mathbb{RP}^4$ with some $U(1)$ bundle this was done in \cite{Witten:2016cio} finding that the phase of the partition function is a root of 16 times a root of 4. This means that
\al{\widetilde \Omega^{\mathsf{Pin}^+}_4 (BU(1)) &= \mathbb Z_4\,.
}
The same result for $\widetilde \Omega^{\mathsf{Pin}^+}_4 (BU(1))$ was obtained in \cite{Davighi:2020kok} using Adams spectral sequence.

Finally we can ask what is the anomaly cancellation condition for any group $G$ on a $\mathsf {Pin}^+$ manifold. The computation of the bordism group $\Omega_4^{\mathsf{Pin}^+}(BG)$ is identical for any simply connected group giving
\al{ \widetilde \Omega^{\mathsf{Pin}^+}_4 (BG) &= \mathbb Z_2\,, \qquad \text{if } \pi_1(G) = 0\,.
}
The generator is again $S^4$ with an instanton.

For the case of non-simply connected gauge groups the results can vary due to the presence of additional gauge bundles allowed in the path integral. The computations become more complicated and we will simply quote some results in the literature:
\begin{itemize}
\item[-] For $G=SO(3)$ it was found in \cite{Wan:2018bns} that  $\widetilde \Omega_4^{\mathsf{Pin}^+}(BSO(3)) = \mathbb Z_4$. The invariant measuring the anomaly in this case is $q(w_2)$ where $w_2$ is the second Stiefel--Whitney class of the $SO(3)$ bundle and $q : H^2(X,\mathbb Z_2) \rightarrow H^4(X,\mathbb Z_4)$ is a quadratic refinement. Such refinements depend on the choice of a $\mathsf{Pin }^+$ structure; one choice of such refinement is the Pontryagin square. See footnote 7 of \cite{Wan:2018bns} for more details on such a refinement;
\item[-] For $G=PSU(3)$ it was found in \cite{Wan:2018bns} that $\widetilde \Omega_4^{\mathsf{Pin}^+}(BPSU(3)) = \mathbb Z_2$. The generator is the same as in the case of $SU(3)$.
\end{itemize}



\newpage

\bibliographystyle{utphys}
\bibliography{Spin7CRT}

\end{document}